\documentclass[aps,superscriptaddress,twocolumn,longbibliography]{revtex4-1}
\usepackage{dcolumn}
\usepackage{bm}
\usepackage{epsfig}
\usepackage{graphicx}
\usepackage{amssymb,amsmath,amsbsy,amsgen,amsfonts,amsthm,mathtools}    
\usepackage{mathrsfs}
\usepackage{latexsym}
\usepackage{array}
\usepackage{color}
\usepackage{amstext}
\allowdisplaybreaks[1]
\usepackage{txfonts}
\usepackage{pifont}
\usepackage{verbatim}
\usepackage[hidelinks]{hyperref}
\usepackage[T1]{fontenc} 
\usepackage{soul}
\usepackage{acronym}
\usepackage{breakcites}
 \usepackage{url}

\newcommand{\bra}[1]{\left\langle{#1}\right\vert}
\newcommand{\ket}[1]{\left\vert{#1}\right\rangle}
\newcommand{\ketbra}[2]{|#1\rangle \langle#2|}

\DeclareSymbolFont{symbols}{OMS}{cmsy}{m}{n}

\begin{document}
\title{Quantum Plasmonic Sensors}

\author{Changhyoup Lee}
\email{changhyoup.lee@gmail.com}
\affiliation{Institute of Theoretical Solid State Physics, Karlsruhe Institute of Technology, 76131 Karlsruhe, Germany}
\affiliation{Quantum Universe Center, Korea Institute for Advanced Study, Seoul 02455, Republic of Korea}

\author{Benjamin Lawrie}
\affiliation{Materials Science and Technology Division, Oak Ridge National Laboratory, Oak Ridge, Tennessee 37831, USA}

\author{Raphael Pooser}
\affiliation{Quantum Information Science Group, Oak Ridge National Laboratory, Oak Ridge, Tennessee 37831, USA}

\author{Kwang-Geol Lee}
\affiliation{Department of Physics, Hanyang University, Seoul 04763, Republic of Korea}

\author{Carsten Rockstuhl}
\affiliation{Institute of Theoretical Solid State Physics, Karlsruhe Institute of Technology, 76131 Karlsruhe, Germany}
\affiliation{Institute of Nanotechnology, Karlsruhe Institute of Technology, 76021 Karlsruhe, Germany}
\affiliation{Max Planck School of Photonics, Germany}

\author{Mark Tame}
\affiliation{Department of Physics, Stellenbosch University, Stellenbosch 7602, South Africa}
\date{\today}

\begin{abstract}
The extraordinary sensitivity of plasmonic sensors is well known in the optics and photonics community. These sensors exploit simultaneously the enhancement and the localization of electromagnetic fields close to the interface between a metal and a dielectric. This enables, for example, the design of integrated biochemical sensors at scales far below the diffraction limit. Despite their practical realization and successful commercialization, the sensitivity and associated precision of plasmonic sensors are starting to reach their fundamental classical limit given by quantum fluctuations of light -- known as the shot-noise limit. To improve the sensing performance of these sensors beyond the classical limit, quantum resources are increasingly being employed. This area of research has become known as `quantum plasmonic sensing' and it has experienced substantial activity in recent years for applications in chemical and biological sensing. This review aims to cover both plasmonic and quantum techniques for sensing, and shows how they have been merged to enhance the performance of plasmonic sensors beyond traditional methods. We discuss the general framework developed for quantum plasmonic sensing in recent years, covering the basic theory behind the advancements made, and describe the important works that made these advancements. We also describe several key works in detail, highlighting their motivation, the working principles behind them, and their future impact. The intention of the review is to set a foundation for a burgeoning field of research that is currently being explored out of intellectual curiosity and for a wide range of practical applications in biochemistry, medicine, and pharmaceutical research.   
\end{abstract}

\maketitle

\tableofcontents

\section{Introduction}

Optical sensors are used in many areas of science and technology -- notable examples include gyroscopes for navigation~\cite{Lai2020}, accelerometers for monitoring structural deformations~\cite{Na2020} and seismic wave activity~\cite{Rong2017}, electrochemical devices for measuring renewable energy storage~\cite{Lao2018}, and density sensors for analyzing seawater in climate change research~\cite{Uchida2019}. Of particular interest to the biological and chemical sciences are sensors that measure the presence and behavior of bacteria, viruses, and proteins~\cite{Leonard2003,Andreescu2004,Pejcic2006,Lazcka2007}. Here, highly sensitive optical sensors provide a deeper understanding of the biochemical processes involved in a given scenario in a non-invasive label-free manner, enabling, for example, drug development for improved health care~\cite{Bahadir2015}. Plasmonic systems based on metal nanostructures have long been used as a basis for optical sensors in this context~\cite{Homola2003,Hoa2007,Abdulhalim2008,Anker2008,Jackman2017}. The surface of a metal supports a quasiparticle known as a surface plasmon (SP)~\cite{Raether88,Maier2007}, which enables electromagnetic field confinement below the diffraction limit~\cite{Takahara97,Takahara09,Gramotnev2010} and allows enhanced sensing compared to conventional optical sensors~\cite{Homola1999a,Homola1999b,Homola2006,Homola2008,Lal2007,Stewart2008,Shalabney2011,Li2015}. Indeed, since the 1990s, surface plasmon resonance (SPR) sensors have been commercialized by several companies~\cite{Homola2008,Baird2001,ibis,reichertspr,biacore}; they are now a vital tool for studying biomolecular interactions for fundamental and applied sciences~\cite{Shankaran2007,Mejia-Salazar2018}.

Despite their successful commercialization, the high sensitivity and precision of plasmonic sensors are beginning to reach a fundamental limit given by quantum fluctuations of light -- known as \textit{the shot-noise limit} (SNL)~\cite{Piliarik2009,Wang2011}. In the past few years, improved sensitivity and precision beyond the SNL has been shown to be achievable for practical optical sensors using concepts from quantum metrology~\cite{Giovannetti2004,Paris2009,Giovannetti2011,Toth2014,Demkowicz2015,Taylor2016,Degen2017,Pirandola2018,Braun2018,Tan2019}. Several impressive experiments have demonstrated the basic working principles of quantum metrology using various types of quantum techniques in bulk optics~\cite{Kuzmich88,Fonseca99,Bouwmeester04,Mitchell04,Walther04}, integrated optics~\cite{Matthews09}, and biological systems~\cite{Taylor2016,Crespi12}. It is natural to ask whether these techniques can also be applied to plasmonic sensors to improve their performance. This question has been further motivated by recent advances in our understanding of the fundamental theory underlying quantum plasmonic systems, which has recently been extensively developed and experimentally studied~\cite{Tame2013,deLeon2012,torma2014strong,Brongersma2015,Pelton2015,Fitzgerald2016,Zhu2016,Marquier2017,Xu2018,Fernandez2018,Hughes2019,Slepyan2020,You2020a}. Advances in quantum plasmonics have enabled research groups to pursue the application of quantum metrology principles to plasmonic sensors in just the past few years. This area of research has become known as `quantum plasmonic sensing'. As the community moves further into this growing area, it is now an ideal time to look back to review and consolidate the past research achievements, as well as to look forward to identify future applications in the biochemical sciences and potential technologies beyond.

The interested reader can find a multitude of review articles on plasmonic sensors that have already been written from various perspectives over the last few decades~\cite{Homola2006, Homola1999a, Homola2003, Lal2007, Hoa2007, Abdulhalim2008, Anker2008, Homola2008, Shalabney2011, Mayer2011, Li2015, Mejia-Salazar2018}. This review, on the other hand, aims to comprehensively introduce for the first time the general framework developed for quantum plasmonic sensors in recent years and covers the basic theory behind the framework. While the goal is to present a broad overview of the research landscape, we also focus on several key works, where we highlight the motivation, provide details of the working principles, and discuss the future impact. We have endeavored to include all relevant works in this exciting and growing multidisciplinary area of research.

This review is aimed at researchers working in the broad fields of plasmonics, biochemistry, photonics, quantum optics, and quantum information science. In order for it to be accessible and beneficial to readers from all these fields, the philosophy has been to structure the review into three distinct sections: In section~\ref{sec:Plasmonic_sensors}, we introduce conventional plasmonic sensing, where we describe the different types of plasmonic sensors that exist in research and industry based on the principles of classical physics, along with their sensing performance. In section~\ref{sec:Quantum_sensors}, we introduce quantum sensing, where we describe the tools and concepts that have taken conventional sensing from the classical to quantum regime, which offers a significantly improved sensing performance. In section~\ref{sec:Quantum_plasmonic_sensors}, we then bring the concepts of sections~\ref{sec:Plasmonic_sensors} and \ref{sec:Quantum_sensors} together, and review recent work that has focused on integrating quantum sensing techniques with conventional plasmonic sensors. We describe how this approach has opened up a new route for improving the performance of plasmonic sensors.

\section{Plasmonic sensors}\label{sec:Plasmonic_sensors}

The sensitivity of a sensor, including that of a plasmonic sensor, is generally defined as 
\begin{align}
{\cal S}_y=\left\vert\frac{\text{d} y}{ \text{d} x }\right\vert,
\label{eq:sensitivity}
\end{align}
where~$x$ represents an implicit parameter to be estimated from a measurement of an explicit quantity~$y$~\cite{Homola2006}. The sensitivity can thus be understood as the extent to which the explicit parameter~$y$ changes for a given change of the implicit parameter~$x$. Such a relation between~$x$ and~$y$ strongly depends on the physical system used to encode the parameter to be estimated, but it also depends on the physical quantities that~$x$ and~$y$ represent. As will be shown in the following sections,
in a chosen plasmonic setting, various physical parameters~$y$ can be measured to estimate an individual parameter~$x$ at stake. An appropriate sensor, therefore, needs to be considered in a way such that the sensitivity expressed in eq~\ref{eq:sensitivity} takes rather large values. Plasmonic sensors, which exploit the interaction of light with the materials to be sensed at the interface between a metal and a dielectric in the sensing process, cope with this requirement very well. In addition to a high sensitivity, another appealing feature of plasmonic systems is the capability to confine the light to a spatial domain well below the diffraction limit, which is not possible with conventional photonic systems~\cite{Takahara97,Takahara09,Gramotnev2010}. Therefore, plasmonic sensors are known to enable compact sub-diffraction-limited sensing with high sensitivity. This also allows the measurement of tiny quantities of analytes, which is a feature that has been prompting their development now for quite some time.    

In this section, we explain the physical basis of plasmonic sensing with typical basic structures. We describe how plasmonic features can be designed to optimize the sensitivity. 
We also introduce a few plasmonic sensors that have attracted intensive interest from various scientific communities. The plasmonic sensors in this section are considered to be operated with classical light, but in following sections we will consider how quantum states of light and quantum measurements can be combined with such plasmonic sensors to further improve their performance.

\subsection{Surface plasmon polaritons}
\begin{figure}[!t]
\centering
\includegraphics[width=0.45\textwidth]{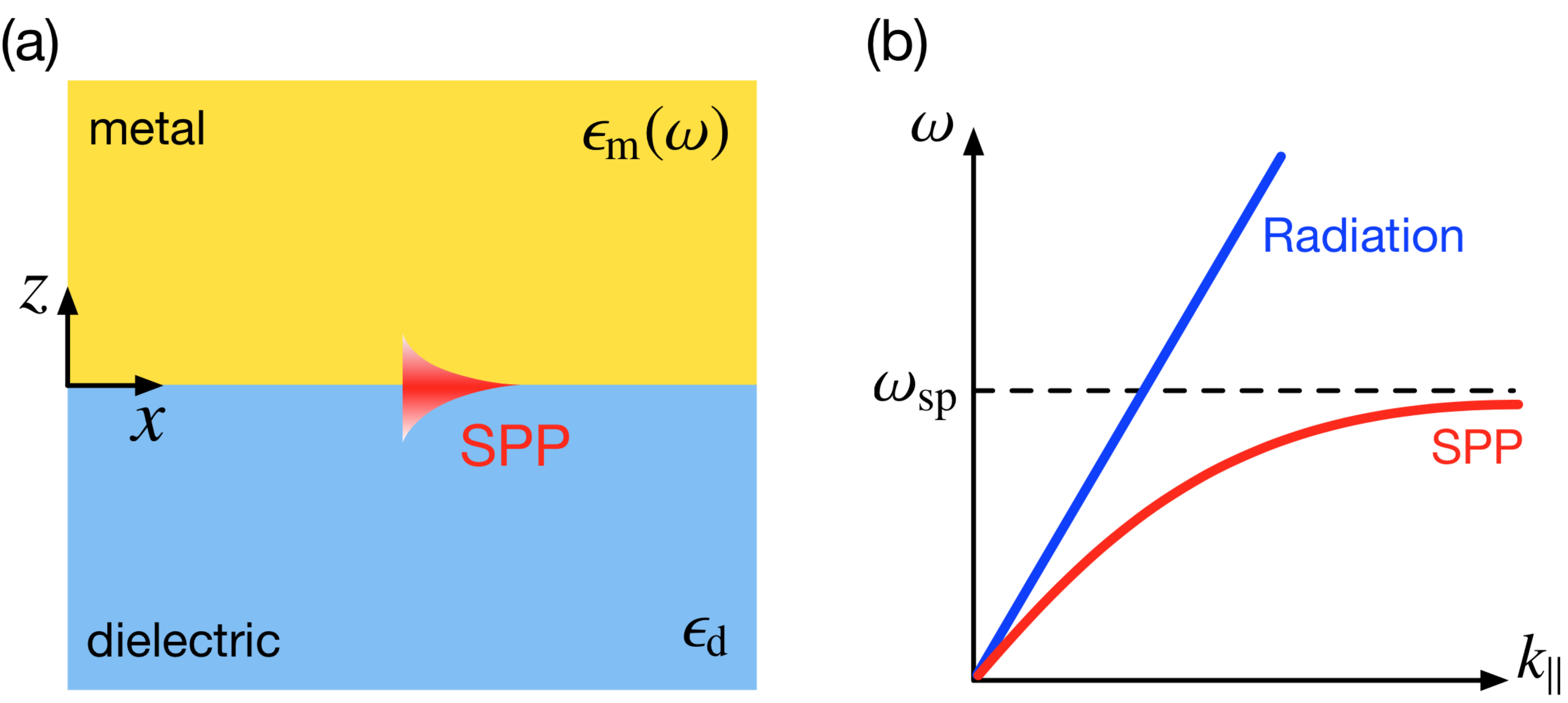}
\caption{
(a) Electric field profile of surface plasma waves, the quanta of which are called SPPs, at the interface between a bulk dielectric and a bulk metal. (b) Dispersion relation for a SPP at the interface between the metal and dielectric -- see eq~\ref{eq:SPP_dispersion} -- and a freely propagating field mode in the dielectric [$k_{||, \text{max}}= \sqrt{\epsilon_\text{d}} (\omega/c)$]. The frequency~$\omega_\text{sp}$~denotes the surface plasma frequency -- the frequency at which surface electrons collectively oscillate.}
\label{SPP}
\end{figure}

According to established notation, a plasmon is a charge density oscillation in a metal. A plasmon polariton is a hybrid excitation where an electromagnetic field is coupled to a plasmon. The additional term `surface' expresses that such an excitation is confined to the interface between a metal and a dielectric. To understand the appearance and the basic physical properties of surface plasmon polaritons (SPPs), we start by considering an interface between a bulk dielectric and a bulk metal, as shown in Figure~\ref{SPP}(a). Here, `bulk' simply implies that the size of the material is much larger than the wavelength of light in all directions. That is, we consider the dielectric and metal as semi-infinite half spaces. 

For a given material distribution in space, various electromagnetic modes can be found by solving Maxwell's equations. The `macroscopic' Maxwell equations in the time domain are written as~\cite{JacksonBook}
\begin{align}
\nabla\cdot{\boldsymbol{D}}({\boldsymbol{r}},t) &= \rho_\text{ext}({\boldsymbol{r}},t),\\
\nabla\cdot{\boldsymbol{B}}({\boldsymbol{r}},t) &= 0,\\
\nabla\times {\boldsymbol{E}}({\boldsymbol{r}},t) &= -\frac{\partial {\boldsymbol{B}}({\boldsymbol{r}},t)}{\partial t},\\
\nabla\times{\boldsymbol{H}}({\boldsymbol{r}},t) &= \boldsymbol{J}_\text{macr}({\boldsymbol{r}},t)+ \frac{\partial {\boldsymbol{D}}({\boldsymbol{r}},t)}{\partial t}.
\end{align}
Note that the electric field  $\boldsymbol{E}({\boldsymbol{r}},t)$ and magnetic induction $\boldsymbol{B}({\boldsymbol{r}},t)$ are experimentally observable fields, while the electric displacement field $\boldsymbol{D}({\boldsymbol{r}},t)$ and magnetic field $\boldsymbol{H}({\boldsymbol{r}},t)$ are auxiliary fields introduced to capture the response of the materials.
$\boldsymbol{J}_\text{macr}({\boldsymbol{r}},t)$ is the macroscopic current density and $\rho_\text{ext}({\boldsymbol{r}},t)$ is the external (free) charge density, respectively. The latter, however, vanishes when considering optical problems and it is assumed to be zero from now on. The macroscopic current density is separated into a conduction and a convection current. The conduction current will depend on the electric field while the convection current depends on the physical motion of matter. The latter is assumed to be vanishing in the systems we look at; there is only a conduction current.  

To solve the Maxwell equations, the relationship between the observable and auxiliary fields needs to be specified. This is done by constitutive relations that are given for the electric and magnetic fields in the time domain by~${\boldsymbol{D}}({\boldsymbol{r}},t)=\epsilon_{0}{\boldsymbol{E}}({\boldsymbol{r}},t)+{\boldsymbol{P}}({\boldsymbol{r}},t)$ and~${\boldsymbol{H}}({\boldsymbol{r}},t)= {\boldsymbol{B}}({\boldsymbol{r}},t)/\mu_{0} -{\boldsymbol{M}}({\boldsymbol{r}},t)$. The electric polarization ${\boldsymbol{P}}({\boldsymbol{r}},t)$ and the magnetization ${\boldsymbol{M}}({\boldsymbol{r}},t)$ can depend, in general, on both the electric and magnetic field. In the time domain and while restricting ourselves to a linear system governed by response theory, the dependency of the polarization (magnetization) can usually be expressed by a convolution of the electric (magnetic) field with some material specific response function.  

In the context of this review it is fully sufficient to consider isotropic, homogeneous, non-magnetic materials [$\boldsymbol{M}=0$, i.e.,~${\boldsymbol{H}}({\boldsymbol{r}},t)={\boldsymbol{B}}({\boldsymbol{r}},t)/\mu_0$], and without electro-magnetic coupling. The latter property would lead to bi-anisotropic effects. This boils down to considering a polarization that depends only on the electric field.

While the expression of the material properties in the time domain based on a convolution is cumbersome, the usual approach in the linear regime is to Fourier-transform all the fields into the frequency domain by making a time-harmonic Ansatz according to~$\boldsymbol{E}(\boldsymbol{r}, t)=\boldsymbol{E}(\boldsymbol{r},\omega)e^{-i\omega t}$. The convolution in the time domain will be a product in the frequency domain and the constitutive relations collapse to some algebraic equations. As the Fourier transformed polarization and the Fourier transformed conduction current density appear on an equal footing, these two terms are usually combined. The response of the individual bound charges (leading to a polarization) and the induced conduction currents in the bulk materials in response to an electromagnetic field are then characterized by the relative electric permittivity in the frequency domain, i.e.,~$\epsilon_\text{d}$ for the dielectric and~$\epsilon_\text{m}(\omega)$ for the metal, and we write~${\boldsymbol{D}}({\boldsymbol{r}},\omega)=\epsilon_0\epsilon_j(\omega){\boldsymbol{E}}({\boldsymbol{r}},\omega)$ in the respective material~$j$. Note that the permittivity of any material (except the vacuum) depends on the frequency, sometimes very well pronounced. Only for dielectric materials in the transparency window can the permittivity ~$\epsilon_\text{d}$ be assumed to be non-dispersive with a real positive value to a good approximation. For the metal, the electric permittivity can be described by the Drude model, written as~\cite{Maier2007}
\begin{align}
\epsilon_\text{m}(\omega)=1-\frac{\omega_\text{p}^{2}}{\omega^{2}+i\gamma \omega},
\label{eq:Drude}
\end{align}
where~$\gamma$ denotes a damping factor and~$\omega_\text{p}$ is the plasma frequency. We thus have~$\epsilon_\text{m}(\omega)=\epsilon_\text{m}'(\omega)+i\epsilon_\text{m}''(\omega)$. One can see that when~$\gamma<\sqrt{\omega_\text{p}^{2}-\omega^{2}}$, the real part of the permittivity is negative, i.e.,~$\epsilon_\text{m}'(\omega)<0$, which applies to typical metals (e.g., gold and silver) at optical frequencies.
An extended Drude model can also be found for the metal, for example in ref~\citenum{Vial2005}, which fits the empirically measured data for the dispersion in the case of gold in the range of wavelength between 500~nm and 1~$\mu$m, as reported in ref~\citenum{Johnson1972}.

In the structure considered above, various electromagnetic modes can be found by considering the wave equation in the frequency domain,~$\nabla^2 \boldsymbol{E}(\boldsymbol{r},\omega)+\frac{\omega^2}{c^2}\epsilon(\boldsymbol{r},\omega)\boldsymbol{E}(\boldsymbol{r},\omega)=0$, that can be derived directly from the Maxwell equations~\cite{JacksonBook,Maier2007}. For a homogeneous space, i.e., $\epsilon(\boldsymbol{r},\omega)=\epsilon(\omega)$,
propagating modes exist in the bulk dielectric and exhibit a dispersion relation linking the wavenumber~$\vert \boldsymbol{k}\vert = k$ of the spatial part~$\boldsymbol{E}(\boldsymbol{r},\omega)={\boldsymbol{{\cal E}}}(\boldsymbol{r})e^{i \boldsymbol{k}\cdot\boldsymbol{r}}$ to the frequency~$\omega$ of the temporal part~$e^{-i\omega t}$, given by~$k=\frac{\omega}{c}\sqrt{\epsilon_\text{d}}$ [straight line in Figure~\ref{SPP}(b) as the permittivity of the dielectric is assumed to be non-dispersive], where~$c$ is the speed of light in vacuum,~$k=\sqrt{k_x^2+k_y^2+k_z^2}$ and the~$k_i$ are the components of the wavevector~$\boldsymbol{k}$ for the mode. In Figure~\ref{SPP}(b) we consider the particular case of the modes propagating in a direction set by the plane of the interface, i.e., the~$x$-$y$ plane, so that~$k_z=0$ and~$k_{||}=\sqrt{k_x^2+k_y^2}=k$. In the metallic half space, evanescent plane waves exist as elementary solutions of the wave equation in free space~\cite{Griffiths99}. 

To find the solution in the more complicated situation of the two semi-infinite half spaces, a usual approach would be to expand the solution in each of the half spaces into the elementary solutions in free space and to match the amplitudes at the interfaces. This requires the specification of the necessary interface conditions. They can be written in the frequency domain as~\cite{JacksonBook}
\begin{align}
\boldsymbol{n}\times (\boldsymbol{E}_\text{m}-\boldsymbol{E}_\text{d})&=0,\label{eq:BC_E}\\
\boldsymbol{n}\cdot (\boldsymbol{B}_\text{m}-\boldsymbol{B}_\text{d})&=0,\label{eq:BC_B}\\
\boldsymbol{n}\cdot (\boldsymbol{D}_\text{m}-\boldsymbol{D}_\text{d})&=0,\label{eq:BC_D}\\
\boldsymbol{n}\times (\boldsymbol{H}_\text{m}-\boldsymbol{H}_\text{d})&=0,\label{eq:BC_H}
\end{align}
where~$\boldsymbol{n}$ is the normal vector characterizing the interface. This implies that the tangential component of the electric and magnetic field are continuous and the normal component of the electric displacement and magnetic induction are continuous.

From such a procedure, propagating modes bound to the metal-dielectric interface as solutions to the wave equation can be found as well~\cite{Maier2007}. These are hybridized modes, in that they correspond to a coupling between the electromagnetic field and longitudinal plasma oscillations, i.e., density oscillations of electrons in the conduction band of the metal. The modes are called surface plasma waves [lower curve in Figure~\ref{SPP}(b)]. In the limit of large~$k_{||}$, the modes have a corresponding small wavelength and an electrostatic approach can be used. The modes can thus be obtained for the electric potential~$\Phi$ as a solution to the Laplace equation~$\nabla^{2} \Phi =0$ for the single interface geometry, leading to the relation~$\epsilon_\text{m}(\omega)+\epsilon_{d}=0$. Substituting the Drude model of eq~\ref{eq:Drude} into~$\epsilon_\text{m}(\omega)$ with $\gamma=0$ for simplicity, one finds the surface plasma frequency~$\omega_\text{sp}=\omega_\text{p} / \sqrt{1+\epsilon_\text{d}}$. The quanta of these surface plasma waves are called SPs~\cite{Maier2007} and were first predicted by Ritchie~\cite{Ritchie1957}. In the limit of small~$k_{||}$, the modes have a corresponding long wavelength, where charge is transported over a considerable distance during the plasma oscillation. The resulting current sets up additional electromagnetic fields that interact back again with the electrons during their oscillation, causing retardation. In this regime, the wave equation needs to be used to obtain solutions for the modes. The quanta of these surface plasma waves are called SPPs and contain SPs in the limit~$k_{||}\to \infty$. The word `polariton' emphasizes the joint interaction between the matter part of the excitation (the electron plasma oscillation, or plasmon) and the light part of the excitation (the electromagnetic field, or photon)~\cite{Tame2013}. In this section, we will use a classical description of SPPs, i.e., surface plasma waves, and show how they enable classical plasmonic sensing, but we keep the quantized name, SPP, as is regularly done in the literature. This also has the benefit that it anticipates their use in the quantum regime, which we cover in detail in section~\ref{sec:Quantum_plasmonic_sensors}.

\begin{figure*}[!t]
\centering
\includegraphics[width=0.85\textwidth]{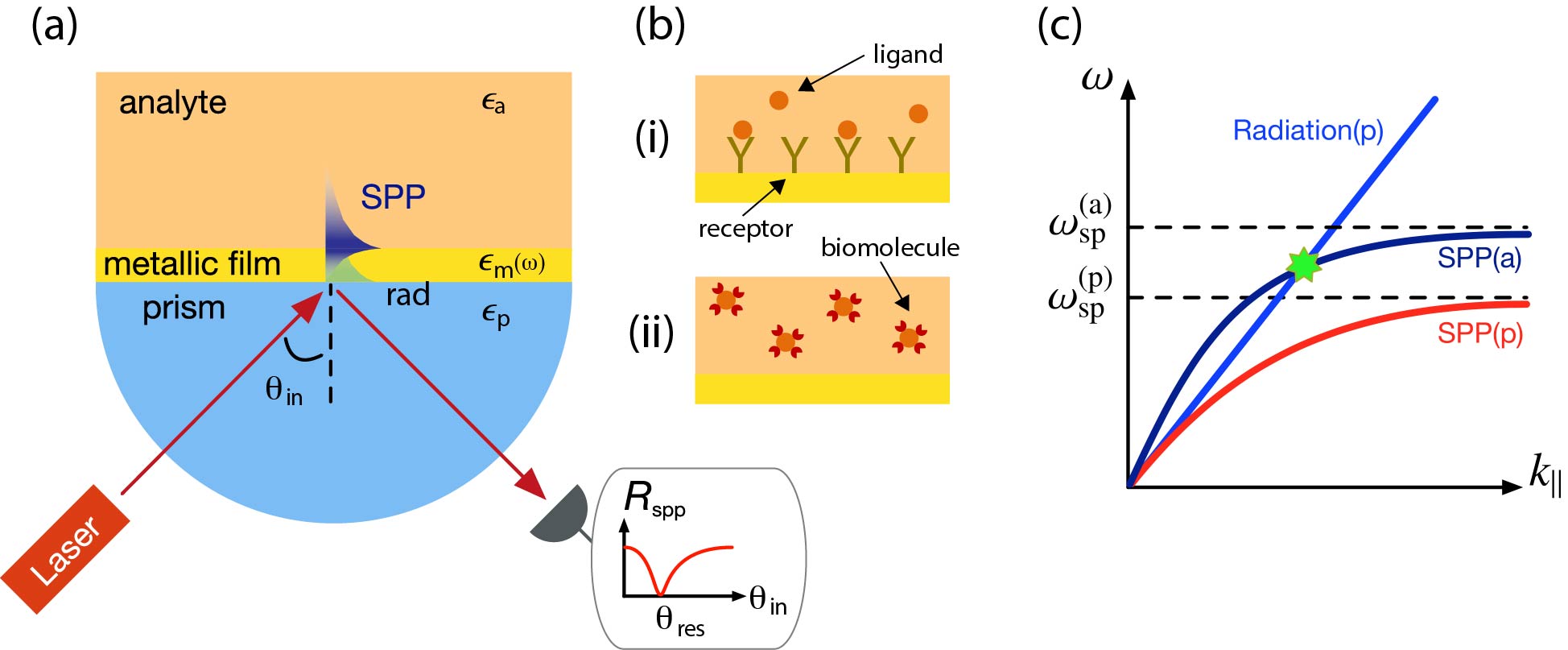}
\caption{
The Kretschmann configuration for plasmonic sensing. (a) The system is composed of three layers, where a thin metallic film is sandwiched by a prism with permittivity~$\epsilon_\text{p}$ and an analyte with permittivity~$\epsilon_\text{a}$. (b) Two examples of an analyte are shown. In (i) a ligand molecule binds to a receptor on the metal surface modifying the local permittivity~$\epsilon_\text{a}$. In (ii) a biomolecule is present near the surface resulting in a change of~$\epsilon_\text{a}$. (c) Dispersion relation for light and SPPs in the system.  The label `SPP(a)' and `SPP(p)' denote the SPP mode at the analyte-metal interface and prism-metal interface, respectively. The corresponding surface plasma frequencies are given as $\omega_\text{sp}^\text{(a)}$ and $\omega_\text{sp}^\text{(p)}$. The incident light in the prism region (straight line) cannot cross the dispersion curve for `SPP(p)', whereas the evanescent field on the opposite side of the prism (the same straight line) can cross the dispersion curve for `SPP(a)'. 
}
\label{Kretschmann}
\end{figure*}

From the wave equation, the electric field of the confined SPP mode at the interface can be written as~\cite{Maier2007} 
\begin{align}
{\boldsymbol{E}}_j({\boldsymbol{r}},t) = {\boldsymbol{{\cal E}}}_j({\boldsymbol{r}}) e^{i (k_\parallel x-\omega t)}e^{ -\kappa_j \vert z\vert}, 
\label{eq:SPP_electric}
\end{align}
where~${\boldsymbol{\cal E}}_j({\boldsymbol{r}})$ represents a vectorial field profile for the mode, the subscript~$j$ denotes either dielectric (d) or metal (m) and we are considering propagation in the~$x$ direction, i.e.,~$k_x=k_{||}$ and~$k_y=0$. The vectorial field profile,~${\boldsymbol{{\cal E}}}_j({\boldsymbol{r}})$, corresponds to a field with directionality in the~$x$ and~$z$ direction, i.e., a transverse magnetic field (in the~$y$ direction). This is due to the imaginary part of the conductivity associated with the Drude model being positive~\cite{Mikhailov2007}. The dispersion relation for the parallel-to-interface wavenumber is written as~\cite{Maier2007}
\begin{align}
k_{\parallel }^\text{SPP}(\omega) =\frac{\omega}{c}\sqrt{\frac{\epsilon_\text{d} \epsilon_\text{m}(\omega)}{\epsilon_\text{d}+\epsilon_\text{m}(\omega)}}.
\label{eq:SPP_dispersion}
\end{align}
This dispersion relation is valid for both real and complex~$\epsilon_\text{m}(\omega)$, i.e., for metals with and without attenuation. When~$\omega$ approaches the surface plasma frequency~$\omega_\text{sp}$, the denominator~$\epsilon_\text{d}+\epsilon_\text{m}(\omega)$ approaches zero, so the wavenumber~$k_{\parallel}^\text{SPP}(\omega)$ diverges. This is another indication that SPs can be understood as the limiting case of SPPs for large wavenumber.

In terms of field confinement, the electrons at the metal surface strongly bind the electromagnetic field of the SPP to the interface, resulting in a huge enhancement of the electromagnetic field near the surface. The field given in eq~\ref{eq:SPP_electric} in the~$z$ direction is generally `sub-wavelength' confined, as the field falls off as $e^{-\kappa_j |z|}$, where $\kappa_j=\sqrt{(k_{||}^\text{SPP})^2-k_0^2\epsilon_j}$ and~$k_0=\frac{\omega}{c}$. On the other hand, the confinement of the field in the~$x$-$y$ plane is no longer limited as it would be in a bulk material, i.e., limited by the usual three-dimensional diffraction limit~\cite{Takahara97,Takahara09,Tame2013}. The field given in eq~\ref{eq:SPP_electric} possesses a plane wave component with respect to a single wavevector $k_{||}$ and in the present discussion has an infinite spatial extent in the $y$ direction. While convenient for mathematical modeling, an actual SPP field will be laterally confined in the $y$ direction and for a fixed frequency $\omega$ it must be made from a sum of plane waves via Fourier synthesis, each wave with different $k_x$ and $k_y$ components. The corresponding spatial extent of such a confined field can be smaller than that allowed in the bulk dielectric. This leads to a SPP to be described as `sub-diffraction' confined on the surface, with the field obeying a more compact two-dimensional diffraction limit for the geometry~\cite{Takahara09}. Such a confinement in space on the surface (sub-diffraction) and perpendicular to the surface (sub-wavelength) is the inherent feature that enables SPP modes to be highly sensitive to the optical properties of the dielectric medium. 

When the metal-dielectric interface is illuminated with light from the dielectric region, the dispersion curve of the radiation mode,~$k^\text{in}$, does not cross the dispersion curve of the SPP,~$k_{||}^\text{SPP}$, for a given permittivity~$\epsilon_\text{d}$ in the dielectric for any frequency~$\omega$~[see Figure~\ref{SPP}(b) for the extreme case where~$k_{||}^\text{in}=k^\text{in}=\frac{\omega}{c}\sqrt{\epsilon_d}$]. In other words, the parallel-to-interface wavenumber~$k_{\parallel}^\text{in}(\omega)$ of the incident light can never be equal to~$k_{\parallel}^\text{SPP}(\omega)$ of eq~\ref{eq:SPP_dispersion} for a fixed frequency.
This means that any light directly incident on the metal cannot excite SPPs. A novel scheme is therefore required to satisfy the excitation condition~$k_{\parallel}^\text{in}(\omega)=k_{\parallel}^\text{SPP}(\omega)$. To this end, various schemes have been demonstrated for the excitation of SPPs at the interface between a bulk metal and a bulk dielectric, e.g., a prism~\cite{Otto1968,Kretschmann1968}, a grating~\cite{Raether88}, a randomly rough surface~\cite{Vinogradov1990,Salomon2002}, or a scanning near-field probe with sub-wavelength aperture~\cite{Bielefeldt1996}. The most widely used scheme is the prism setup, with two typical configurations: the Otto configuration~\cite{Otto1968} and the Kretschmann configuration~\cite{Kretschmann1968}. The former configuration requires dedicated techniques in practice, while the latter configuration has led to many successful applications in plasmonic sensing~\cite{Bahadir2015}, including commercialized versions. Therefore, we will focus on the Kretschmann configuration as a plasmonic sensing platform and elaborate on various assessments of its sensing performance in the next section.

\begin{figure*}[!t]
\centering
\includegraphics[width=1\textwidth]{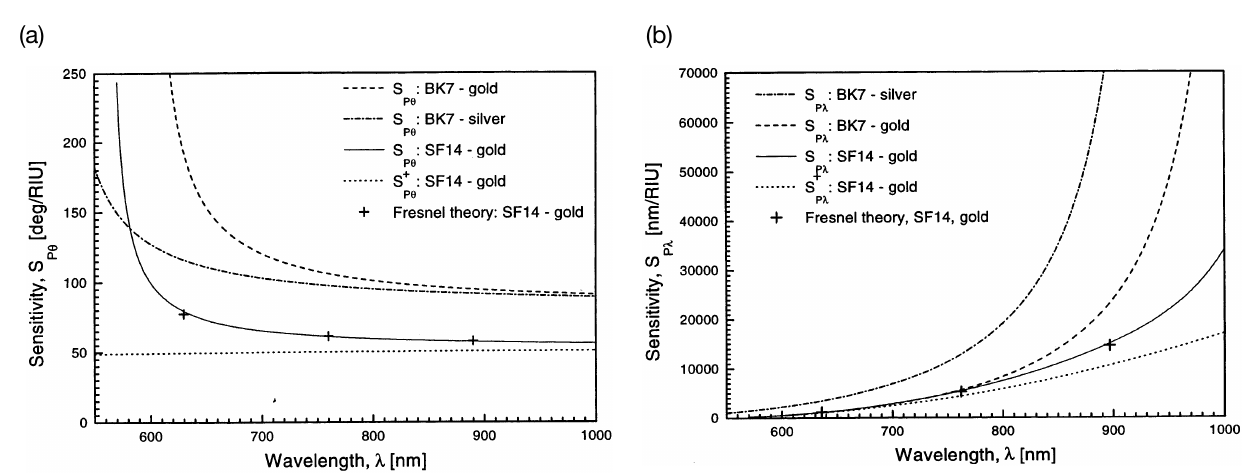}
\caption[Caption for LOF]{
The angular sensitivity in (a) and the spectral sensitivity in (b) are investigated as a function of the wavelength injected into the prism coupler. Example setups consider glass (SF14 or BK7), metal (gold and silver) and analyte ($n_\text{a}=1.32$), for which the sensitivities of eqs~\ref{eq:S_theta} and~\ref{eq:S_lambda} are evaluated (see solid, dashed, dot-dashed curves, whose details are provided in the legend) in (a) and (b), respectively, together with an asymptotic sensitivity in the limit of long wavelengths (dotted line) and sensitivities calculated numerically via Fresnel equations (crosses) for a structure with SF14-prism and~$50$~nm thick gold film. Reprinted with permission from ref~\citenum{Homola1999b}. Copyright 1999 Elsevier.
}
\label{SenActBChemHomola}
\end{figure*}

\subsection{Surface plasmon resonance sensing}\label{sec:SPR_Sensing}
SPPs at the metal-dielectric interface can only be excited when the excitation condition is satisfied, i.e., the mode matching condition~$k_{\parallel}^\text{in}(\omega) = k_{\parallel}^\text{SPP}(\omega)$. However, realizing the excitation condition is practically not easy because the dispersion relation of eq~\ref{eq:SPP_dispersion} is sensitive to the optical properties of the interfaced medium, i.e., the permittivity of the dielectric~$\epsilon_\text{d}$. Thus, when the aforementioned schemes are used to excite SPPs for a given structure, the system parameters need to be finely tuned and carefully stabilized in a controlled manner to satisfy the resonance condition. This might be undesirable and impractical from a general point of view, but paradoxically it is the desired feature for sensing in general. This is the basic principle of SPR sensing. 

We consider the most paradigmatic setup for the excitation of SPPs, the so-called Kretschmann configuration, that consists of three layers: the first layer is a prism ($\epsilon_{1}=\epsilon_\text{p}$), the second layer is a metal film [$\epsilon_{2}=\epsilon_\text{m}(\omega)$], and the third layer is a medium of an analyte ($\epsilon_{3}=\epsilon_\text{a}$), as shown in Figure~\ref{Kretschmann}(a). Examples of the analyte medium are given in Figure~\ref{Kretschmann}(b). Suppose that light is injected towards the metal interface from the prism region below, with an incident angle~$\theta_\text{in}$. The parallel-to-interface wavenumber of light impinging on the metal film is given as~$k_{\parallel}(\omega)=\sqrt{\epsilon_\text{p}}(\omega/c)\sin\theta_\text{in}$ and as mentioned, it is always smaller than the wavenumber of a SPP at the interface between the prism and the metal film for a fixed frequency~$\omega$, i.e., SPPs cannot be excited via direct illumination. However, when the incident angle is greater than the critical angle, this leads to total internal reflection and subsequently it causes the excitation of an evanescent field on the opposite side of the prism. The parallel-to-interface wavenumber of the evanescent field is still the same as the incident one, as it is a conserved quantity because of the translation invariance of the interface. When the thickness of the metal film is small, the decaying tale of the evanescent field can reach the interface between the metal film and the analyte region, where the wavenumber of the SPP is given by~$k_{\parallel}^\text{SPP}(\omega)$ with~$\epsilon_\text{d}=\epsilon_\text{a}$ being smaller than~$\epsilon_\text{p}$. Consequently, for a SPP at the metal-analyte interface, the resonant condition~$k_{\parallel}(\omega)=k_{\parallel}^\text{SPP}(\omega)$ can be met for a specific incidence angle called the resonance angle~$\theta_\text{res}$. This is shown in Figure~\ref{Kretschmann}(c) as a crossing point (star) between the dispersion curves. The SPP at the metal-analyte interface is thus excited via so-called evanescent field coupling. For a full resonant excitation of SPPs, the spatial mode of the evanescent field should have a significant overlap with that of the SPP. Furthermore, a finite beam width of the incoming light and the radiative and damped nature of the SPP mode need to be considered to include practical aspects involved in the conversion process~\cite{Vinogradov2018}.

When all the coupling conditions are fulfilled, the total internal reflection is attenuated, an effect known as attenuated total internal reflection (ATR). Observing that the reflectance drops down from near-unity to zero, one can indirectly certify that the incident light is converted into a SPP mode. To see this in theory, one can solve Maxwell's equations for the three-layer system and obtain the reflection coefficient written as~\cite{Raether88}
\begin{align}
r_\text{spp}=\frac{e^{ i 2 k_{2} d} r_{23}+ r_{12}}{e^{ i 2 k_{2} d} r_{23}r_{12} + 1},
\label{eq:rspp}
\end{align}
where $d$ is the thickness of the metal film, $r_{uv}=\left(\frac{k_{u}}{\epsilon_{u}} - \frac{k_v}{\epsilon_v}\right)\Big/\left(\frac{k_{u}}{\epsilon_{u}}+\frac{k_v}{\epsilon_v}\right)$ for~$u,v\in \{1,2,3 \}$,~$k_{u}=\sqrt{\epsilon_{u}}(\omega/c)[1-(\epsilon_1/\epsilon_u)\sin^{2}\theta_\text{in}]^{1/2}$ denotes the normal-to-surface component of the wave vector in the~$u$th layer and~$\epsilon_{u}$ is the respective permittivity. 
The resonant excitation of SPPs can be identified by a dip in the reflectance~$R_\text{spp}=\vert r_\text{spp}\vert^{2}$ measured in terms of the incident angle~$\theta_\text{in}$, the wavelength~$\lambda$, or frequency~$\omega$ [see inset of Figure~\ref{Kretschmann}(a)]. The reflectance dip is highly sensitive to the permittivity~$\epsilon_\text{a}$ of an analyte, so analysis of the reflected light enables the estimation of various kinds of implicit parameters that characterize the optical properties of an analyte, or its relevant effects. For example, SPR sensors can be used to characterize kinetic parameters~\cite{Malmqvist1993, Lundstrom1994, Karlsson1997, Scarano2010}, such as the equilibrium constant, dissociation constant, and association constant.  They can also be used for sensing of structural or material parameters~\cite{Pockrand1978, Peterlinz1996, Homola2006}, such as the thickness of adsorbed molecules and their refractive index. Interestingly, these examples can be understood by a single phenomenological and macroscopic parameter -- a refractive index ($n_\text{a}=\sqrt{\epsilon_\text{a}}$) that changes. Therefore, for the sake of simplicity, but without loss of generality, we will focus on refractive index sensing. Furthermore, specificity, i.e. the response of the sensor to changes in a specific type of environment, can be achieved in SPR sensors by coating the metal surface with an appropriate receptor~\cite{Homola2008}, as shown in Figure~\ref{Kretschmann}(b)(i). This means that a specific ligand that is desired to be sensed will bind to it and effectively change the refractive index of the medium above the metal.

\subsubsection{Angular interrogation}
When light is converted to a SPP in a prism setup such as the Kretschmann configuration, the reflectance is minimized, which ensures the resonant excitation of a SPP. By varying the incidence angle~$\theta_\text{in}$ for a fixed frequency~$\omega$ of light in the prism setup, one can find the reflectance dip at the resonant incident angle~$\theta_\text{res}$, at which the parallel-to-interface wavenumber of the exciting evanescent field is equal to that of the SPP. Mathematically we have
\begin{align}
k_0 n_\text{p} \sin\theta_\text{res} = k_0 \sqrt{\frac{n_\text{a}^{2} \epsilon_\text{m}'(\omega)}{n_\text{a}^{2}+\epsilon_\text{m}'(\omega)}},
\label{eq:Res_Cond}
\end{align}
where~$n_\text{a(p)}=\sqrt{\epsilon_\text{a(p)}}$ denotes the refractive index of the analyte~(prism). It is clear to see that the resonance angle~$\theta_\text{res}$ changes with the refractive index~$n_\text{a}$ of the analyte, i.e., the reflectance curve shifts when the refractive index~$n_\text{a}$ changes. The sensitivity of the resonance angle with respect to changes in~$n_\text{a}$ can be obtained from eq~\ref{eq:Res_Cond} and is written as~\cite{Homola1999b}
\begin{align}
{\cal S}_\theta
=\left\vert \frac{\text{d} \theta_\text{res}}{\text{d} n_\text{a}}\right\vert 
= \frac{\epsilon_\text{m}'(\omega)\sqrt{-\epsilon_\text{m}'}}{\left[\epsilon_\text{m}'(\omega)+n_\text{a}^{2}\right]\sqrt{\epsilon_\text{m}'(\omega)\left[n_\text{a}^{2}-n_\text{p}^{2}\right]-n_\text{p}^{2}n_\text{a}^{2}}}.
\label{eq:S_theta}
\end{align}
The angular sensitivity~${\cal S}_\theta$ is given in degrees per refractive index unit (RIU) and monotonically increases with decreasing wavelength [see Figure~\ref{SenActBChemHomola}(a)]. On the other hand, it diverges in the short wavelength regime because the sensitivity~${\cal S}_\theta$ becomes singular when~$\epsilon_\text{m}'(\omega)=n_\text{p}^{2}n_\text{a}^{2}/\left[n_\text{a}^{2}-n_\text{p}^{2}\right]$. This indicates that the sensitivity of the Kretschmann configuration using gold is better than that using silver in the typical range of wavelengths of interest in optics [see Figure~\ref{SenActBChemHomola}(a)] because the plasma frequency of gold is smaller than that of silver. Furthermore, a smaller contrast between~$n_\text{a}$ and~$n_\text{p}$ is helpful for increasing the angular sensitivity, so for example, using BK7-glass is better than using SF14-glass for a given analyte with~$n_\text{a}=1.32$. As can be seen from Figure~\ref{SenActBChemHomola}(a), typical sensitivities of a Kretschmann plasmonic sensor using the angular interrogation method are in the range $10$-$10^3$ degrees/RIU.

\subsubsection{Spectral interrogation}
The reflectance curve can also be measured as a function of wavelength (or frequency) of the incident light that is injected into the prism setup with a fixed incidence angle. In this scenario, the reflectance dip is obtained at a resonance wavelength~$\lambda_\text{res}$, which satisfies the resonant condition of eq~\ref{eq:Res_Cond}, and shifts with the change of the refractive index~$n_\text{a}$ of an analyte. The spectral sensitivity can be obtained in the same way as above and is written as~\cite{Homola1999b}
\begin{align}
{\cal S}_\lambda
=\left\vert \frac{\text{d} \lambda_\text{res}}{\text{d} n_\text{a}}\right\vert 
= \frac{[\epsilon_\text{m}'(\omega)]^{2}}{\frac{n_\text{a}^{3}}{2}\left\vert \frac{\text{d} \epsilon_\text{m}'(\omega)}{\text{d} \lambda_\text{res}}\right\vert+\left[\epsilon_\text{m}'(\omega)+n_\text{a}^{2}\right]\epsilon_\text{m}'(\omega)\frac{\text{d} n_\text{p}}{\text{d} \lambda_\text{res}}\frac{n_\text{a}}{n_\text{p}}}.
\label{eq:S_lambda}
\end{align}
Like the angular sensitivity~${\cal S}_\theta$, the spectral sensitivity~${\cal S}_\lambda$ exhibits a singularity but in the limit of long wavelengths, i.e., the spectral sensitivity~${\cal S}_\lambda$ monotonically increases with the wavelength [see Figure~\ref{SenActBChemHomola}(b)]. However, the spectral sensitivity~${\cal S}_\lambda$  increases with~$\vert \epsilon_\text{m}'(\omega)\vert$, so the Kretschmann configuration using silver is more sensitive than one using gold, which is contrary to the angular sensitivity~${\cal S}_\theta$. Here again, using BK7-glass is better than using SF14-glass for a given analyte with~$n_\text{a}=1.32$ since the smaller the contrast between~$n_\text{p}$ and~$n_\text{a}$ is, the more sensitive the Kretschmann configuration setup is to the change of~$n_\text{a}$. As can be seen from Figure~\ref{SenActBChemHomola}(b), typical sensitivities of a Kretschmann plasmonic sensor using the spectral interrogation method are in the range $10^3$-$10^5$ nm/RIU.

\subsubsection{Limit of detection}\label{sec:Limit_of_detection}
The higher the sensitivity~${\cal S}_y$ of a sensor, the more sensitively an explicit observable parameter~$y$ changes with respect to the change of an implicit parameter~$x$ of an analyte. However, the noise inevitably involved in the measurement limits the minimum detectable range~$\Delta y_\text{min}$ of the parameter~$y$ being measured, even though the sensor may be highly sensitive. An overall figure of merit for sensing quality needs to take into account both the sensitivity~${\cal S}_{y}$ and the minimum detectable range~$\Delta y_\text{min}$, or equivalently the value of the noise level. This leads to the definition of a `limit of detection' (LOD)~\cite{Homola2006,Shalabney2011}, also known as the resolution~\cite{Piliarik2009} of the sensor, written as
\begin{align}
\text{LOD}=\frac{\Delta y_\text{min}}{S_y}.
\label{LOD}
\end{align}
Current state-of-the-art classical plasmonic sensors can achieve a minimum LOD of $\sim 10^{-6}$-$10^{-7}$ RIU, which covers a wide range of optical designs, interrogation methods, and operating
wavelengths~\cite{Piliarik2009}. Examples include the detection of nucleic acids identifying specific bacterial pathogens~\cite{Piliarik2009b} and the monitoring of protein multilayer systems~\cite{Stemmler1999}. Although we focus on the refractive index change for the sensitivity, the LOD may be given in other units better suited to the specific application, depending on the quantity $x$ being measured. Other commonly used units for the LOD in biochemical plasmonic sensors are ng/mL and nM, which correspond to measuring concentrations of substances. For consistency we will continue to use RIU, as it covers these cases also up to a functional relation. More generally, the above formula in eq~\ref{LOD} clearly shows that a good sensor requires the LOD to be reduced via the enhancement of the sensitivity~${\cal S}_y$ and a reduction of the noise~$\Delta y_\text{min}$. As will be discussed in section~\ref{sec:Quantum_plasmonic_sensors}, the LOD can be significantly reduced when the quantum resources described in section~\ref{sec:Quantum_sensors} are exploited together with plasmonic systems. In general, the sensitivity and specificity of the plasmonic sensor are not modified in the quantum scenario as they depend mainly on the physical setup; it is the reduction of the noise $\Delta y_\text{min}$, for a given integration time and intensity, that is the crucial feature quantum sensing provides. 
In the literature, LOD, resolution, and sensitivity have often been interchangeably used in the sense that a small LOD (or resolution) implies a large sensitivity for a fixed $\Delta y_\text{min}$ and vice versa. For a classical plasmonic sensor, the main and inevitable contribution to the noise stems from `shot noise', as will be discussed in detail in section~\ref{sec:Quantum_sensors}. On the other hand, for a quantum plasmonic sensor, the noise can be reduced below that of the shot noise.

The noise reduction offered by the use of quantum resources opens up a route to reducing the LOD below what is possible classically, which enables the precision of a plasmonic sensor to be improved so that it can detect smaller changes in the implicit parameter $x$ of an analyte. Such an improvement is beneficial in many applications of plasmonic sensors, for instance the detection of pathogens in small quantities in the early stages of a disease~\cite{Shankaran2007}, or the contamination of food and water by minute amounts of a substance~\cite{Leonard2003}.

\subsection{Intensity vs phase sensing}\label{sec:Intensity_vs_phase}
When SPPs are excited at the resonance angle in the Kretschmann configuration, the total internal reflection of light is maximally attenuated. Around the resonance a steep curve in the intensity of the reflected light is exhibited as the incident angle is varied, as shown in the inset to Figure~\ref{Kretschmann}(a). The change of the intensity is often analyzed to infer the refractive index of an analyte medium or other relevant properties. In angular interrogation, the shift in the resonance angle corresponding to the minimum of the reflected intensity is measured. Another intensity-based sensing approach, known as `intensity modulation', is to fix the incident angle at the steepest point of the reflection curve (off-resonance), i.e., the inflection point, and monitor the change in the reflected intensity as the entire resonance curve shifts. Similar LODs to angular interrogation can be achieved using this method~\cite{Piliarik2009}. On the other hand, the phase of the reflected light also changes across the resonance curve and abruptly at the resonance angle, which can be detected in an interferometric setup. This indicates that detecting the phase or phase change of the reflected light can be exploited as an alternative sensing method to intensity sensing. Such behavior has motivated the study of various phase-sensitive plasmonic or SPR sensors linked with an interferometric system to measure a relative phase shift~\cite{Huang2012, Kashif2014, Deng2017}, including multi-pass interferometers~\cite{Ho2007}, imaging interferometers~\cite{Xinglong2005, Yesilkoy2018}, Mach-Zehnder interferometers (MZIs)~\cite{Ho2002, Wu2004, Ho2003, Nemova2008, Gao2011}, and heterodyne interferometers~\cite{Wu2003, Kuo2011}.

\begin{figure*}[!t]
\centering
\includegraphics[width=0.8\textwidth]{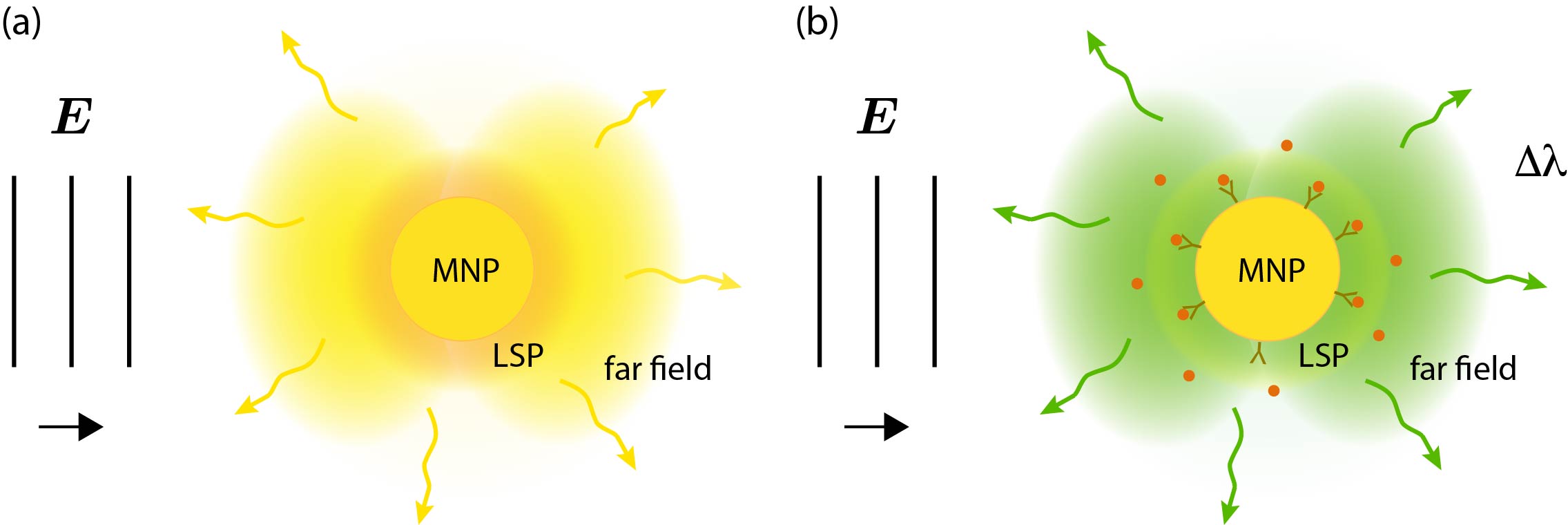}
\caption{
MNP supporting a LSP excitation when illuminated with an external electric field. (a) The MNP supports a LSP which is damped by internal ohmic loss and external radiation into the far field. (b) An analyte, consisting of a ligand binding to a receptor, changes the background permittivity of the MNP. This causes a change in the resonance position of the system with respect to the wavelength of the exciting field.
}
\label{nanoparticle}
\end{figure*}

Aside from the issue of noise in the measurement, a natural question arises: Which one is more sensitive, intensity sensing or phase sensing? There have been many studies comparing these two canonical sensing schemes. A number of works have shown that the sensitivity of phase sensing schemes is a few orders of magnitude higher than schemes based on the detection of the intensity change~\cite{Kabashin1997, Kabashin1998a, Kochergin1998a, Kochergin1998b, Kabashin1998b, Shen1998, Grigorenko1999, Nikitin1999, Kabashin1999, Nikitin2000, Xinglong2003, Sheridan2004, Wu2004, Wu2004b, Kuo2011}. Although contradictory observations may be found in the literature, e.g., ref~\citenum{Ran2006}, their paradoxical conclusion has been rebutted by other studies, for instance those mentioned in the dedicated discussion by Kabashin \textit{et al.}~\cite{kabashin2009phase}, who report a Kretschmann plasmonic sensor with a LOD of $10^{-8}$ using phase sensing with measurement noise included, representing an improvement in the LOD of two orders of magnitude compared to intensity sensing.

Regardless of which one is more sensitive, intensity and phase are the canonical conjugate variables for light. One can almost always decompose plasmonic sensors, or even more generally, photonic sensors into the two types: phase sensors and intensity sensors (often called amplitude sensors). Each type of plasmonic sensor works differently and consequently offers distinct functionalities and respective advantages~\cite{kabashin2009phase}. Such a distinction is also used to categorize quantum optical sensors, where useful quantum properties are different depending on the type of sensing being performed, as will be discussed in section~\ref{sec:Quantum_sensors}. Thus, for quantum plasmonic sensors, intensity and phase sensing will also be treated separately in section~\ref{sec:Quantum_plasmonic_sensors}. One may wish to have a combined version, but again they are complementary variables, so both parameters cannot be precisely estimated simultaneously due to the Heisenberg principle. This is also discussed in section~\ref{sec:Quantum_sensors}.

\subsection{Localized surface plasmon resonance sensing}
The SPPs discussed above are propagating surface waves and exist at a planar interface between a dielectric and a metal. On the other hand, non-propagating SPs can also be found in spatially localized metallic structures with a size comparable to or smaller than the wavelength of light. In this case, the conduction electrons oscillate coherently with a frequency that depends on the size and the shape of the metallic structure, the density and the effective mass of the electrons, and the medium in which the structure is embedded~\cite{Kelly2003}. The oscillation of the electron density is called a localized surface plasmon~(LSP)~\cite{Boardman1982, Kreibig1995}. Typical examples include metallic nanoparticles (MNPs)~\cite{Faraday1857,Mie1908}, where the excited LSPs with discrete resonance modes lead to sharp spectral absorption and scattering profiles, but also strong electromagnetic near-field enhancement in the proximity of metallic structures. Since LSPs can be excited in metallic structures with a size smaller than the diffraction limit, they have often been used for super-resolution imaging~\cite{Zhang2008,Kawata2009,Eghlidi2009,Rotenberg2014,Willets2017}.

As the simplest example, consider a single spherical MNP with a radius~$r_0$ that is embedded in a medium with a dielectric constant~$\epsilon_\text{d}$, as shown in Figure~\ref{nanoparticle}(a). When the MNP is illuminated by a single-mode electromagnetic field whose wavelength is much longer than the size of the MNP, i.e.,~$\lambda\gg r_0$, the electrostatic (or non-retarded) approximation can be applied. This approximation allows the problem to be solved using Laplace's equation~$\nabla^{2} \Phi(\boldsymbol{r}) =0$ for the spatial dependent electric potential~$\Phi(\boldsymbol{r})$. This is known as the quasi-static approximation as the spatial dependency of the field is governed by the same equation as in the electrostatic case but the fields continue to oscillate in time at the high frequencies corresponding to the visible part of the spectrum. The solution of Laplace's equation inside and outside the MNP reads
\begin{align}
\Phi_\text{in}(r,\theta,\phi)&=\sum_{l=0}^{\infty}\sum_{m=-l}^{l} a_{lm} r^{l} Y_{l}^{m}(\theta,\phi), ~~~~~~~0 \le r \le r_0, \\
\Phi_\text{out}(r,\theta,\phi)&=\sum_{l=0}^{\infty}\sum_{m=-l}^{l} b_{lm} r^{-(l+1)} Y_{l}^{m}(\theta,\phi), ~~~~r_0 \le r,
\end{align}
where~$\theta$ ($\phi$) represents the polar (azimuthal) angle in a spherical coordinate system, $Y_{l}^{m}(\theta,\phi)$ is a spherical harmonic function of degree~$l$ and order~$m$, and the $a_{l m}$ and $b_{l m}$ are amplitudes.
The interface conditions of eqs~\ref{eq:BC_E} and~\ref{eq:BC_D} impose the continuities of $\partial \Phi/\partial r$ for the tangential components and~$\epsilon_i\partial \Phi / \partial r$ for the normal components across the MNP's interface at~$r=r_0$. They lead to the explicit expression for the resonance frequencies of the LSPs, written as 
\begin{align}
\frac{\epsilon_\text{m}'(\omega)}{\epsilon_\text{d}}+\frac{l+1}{l}=0,
\label{eq:LSP_dispersion}
\end{align}
where~$l$ denotes the mode index of angular momentum of the LSP~\cite{Zayats2003}. Neglecting damping for simplicity, i.e., using the Drude permittivity of eq~\ref{eq:Drude} with~$\gamma=0$, one can obtain the resonance frequencies of LSPs, written as
\begin{align}
\omega_{l}=\omega_\text{p} \left[ \frac{l}{\epsilon_\text{d}(l+1)+l} \right]^{1/2}.
\label{eq:Res_Freq_LSP}
\end{align}
For small spheres, the dipolar excitation of LSPs is the most dominant, i.e.,~$l=1$, so that eq~\ref{eq:LSP_dispersion} gives the resonant condition for the dipolar LSP excitation as~$\epsilon_\text{m}'(\omega)=-2\epsilon_\text{d}$, leading to the resonant frequency~$\omega_1=\omega_\text{p}(1+2\epsilon_\text{d})^{-1/2}$ using eq~\ref{eq:Res_Freq_LSP}, known as the Fr\"ohlich condition. If the background permittivity~$\epsilon_\text{d}$ is modified, for instance by a ligand binding to a receptor, as shown in Figure~\ref{nanoparticle}(b), then the resonance position shifts by~$\Delta \omega$ (correspondingly by~$\Delta \lambda$ for wavelength). Thus, the resonance position can be used to sense a change in the environment of the MNP.

Such resonant features of the dipolar LSP excitation can also be seen in the expression of the scattering and absorption cross sections written as~\cite{Bohren1998}
\begin{align}
\sigma_\text{sca}(\omega)&=\frac{2\omega^{4}\epsilon_\text{d}^{2}V^{2}}{c^{4}}\frac{[\epsilon'_\text{m}(\omega)-\epsilon_\text{d}]^2+[\epsilon''_\text{m}(\omega)]^{2}}{[\epsilon_\text{m}'(\omega)+2\epsilon_\text{d}]^{2}+[\epsilon_\text{m}^{''}(\omega)]^{2}}, \label{eq:cross_section_sca}\\
\sigma_\text{abs}(\omega)&=\frac{9\omega \epsilon_\text{d}^{3/2}V}{c}\frac{\epsilon''_\text{m}(\omega)}{[\epsilon_\text{m}'(\omega)+2\epsilon_\text{d}]^{2}+[\epsilon_\text{m}^{''}(\omega)]^{2}},\label{eq:cross_section_abs}
\end{align}
where~$V$ is the particle volume. Note that LSPs can be resonantly excited at the frequencies~$\omega_l$ regardless of the wavevector, i.e., the resonance is independent of the illumination direction due to the full spherical symmetry of the MNP~\cite{Bohren1998}. This is in contrast to the excitation of SPPs, which are only excited when both the frequency and the wavenumber of the incident light equal those of the SPP for a given structure.

For larger spheres or in situations where the spheres are excited by some emitter in close proximity, higher-order modes with~$l\ge 2$ need to be taken into account in the description. If the size of the spheres gets even larger, such that the quasi-static approximation no longer holds, i.e., if the phase retardation of the incident field across the MNP can no longer be neglected, the cross sections caused by multipolar LSP excitations can be calculated using Mie theory in an analytical manner~\cite{Mie1908}. 
LSP excitations at elliptical particles can also be investigated using Gans theory~\cite{Gans1912}, providing an analytical solution where the aspect ratio of the ellipsoid plays a role. Of particular interest are spheroids, since they lead to a double resonance behavior, corresponding to electron oscillations along the major and minor axes~\cite{Bohren1998}. Behaviors of LSP excitation at nano-rods with thickness comparable to the skin depth of the metal can also be described and thus understood in an analytical way~\cite{Novotny2007}. More complex structures, such as multilayered spheres or ellipsoids with different materials, or non-spherical/non-ellipsoidal MNPs, can be considered, but one should use numerical electromagnetic methods such as finite-difference time-domain and finite-element methods to obtain approximate solutions, since an analytical solution for arbitrary structures cannot be found~\cite{Taflove1988,Barber1990,Kahnert2003,Zhao2008}. Complex structures are known to exhibit intriguing features and the potential for interesting optical applications in plasmonic sensing~\cite{Kerker1969, Schaubert1984, Martin1995, Mishchenko1996, Wriedt1998, Sosa2003, Kolokolova2003, Kahnert2010}. In section~\ref{sec:intmultimode} we will discuss quantum plasmonic sensing using LSPs at various types of nanostructures.

Interestingly, in the limit of a very large MNP, i.e.,~$l \rightarrow \infty$, with the electrostatic approximation ($r_0 \ll\lambda$) holding, the dispersion relation of eq~\ref{eq:LSP_dispersion} and the resonant frequency of eq~\ref{eq:Res_Freq_LSP} lead to the relations~${\epsilon_\text{m}'(\omega)=-\epsilon_\text{d}}$ and~$\omega_{l\rightarrow \infty}=\omega_\text{p}(1+\epsilon_\text{d})^{-1/2}$, respectively. These are equal to those for SPs at the interface between a dielectric and a metal. Such a consideration clearly identifies the inherently different nature of LSPs compared to SPPs. SPPs are a hybrid mode comprised of an electromagnetic field and electron oscillations, while LSPs are the excitation of electron oscillations for a given illumination by an electromagnetic field. In other words, SPP modes are bound solutions to the wave equation and once excited they exist without needing reference to the field that caused them, while LSP modes are responsive modes that are highly damped and can couple back into the far field, therefore they are inherently transient for a certain timescale when an external driving field is illuminating the MNP.

\subsubsection{Sensitivity and figure of merit} 

The features of LSPs described above can be exploited for sensing. In particular, the spectral sensitivity can be defined as~${\cal S}_{\lambda,\text{LSP}} =\left\vert\text{d} \lambda_\text{res}/\text{d} n_a\right\vert$ for refractive index sensing. Sensitivities of LSPs at single MNP of various shapes are on the order of $10^2$-$10^3$~nm/RIU~\cite{Kvasnicka2008}. 
However, due to the sharp resonance peaks of LSPs~\cite{Haes2004}, most LSP sensors employ a particular figure of merit, FOM, to quantify the ability of the sensor to resolve small refractive index changes, defined as~\cite{Sherry2005}
\begin{align}
\text{FOM}=\frac{{\cal S}_{\lambda,\text{LSP}}}{\Gamma_\lambda},
\label{FOM}
\end{align}
where~$\Gamma_\lambda$ is the resonance linewidth (or full width at half maximum). The FOM is mainly used to compare the sensing potential of various sensing schemes. From eq~\ref{FOM}, it can be seen that a good LSP sensor requires the FOM to be enhanced by reducing the linewidth and increasing the sensitivity.
\begin{figure}[!b]
\centering
\includegraphics[width=0.43\textwidth]{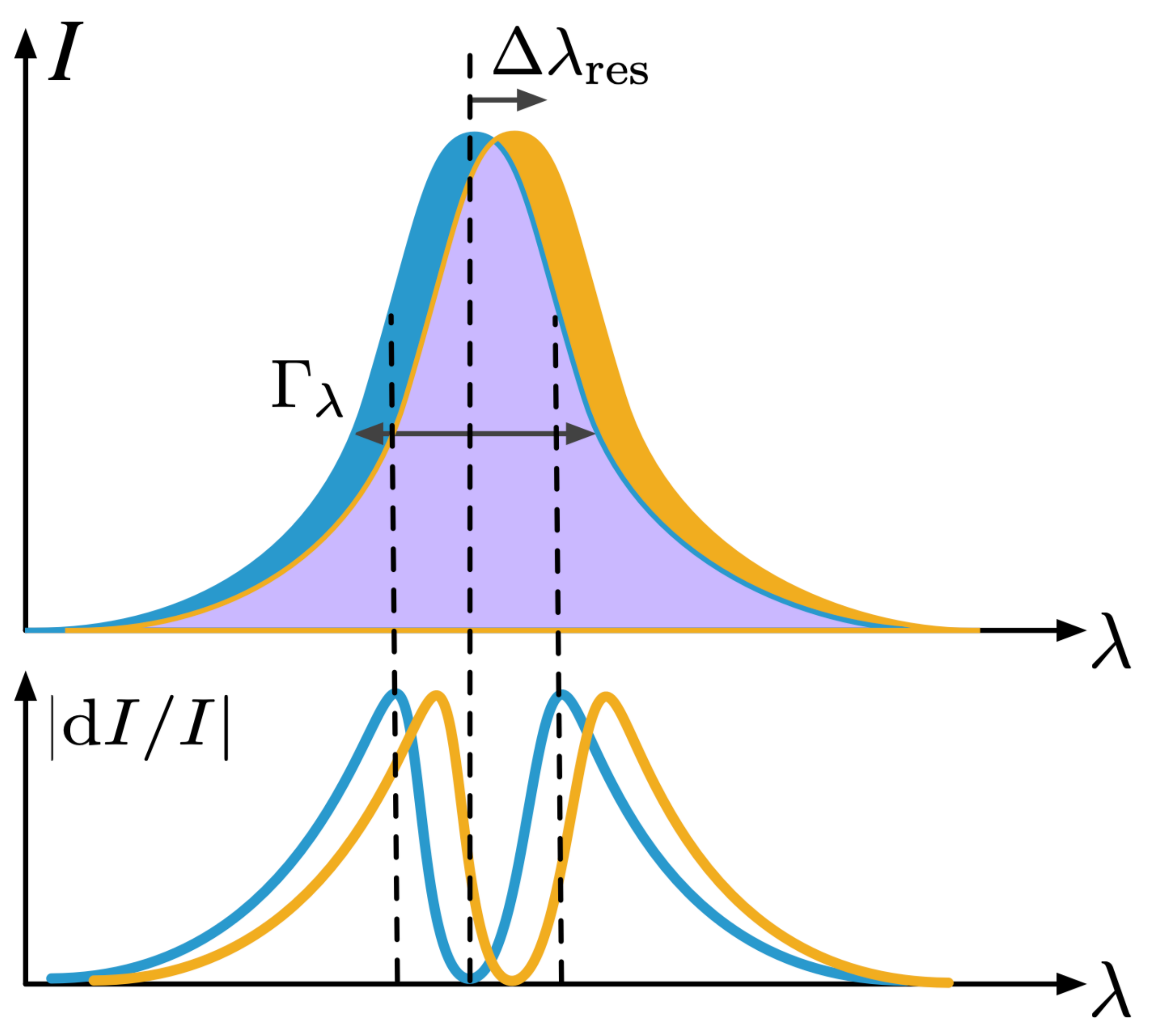}
\caption{(Top) The scattered light, i.e., the intensity $I$, from metallic nanostructures shifts with the refractive index change~$\Delta n_\text{a}$, consequently causing a shift of the resonance wavelength~$\Delta \lambda_\text{res}$. A minute change of the resonance wavelength can only be resolved by a sensor offering a high sensitivity of~$S_{\lambda,\text{LSP}}$. The FOM is defined as the ratio of~$S_{\lambda,\text{LSP}}$ to the full width at half maximum~$\Gamma_\lambda$, when the latter can be well defined. When this is not the case, an alternative version needs to be used, the~$\text{FOM}^{*}$, which is defined as the ratio of the relative intensity change~$\vert \text{d}I/I\vert$ (Bottom) to the refractive index change~$\text{d} n_\text{a}$ of the analyte. 
}
\label{LSPsensing}
\end{figure}

When the resonance spectrum cannot be modeled by a simple Lorentzian shape, e.g., in complex plasmonic structures such as metamaterials~\cite{Liu2010a}, the linewidth~$\Gamma_\lambda$ is ill defined and an alternative figure of merit needs to be defined. In this case, one can use the~$\text{FOM}^{*}$ defined as~\cite{Becker2010}
\begin{align}
\text{FOM}^{*}=\max_{\lambda}\left(\frac{\left\vert \frac{\text{d} I}{\text{d} n_\text{a}}\right\vert}{I}\right)
=\max_{\lambda}\left(\frac{{\cal S}_{\lambda,\text{LSP}}\times\left\vert \frac{\text{d} I}{\text{d} \lambda}\right\vert}{I}\right).
\label{eq:FOMstar}
\end{align}
The~$\text{FOM}^{*}$ accommodates the relative intensity change~$\vert \text{d}I/I\vert=|(I(\lambda+\Delta \lambda)-I(\lambda))/I(\lambda)|$ with respect to the change of the refractive index~$\text{d}n_\text{a}$ for the optimal wavelength~$\lambda$ that maximizes the quantity~$\vert \text{d} I / \text{d} n_\text{a}\vert/ I$, as illustrated in~Figure~\ref{LSPsensing}.

LSP sensors are an interesting approach for measuring refractive index changes, as they provide an opportunity to directly measure molecular binding events~\cite{Homola2003}. However, the FOM for LSP sensors using spherical MNPs is lower than that of SPP sensors by about an order of magnitude~\cite{Kvasnicka2008}, with a corresponding difference in the LOD. For a comparable FOM (or FOM$^{*}$), LSPs at non-spherical MNPs or as unit cells of metamaterials need to be considered~\cite{Chen2012}.

\subsubsection{Metamaterials}

Plasmonic metamaterial sensors are a recent advancement of LSP-based sensors that can lead to an improvement in sensing performance. These sensors consist of a periodic array of unit cells made from metallic nanostructures supporting LSPs, where each unit cell is smaller than the wavelength of the incident light that is used as the sensing probe~\cite{Cai2010}. The collective action of the unit cells gives rise to a macroscopic response of the periodic array, or `bulk' material. These artificially constructed materials are usually in the form of a 2d array, where they are called metasurfaces~\cite{Meinzer2014}, or in a 3d array as metamaterials\cite{Soukoulis2011}. Metamaterials can be made to have rather exotic properties that are not normally available in nature, e.g., negative refraction~\cite{Soukoulis2007,Shalaev2007}, optical cloaking~\cite{Chen2010b} and giant nonlinearity~\cite{Lapine2014,Litchinitser2018,Krasnok2018}. Recent work has started to exploit such unusual properties of plasmonic metamaterials in order to gain improvements in optical sensing functionality~\cite{Chen2012,Wang2019b}. Advantages compared to traditional SPP sensors are similar to those of LSP sensors, in that they offer the potential for improving operational ranges, as well as the integration of sensing components and ultimately miniaturization~\cite{Hoa2007,Caucheteur2015}.

Metamaterial sensors differ from ensemble-based LSP sensors in that for a given change in an analyte being sensed they aim to modify the macroscopic optical properties of the material. However, the distinction between the two is sometimes not well defined. Generally, metamaterial sensors have more flexibility than ensemble LSP sensors in that they allow for the possibility of coupled nanostructures within the unit cells and even the coupling between unit cells in order to influence the bulk response~\cite{Alu2011}. This can give advantages in sensing performance, as we briefly mention below. 

In a study by Kabashin \textit{et al.}~\cite{Kabashin2009}, an array of gold nanorods on a glass substrate was used to construct a plasmonic metamaterial biosensor. Surprisingly, the individual nanorod size deviations due to fabrication had little influence on the optical properties of the material, which was modelled well by effective medium theory. The robustness was due to the size of the nanorods and their spacing with respect to the wavelength of the incident probe field so that only average values of parameters are important. The overall function of the metamaterial sensor was essentially the same as a SPP sensor, with a prism excitation method used. A sensitivity of 32,000~nm/RIU was observed in the experiment, exceeding the sensitivity of LSP sensors by about 2 orders of magnitude. The associated FOM was found to be 330, one order of magnitude larger than that of SPP sensors and two orders larger than that of LSP sensors, leading to corresponding improvements in the LOD. The origin of the sensitivity improvement was found to be due to the increased surface area provided by the nanorods compared to a flat surface for SPP sensors, as well as plasmon-plasmon interactions between the nanorods of the unit cells of the metamaterial.

A different approach to plasmonic metamaterial sensors is the so-called `thin film' approach, which does not require prism coupling but analyzes the transmitted/reflected field~\cite{Liu2010a}. For example, in the work by Chang \textit{et al.}~\cite{Chang2010} the authors reported a thin film plasmonic metamaterial sensor based on unit cells made from split-ring resonators. They found that different resonances in the reflectance spectra could be used for the parallel sensing of biological interactions at different length scales. This work shows that not only can plasmonic metamaterials enhance the sensitivity, but they also provide novel capabilities in terms of parallel sensing. This strategy has recently been used for dual sensing of conformational and binding parameters~\cite{Cao2013}.

The main challenges for further improving the sensitivity of plasmonic metamaterial sensors are in finding the optimal way to reduce the noise associated with the adsorption and desorption of analytes, as well as improving the fabrication of accurate small size features for the nanostructures of the unit cells. More information on the current state of metamaterial sensors can be found in the reviews by Chen \textit{et al.}~\cite{Chen2012}, Wang \textit{et al.}~\cite{Wang2019b}, Tseng \textit{et al.}~\cite{Tseng2020} and Hassan \textit{et al.}~\cite{Hassan2020}, with details on the use of core-shell coatings, nanotubes and dielectric structures for the unit cells, in addition to hyperbolic metamaterials.

\subsubsection{Miniaturization}

While metamaterials provide a novel route for the miniaturization and integration of plasmonic sensors, several other methods exist. In this respect, it is important to note that the majority of current commercial plasmonic biosensors are still based on a traditional SPP sensing approach~\cite{Hoa2007,Caucheteur2015} and their designs do not differ significantly from the Kretschmann configuration. Current research is focused on developing platforms that can provide a more integrated, low-cost, reusable and sensitive biosensor. Fiber and waveguide-based SPP sensors are one such promising platform, for which an LOD of $10^{-6}$~RIU has been achieved using a polarization maintaining fiber~\cite{Piliarik2003}. However, integrating fiber sensors with other necessary components, such as microfluidic channels, is challenging due to the need for precise alignment and complex assembly~\cite{Niu2015}. Alternatives here include the use of a planar optical waveguide~\cite{Stocker2004}, distributed Bragg reflector~\cite{Lin2006}, MZI~\cite{Sepulveda2006}, or even gold coating the capillary tube used for the fluidics in order to enable it to operate as a waveguide itself~\cite{Chinowsky1999}.

Despite the obvious compactness of fiber and waveguide-based plasmonic sensors, the sensitivity and LOD of these integrated sensors realised so far is at best comparable to that of traditional SPP sensors~\cite{Hoa2007}. This has motivated researchers to also focus on the miniaturization of SPP sensing using low cost LEDs and diffractive mirrors~\cite{Ghosh1999,Thirstrup2004}, micro-prisms~\cite{Piliarik2003,Kurihara2004} and metamaterials with fibers~\cite{Consales2020}, D-shaped fiber optic SPR sensors~\cite{Caucheteur2015} and photonic crystal fiber plasmonic sensors~\cite{Li2020e}.

Progress in the miniaturization and integration of plasmonic sensors will benefit greatly from advancements in optoelectronics, integrated optics, surface functionalization, and microfluidic and nanofabrication techniques. Many of the sensors proposed in the literature have passed the proof-of-principle stage and are now being developed into commercial products. Further details of the current state-of-the-art and LOD values for compact plasmonic sensors can be found in the reviews by Hoa \textit{et al.}~\cite{Hoa2007} and Caucheteur \textit{et al.}~\cite{Caucheteur2015}, and can be compared with state-of-the-art compact photonic sensors~\cite{Washburn2010,Yavas2017}. All of these classical sensors have their LOD limited by various types of noise, with the ultimate one being the shot noise. As we will show in section~\ref{sec:Quantum_sensors}, quantum techniques can reduce this noise and therefore could potentially improve the performance of more compact plasmonic sensors if they are designed in such a way that preserves the quantum effects being exploited.

\subsubsection{Plasmon-enhanced fluorescence and Raman scattering}
\label{sec:PEFandSERS}
For LSPs, in addition to the conventional colorimetric detection scheme, i.e., measuring the resonance shift due to the change of the refractive index, other sensing principles have widely been studied and used. Two representatives are plasmon-enhanced fluorescence (PEF) sensing/imaging~\cite{RecentProgressonPlasmonEnhancedFluorescence,Bauch2014} and surface enhanced Raman scattering (SERS)~\cite{KneippSERS}. Here, the plasmon merely enhances the signal of the Raman or fluorescence signal, but is itself not used for sensor transduction. Despite this, the enhancement of the signal by the plasmon can improve the signal-to-noise ratio (SNR) which is basically proportional to the signal intensity and inversely proportional to the LOD. Moreover, the sub-diffraction scale of the supporting metal nanostructure provides increased spatial resolution that can be used for imaging. A variety of biochemical sensing applications based on PEF and SERS can be found in relevant review papers~\cite{ Bauch2014, Li2015, Mejia-Salazar2018, Jeong2018}. This section explores the state-of-the-art for classical PEF and SERS sensing modalities that are now being re-envisioned as quantum plasmonic sensors, as will be discussed in more detail in section~\ref{sec:Quantum_plasmonic_sensors_with_emitter}.

\begin{figure*}[!t]
\centering
\includegraphics[width=0.9\textwidth]{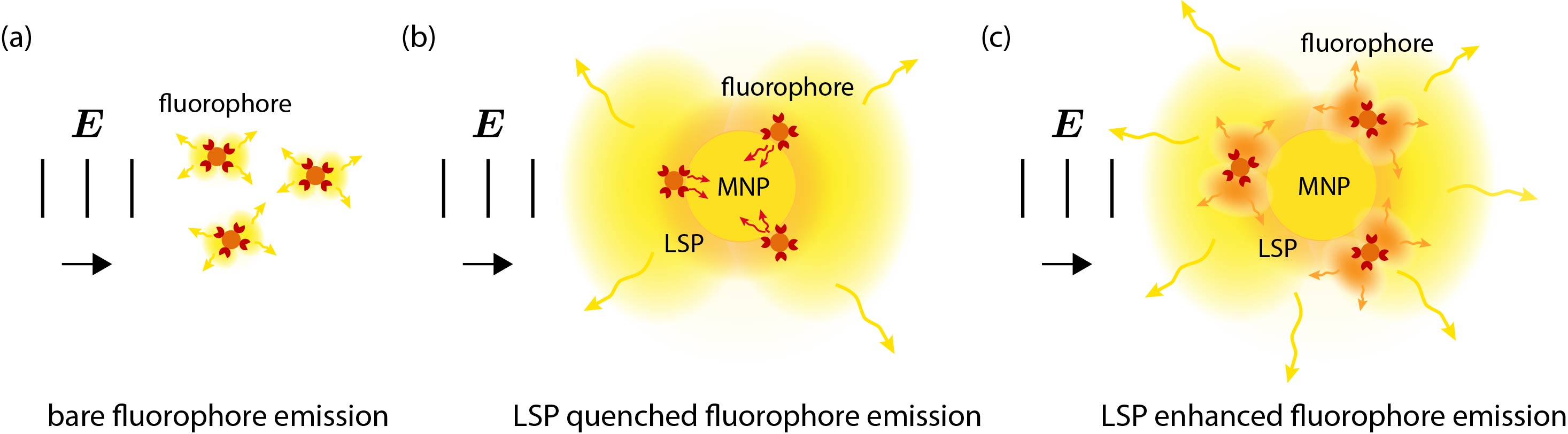}
\caption{
MNP enhancement of fluorescence. (a) Fluorophores are excited by an external field and emit into the far field. (b) Fluorophores close to the surface of a MNP have their emission quenched by coupling to LSPs whose higher-order modes are strongly damped. (c) Fluorophores at a distance from the surface of a MNP couple their emission mainly into the dipolar LSP mode which is not as strongly damped and the fluorescence is therefore enhanced.
}
\label{nanofluor}
\end{figure*}

Although fluorescence signals enable low-background detection by filtering the excitation beam, the fluorescence of nanometer sized fluorophores is inefficient due to the weak light-matter interactions that originate from a large mismatch between the physical size of the fluorophores and the wavelength of visible light, as shown in Figure~\ref{nanofluor}(a). This can lead to photobleaching when a high excitation power is used to extract a measurable signal. In practice, the fluorescence of fluorophores can either be enhanced or quenched depending on the absorption and scattering characteristics of a MNP. A largely enhanced electric field can be formed at metallic nanostructures when the incident wavelength of light matches with their LSP resonance. This enhanced electric field can excite fluorophores placed within the evanescent decay depth of the LSP. This can be efficient when the absorption spectrum of the fluorophore matches with the LSP resonance. The enhancement in the excitation rate of a fluorophore with an absorption dipole moment~$\boldsymbol{d}$ is given by~$\kappa=\left\vert \boldsymbol{d}\cdot\boldsymbol{E}_{\mathrm{LSP}}\right\vert^2/\left\vert\boldsymbol{d}\cdot\boldsymbol{E}_{\mathrm{inc}}\right\vert^2$, where~$\boldsymbol{E}_{\mathrm{LSP}}$ and~$\boldsymbol{E}_{\mathrm{inc}}$ are the electric field at the fluorophore position with and without the plasmonic structure, respectively.~\cite{Kuhn2008} 

In addition to the enhancement of the excitation of the fluorophore, LSPs can also modify the fluorophore emission dynamics. The typical fluorescence lifetime of an isolated and excited fluorophore is on the order of a few ns up to tens of ns. This has to be compared to the almost instantaneous decay of an excited LSP mode in metal nanostructures when the external driving field is switched off~\cite{PhysRevLett.84.4721,PhysRevLett.86.4688}. When a fluorophore is coupled to a metallic nanostructure, the excited state of the fluorophore decays faster while transferring the excited energy to the metallic nanostructure. This effect happens if the emission wavelength of the fluorophore is resonant with the LSP modes. When the fluorophore is less than~$\sim$5~nm away from the metallic nanostructure, the transferred energy is mostly dissipated into Ohmic losses due to the dominant presence of higher-order modes that do not radiate into the far-field, causing quenching instead of enhancing the fluorescence~\cite{Schneider2006}, as shown in Figure~\ref{nanofluor}(b). For moderate distances of approximately 10~nm to 30~nm, the dipolar LSP mode can have a stronger presence than the higher-order modes, possibly enhancing the fluorescence, as shown in Figure~\ref{nanofluor}(c). The shortened fluorescence lifetime for the moderate-distance case can be understood by the plasmonically enhanced local density of states, which is proportional to~$\left|\boldsymbol{E}_\mathrm{LSP}\right|^2$. The dipolar mode of the LSP can radiate into the far-field ($\gamma_{r,\mathrm{LSP}}$), but it also suffers from nonradiative Ohmic losses ($\gamma_{\mathrm{nr},\mathrm{LSP}}$). In general,~$\gamma_{\mathrm{r},\mathrm{LSP}}$ and~$\gamma_{\mathrm{nr},\mathrm{LSP}}$ are much larger than the radiative decay rate ($\gamma_\mathrm{r}$) and nonradiative decay rate ($\gamma_\mathrm{nr}$) of the bare fluorophore, without the plasmonic structure. Therefore, when the excited state of the fluorophore decays mostly into LSP modes, i.e., the fluorescence decay lifetime reduction is significant, the quantum efficiency (QE) of the fluorophore coupled to the LSP can be obtained in a simplified form of 
\begin{figure}[!b]
\centering
\includegraphics[width=0.45\textwidth]{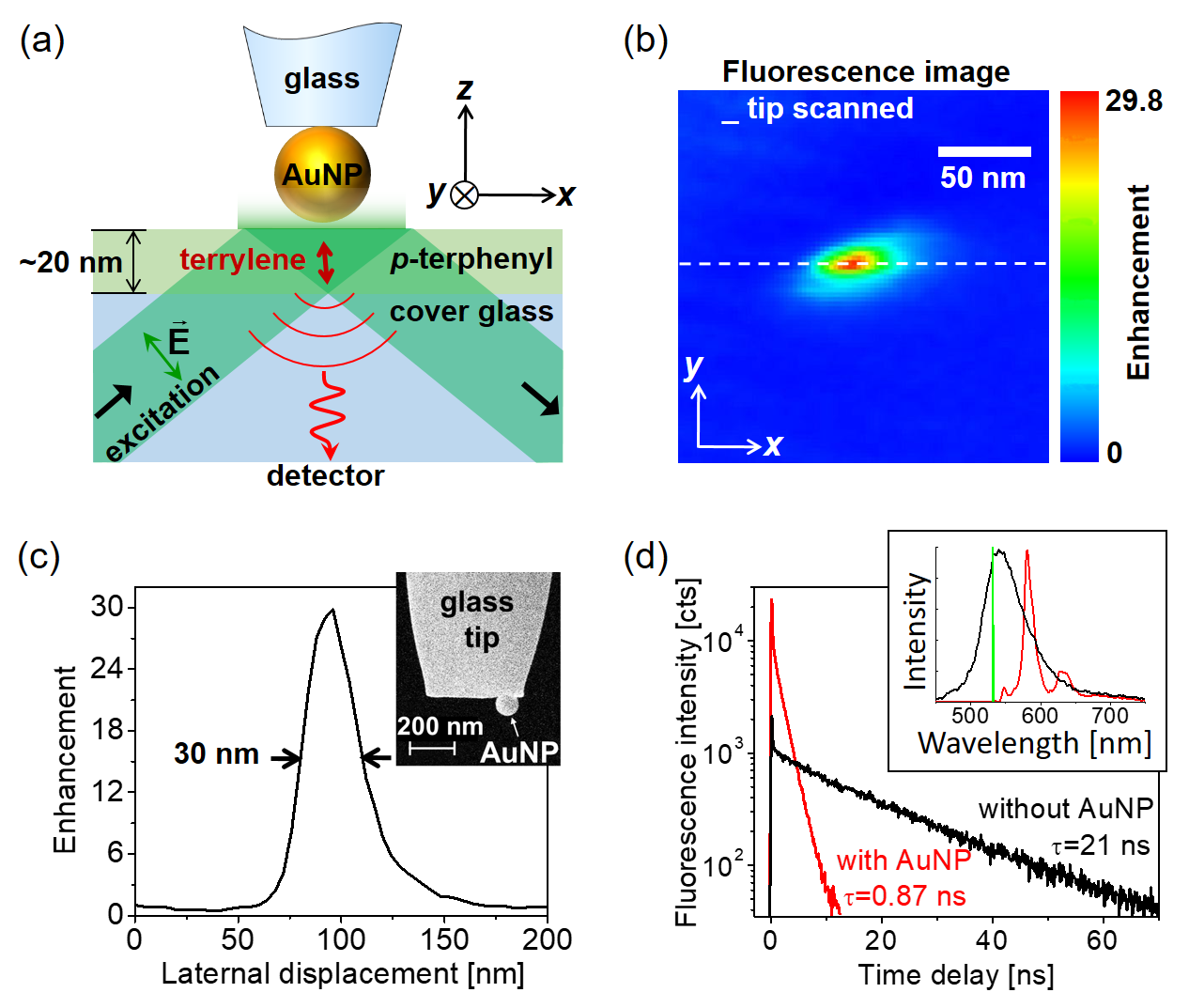}
\caption{(a) Schematic of an experimental setup used to enhance the fluorescence of a single terrylene molecule using a gold nano particle (AuNP). Terrylene molecules are embedded in a thin crystalline p-terphenyl film (20 nm) and are illuminated under total internal reflection. The same objective collects fluorescence of individual molecules while a AuNP attached to a glass fiber tip is scanned across. (b) Near-field fluorescence image using a AuNP (80 nm in diameter) attached tip. (c) Cross-sectional profile of the white dashed line in (b). The inset shows a SEM image of the AuNP attached to the tapered-fiber tip. (d) Fluorescence decay lifetime of a terrylene molecule with (black) and without (red) employing a AuNP (80~nm in diameter). The inset shows a plasmon spectrum of a AuNP (black), the fluorescence spectrum of a single terrylene molecule (red), and the excitation laser line (green). Reproduced and adapted with permission from refs~\citenum{Eghlidi2009}~and~\citenum{Chen2015}. Copyright 2009 American Chemical Society and copyright 2015 The Optical Society.
}
\label{Plasmon_Enhanced_Fluorescence}
\end{figure}
$\eta_\mathrm{coupled} \sim \gamma_{\mathrm{r},\mathrm{LSP}}/(\gamma_{\mathrm{r},\mathrm{LSP}}+\gamma_{\mathrm{nr},\mathrm{LSP}})$. If the intrinsic QE of the fluorophore [$\eta_0=\gamma_\mathrm{r}/(\gamma_\mathrm{r}+\gamma_\mathrm{nr})$] is high, the LSP will reduce the QE of the hybrid system. However, when the original fluorophore QE is very low ($\sim$a few percent), the QE can be highly enhanced by coupling to LSPs. This enhancement in the emission process, when combined with the excitation enhancement, leads to a huge fluorescence enhancement, over a thousand times, becoming possible~\cite{Kinkhabwala2009}. The fluorescence enhancement, well below the saturation limit of the excited population of the fluorophore, is given by~$\kappa \eta_\mathrm{coupled}/\eta_0$. It has been shown that the fluorescence of a vertically aligned terrylene single molecule is enhanced by 30~times when coupled to a spherical gold nanoparticle with~$\kappa\sim80$,~$\eta_\mathrm{coupled}\sim0.3$, and~$\eta_0\sim0.8$~\cite{Eghlidi2009}, as shown in Figure~\ref{Plasmon_Enhanced_Fluorescence}. It should be noted that, in addition to the excitation and the emission processes, the fluorescence emission pattern also can be strongly modified by the antenna modes of metallic structures.~\cite{Kuhn2008, Taminiau2008, Curto2010}

\begin{figure*}[!t]
\centering
\includegraphics[width=0.9\textwidth]{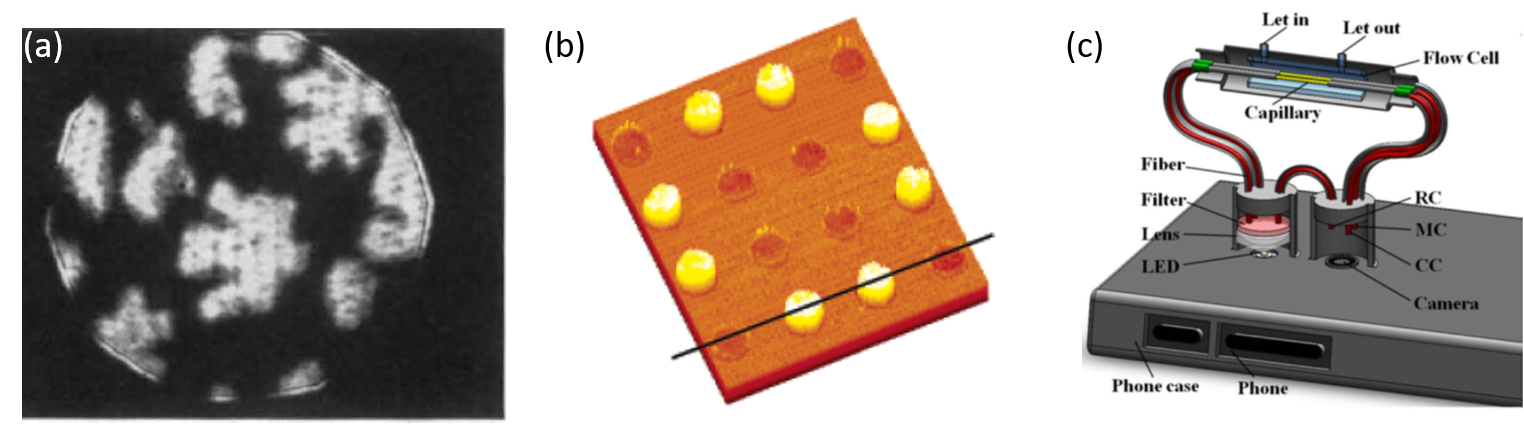}
\caption{(a) Seminal spatially resolved SPR imaging of dimyristoylphosphatidic acid monolayer utilizing a Kretschmann readout scheme at a fixed angle of incidence of 47.2$^\circ$, (b) An early demonstration of spatially resolved biomolecular sensing with Kretschmann-based SPR imaging schemes demonstrating that only appropriately functionalized gold pads give a response to single strand DNA, and (c) a schematic representation of SPR imaging with a smartphone camera. Reproduced with permission from refs~\citenum{hickel1989surface,wark2005long,liu2015surface}. Copyright 1989 Springer Nature, copyright 2005 American Chemical Society, and copyright 2015 Springer Nature.
}
\label{SPRimaging}
\end{figure*}

For a spherical MNP, the absorption of light is proportional to the volume $V\sim r_{0}^3$ [see eq~\ref{eq:cross_section_abs}], while the scattering is proportional to~$V\sim r_{0}^6$ [see eq~\ref{eq:cross_section_sca}], where~$r_{0}$ is the diameter of the MNP~\cite{Menzel:14}. As a result, for a smaller MNP, the radiating power decreases faster than the absorption, thus reducing the QE. Therefore, small nanostructures ($<20$~nm) are usually used to quench fluorescence~\cite{Agio2013}. Larger particles can act as efficient scattering centers. In this case, the excited energy of the fluorophore can be radiated into the far field more efficiently with the help of LSPs that act as an antenna at optical frequencies~\cite{Kuhn2006, Anger2006}. When the diameter of a spherical MNP becomes too large (>100~nm), the dipolar field approximation does not hold anymore, and higher-order modes start to contribute significantly, reducing the QE and the radiation efficiency~\cite{Kolwas2009,Fernandez-Corbaton:15,AlaeeReview}. In addition, the LSP mode volume becomes larger for a larger MNP, reducing the excitation field enhancement. Thus, in PEF-based imaging, the spatial resolution is better when a smaller nanostructure is used~\cite{Eghlidi2009}. 
PEF can be more efficient for other shapes of metallic nanostructures, including rods~\cite{ Mohamed2000}, stars~\cite{ Theodorou2018}, dimers~\cite{LeeKG2012}, and arrays~\cite{Jianga2018}. Sharp structures can allow a higher field enhancement compared to single spheres, and nanogaps between MNPs can support even higher field enhancement together with a high QE~\cite{Mayer2011,davidson2016ultrafast}. The enhanced detection sensitivity by PEF has been demonstrated for detecting many different bio-samples~\cite{Bauch2014, Li2015, Mejia-Salazar2018, Jeong2018}. For example, silver-nanoparticle-assisted PEF has been applied for the detection of streptavidin and human IgE, reaching a LOD of 0.25 ng/mL~\cite{Li2012}. The hot-spots formed in a gold nanorod array have been employed for detecting single-strand DNA with a LOD of 10 pM~\cite{Mei2017}. Although the modification of the fluorescence signal is significant in the given scenarios in this section, they can still be described in the weak-coupling regime. In the strong-coupling regime between the emitter and plasmonic modes of metallic systems, however, exotic functionalities of quantum plasmonic sensing schemes can be utilized, as will be described in section~\ref{sec:Quantum_plasmonic_sensors_with_emitter}.  

The LSP-induced local field enhancement can also contribute to SERS. The overall enhancement in the SERS signal is attributed to the combined effects of the enhanced field intensity as in PEF and the chemical enhancement due to the charge transfer between the metallic nanostructure and the target molecule~\cite{Wang2012,ThomasSERS}. Here, the local field enhancement is frequently considered as the dominating factor for the enhanced SERS. In a back-of-the-envelope estimation, the enhancement of the SERS signal, thanks to the supporting plasmonic structure, is proportional to the fourth power of the electric field at the spatial location where the molecules are placed. Therefore, a major aim is frequently to tune the geometrical properties of the plasmonic structure such that a plasmon resonance is supported at the frequency used in the experiments. Also, the field enhancement has to occur in the spatial region where the molecules are placed.

\subsection{Surface plasmon resonance imaging}\label{sec:SPR_imaging}
Improving the sensitivity of plasmonic sensors can enable the detection of smaller concentrations of materials of interest in the same integration time, or it can enable the detection of the same concentration of materials in a shorter integration time. However, when a sensor needs to detect many different molecules, most plasmonic sensors are operated in an iterative fashion, resulting in slow measurements. SPR imaging attempts to reduce measurement times by parallelizing SPR sensors on a single chip. Whereas conventional LSP and SPP sensors utilize angle-resolved or spectrally-resolved measurements, SPR imaging systems utilize intensity-resolved measurements: they sit at the inflection point of a plasmonic resonance and monitor spatially-resolved changes in intensity for an array of plasmonic sensors, or for a continuous plasmonic sensor. This concept was first introduced in 1988 as SP microscopy~\cite{rothenhausler1988surface}. In general, the ability to spatially map variations in the index of refraction with SPR imaging is a powerful tool for fundamental materials science, chemistry and biology.  When arrays of sensors are functionalized to detect different molecules, SPR imaging can enable substantial scaling of SPR sensors for applied sensing of enzyme-substrate interactions, DNA hybridization, antibody-antigen binding, and protein interaction dynamics~\cite{smith2003surface,steiner2004surface,wark2005long,wong2014surface,jordan1997surface,piliarik2007towards}.

\begin{figure*}[!t]
\centering
\includegraphics[width=0.75\textwidth]{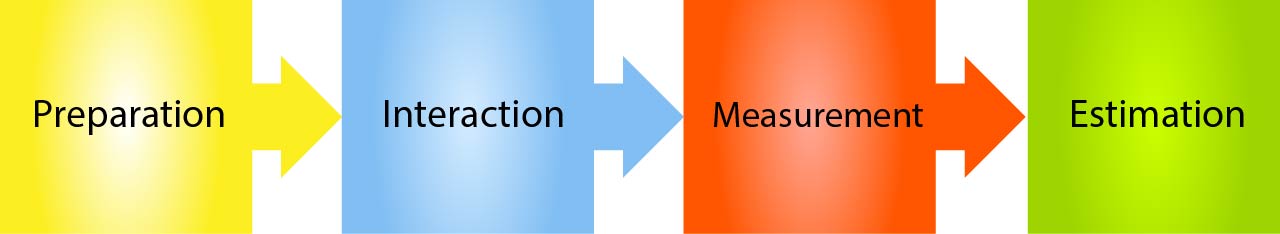}
\caption{The process for the estimation of a parameter in sensing is divided into four key steps: Preparation, Interaction, Measurement and Estimation.
}
\label{estimationprocess}
\end{figure*}

Figure~\ref{SPRimaging}(a) illustrates early SPR imaging of monolayers of dimyristoylphosphatidic acid with a spatially resolved Kretschmann sensor~\cite{hickel1989surface}. This work, along with other early research\cite{rothenhausler1988surface}, illustrated the potential of SPR imaging for sub-diffraction-limited, high sensitivity imaging of low-contrast samples. Since then, substantial technical development has gone into the improvement of SPR imaging systems for parallelized detection of different molecular systems. As shown in Figure~\ref{SPRimaging}(b), SPR imaging of gold pads functionalized with appropriate probes allows for the spatially selective detection of target DNA that is complementary to the functionalization group~\cite{wark2005long}. By functionalizing each element of the plasmonic sensor array differently, it is therefore possible to achieve highly beneficial scaling for the detection of large numbers of different molecules. Figure~\ref{SPRimaging}(c) illustrates the degree to which SPR imaging platforms can be deployed in the field through integration with ubiquitous imaging technologies like smartphone cameras~\cite{liu2015surface}.  However, a common challenge in SPR imaging is that off-the-shelf imaging systems introduce substantial noise compared with single pixel detectors. While it is possible to reach the fundamental classical limit given by the shot noise with appropriate experimental designs, further improvements could be made possible with quantum imaging schemes, as will be discussed in section~\ref{sec:Quantum_plasmonic_sensors}.

\section{Quantum sensors}\label{sec:Quantum_sensors}

Photonic devices, such as the plasmonic sensors introduced in the previous section, exploit light as a probe for their operation. While plasmonic sensors offer an improved sensitivity compared to other types of photonic devices, they share a common problem -- random fluctuations in the measured signal due to the statistical nature of light. The origin of these fluctuations can be derived using a classical theory to some extent by considering light to be made from discrete particles~\cite{Saleh2019}. 
However, once the wave properties of light are included, the fluctuations can only be understood from a more fundamental quantum theory~\cite{Loudon2000}. 
When a coherent state of light -- thought of as the quantum state that most closely describes the light from a laser~\cite{Wiseman2016} -- is employed in a sensor, the noise in the signal arises from the fact that the coherent state consists of photon number states with a weighting that follows a Poisson distribution. This noise will obviously affect the LOD for a given plasmonic sensor, as mentioned in section~\ref{sec:Limit_of_detection}, limiting its resolution. 
However, if the noise can be reduced, then the sensor can provide a better resolution, enabling more precise measurements to be made~\cite{Caves1981}. Reducing this noise involves the use of quantum noise reduction, or squeezing. To understand how the noise, commonly known as `shot noise', is present in plasmonic sensors and to see how it can be reduced by using specialized quantum techniques, we kick off this section by offering a brief overview of parameter estimation theory. This is a theory that has been widely considered in the field of quantum metrology~\cite{Cramer1946, Helstrom1976, Kay1993}, and it is vital for determining fundamental bounds on how well parameters can be estimated in both the classical and quantum regime. After an introduction to parameter estimation theory, we look at a few paradigmatic examples of classical optical sensors and corresponding quantum sensors, while leaving more extensive details to other review articles dedicated to quantum metrology and imaging~\cite{Giovannetti2004, Paris2009, Giovannetti2011, Toth2014, Degen2017, Pirandola2018, Tan2019, Sidhu2020}.

\subsection{Parameter Estimation Theory: Cram{\'e}r-Rao bound}\label{sec:parameter_estimation_theory}
\subsubsection{Single-parameter estimation}
The complete process for the estimation of a parameter in sensing can be divided into four key steps: (i) Input state preparation; (ii) Interaction for parameter encoding; (iii) Measurement; and (iv) Estimation based on the measurement outcomes, as depicted in Figure~\ref{estimationprocess}. In the final step, the estimation of a parameter is made over multiple repetitions of a measurement or time-integration. The number of repeated measurements determines the finite size~$\nu$ of a sample. Each sample with a size~$\nu$ is used by an estimator~$\hat{x}$ to yield an estimate~$x_\text{est}$ of the parameter, and the quality of the estimator~$\hat{x}$ can be assessed by several statistical features. The expectation value of the estimate,~$\langle x_\text{est}\rangle$, is compared with the parameter's true value~$x$ from the underlying population being sampled. Their difference indicates a bias of the estimator and can be interpreted as an estimation accuracy. When the difference is zero, i.e.,~$\langle x_\text{est}\rangle-x=0$, the estimator is said to be unbiased and equivalently the accuracy is perfect. Here,~$\langle .. \rangle$ denotes the average over all possible configurations of the sample, which strictly speaking can only be theoretically considered, but it may be well approximated by an actual repetition of sampling in an experiment, i.e.,~collecting a large number of samples. Another quantity of great importance is the mean-squared-error (MSE) defined as~$\text{MSE}[\hat{x}]=\langle (x_\text{est}- x)^2 \rangle= \langle (x_\text{est}-\langle x_\text{est}\rangle)^2 \rangle+ (\langle x_\text{est}\rangle-x)^2~$, where the first term is the variance~$\Delta x_\text{est}^2$ of the estimator and the second term is the squared bias of the estimator. The MSE and the variance are equivalent when an estimator is unbiased. The variance is often called an estimation uncertainty or interpreted as an estimation precision. The latter implies that the estimate~$x_\text{est}$ would vary over the repetition of an identical and independent sampling, or equivalently, over different configurations of the sample with a size~$\nu$. In this review, we take the square root of the variance,~$\Delta x_\text{est}$, as the estimation uncertainty or estimation precision, as is often done in experiments. In the literature, $\Delta x_\text{est}$ is sometimes called sensitivity/resolution with the interpretation that a large $\Delta x_\text{est}$ limits the capability of a sensor to sense/resolve a minute change of an observable quantity. In a qualitative sense, these terms can be used interchangeably according to their close relation, but in a quantitative sense, $\Delta x_\text{est}$ is the uncertainty of an estimator according to parameter estimation theory, which we introduce in this section. The LOD is then simply given by $\Delta x_\text{est}$ divided by the sensitivity.

Let us consider a simple example of the above definitions, where one aims to estimate the parameter given by the population mean, whose true value is~$\mu$, by using the sample mean as an estimator, i.e.,~$\hat{x}=\bar{x}=\sum_{j=1}^{\nu}x_{j}/\nu$. The value obtained by a sample with a size~$\nu$ is then~$x_\text{est}$ and because the sample mean is an unbiased estimator for any population~\cite{Hogg2013}, we have~$\langle x_\text{est}\rangle=\mu$. However,~$x_\text{est}$ for each sample will vary and the variance of the sample mean is given by~$\Delta x_\text{est}^2=\sigma_{x}^2/\nu$, where~$\sigma_{x}^2$ is the population variance, which holds for any size~$\nu$ without requiring the approximation imposed by the central limit theorem~\cite{Kay1993}. It is clear that the standard deviation~$\Delta x_\text{est}$ is inversely proportional to~$\sqrt{\nu}$ when the sample mean is taken to estimate the population mean, i.e., the estimation becomes more precise as the sample size~$\nu$ increases. Besides the statistical scaling with $\nu$, the population variance, $\sigma_x^2$, also plays an important role in reducing the standard deviation~$\Delta x_\text{est}$, which is related to the role of the quantum resource in quantum metrology, as we will see in this section. 

Now consider a more general estimation scenario where an unbiased estimator~$\hat{x}$ is used to estimate a true value~$x$. In this case, the MSE is simply equal to the variance~$\Delta x_\text{est}^2$, and it is known that a lower bound to the standard deviation~$\Delta x_\text{est}$ exists and is given by
\begin{align}
\Delta x_\text{est} \ge \frac{1}{\sqrt{\nu F(x)}},
\label{eq:CRinequality}
\end{align}
which is called the Cram{\'e}r-Rao (CR) inequality~\cite{Cramer1946,Hogg2013}. Here~$F(x)$ denotes the Fisher information (FI) defined as~\cite{Fisher1925}
\begin{align}
F(x) = \int dy \frac{1}{p(y\vert x)}\left(\frac{\partial p(y\vert x)}{\partial x}\right)^2,
\label{eq:FisherInformation}
\end{align}
with~$p(y\vert x)$ being an underlying conditional probability density of obtaining the measurement outcome within the interval ~$y$ and~$y+dy$ when the true value is~$x$ (for a finite distribution we have~$\int dy\to\sum_y$). The FI represents a sort of measure of the amount of information that a measurement outcome~$y$ carries on average about the true value~$x$. A higher value of the FI is obviously more advantageous for obtaining a better estimation precision. The lower bound of eq~\ref{eq:CRinequality}, called the CR bound, is asymptotically saturable in the limit~$\nu\rightarrow \infty$ when the maximum-likelihood estimator is employed~\cite{Fisher1925,Hogg2013}. 

Importantly, the FI of eq~\ref{eq:FisherInformation} depends on the physical scenario, i.e., the probe state, the parameter encoding, and the measurement for a fixed true value~$x$. This implies that the FI may be increased further by changing the physical scenario, for instance, by going from a classical scenario to a quantum scenario. Thus, a further lower bound to eq~\ref{eq:CRinequality} can exist and be achieved when a more optimal scenario is chosen. The CR inequality can be developed to~\cite{Braunstein1994a,Braunstein1996}
\begin{align}
\Delta x_\text{est}  \ge \frac{1}{\sqrt{\nu F(x)}}\ge \frac{1}{\sqrt{\nu H}},
\label{eq:QCRinequality}
\end{align}
where~$H$ represents the quantum Fisher information (QFI), defined as
\begin{align}
H=\max_{\{\hat{\Pi}_y\}} F(x).
\label{eq:QuantumFisherInformation}
\end{align}
The QFI is essentially the maximized FI over all possible quantum measurements, or put more formally, over all positive-operator valued measures (POVMs)~$\{\hat{\Pi}_y\}$, such that~$\hat{\Pi}_y \ge 0$ (positivity) and~$\int dy \hat{\Pi}_y =\boldsymbol{1}$ (completeness). Here, the~$y$ correspond to the possible outcomes from the quantum measurement, similar to the outcomes~$y$ in the classical case in eq~\ref{eq:FisherInformation}, with~$p(y|x)={\rm Tr}(\hat{\rho}_x\hat{\Pi}_y)$ and~$\hat{\rho}_x$ as the parameter encoded state. When the parameter encoding is caused by a unitary process, the QFI is independent of the value~$x$ (see examples discussed in section~\ref{sec:sub_shot_noise_phase_sensing}), since the same type of unitary process could be implicitly included as part of the optimal measurement~\cite{Nielsen2010}, such that the value~$x$ can be tuned to a certain value for which~$F$ is fully maximized to be $H$. This is not the case when the parameter is encoded through a non-unitary process, i.e., the QFI is given as a function of~$x$ (see examples discussed in section~\ref{sec:sub_shot_noise_intensity_sensing}). In either case, the QFI still depends on the probe state and the encoding process of the parameter, implying that the QFI can be maximized by using an optimal probe state for a given encoding process. The lowest bound given by the maximized QFI over all input states is often called the ultimate quantum limit. In this sense, a number of studies have investigated the optimal probe states in various sensing or estimation scenarios~\cite{Paris2009,Giovannetti2011,Demkowicz2015,Degen2017}. 

The lower bound of eq~\ref{eq:QCRinequality} is called the quantum Cram{\'e}r-Rao (QCR) bound and it is regarded as the fundamental limit in the estimation precision for a given input state and encoding process. It can be shown that for a probe state with the parameter~$x$ encoded, given by the state~$\hat{\rho}_x$, the optimal measurement with the POVM~$\{\hat{\Pi}_{y}\}$ that reaches the QCR bound has to satisfy the following two conditions~\cite{Braunstein1994a,Paris2009}:
\begin{align}
\text{Im}[ \text{Tr} (\hat{\rho}_x \hat{\Pi}_{y}\hat{\cal L}_x) ]&=0, \\
\frac{\hat{\Pi}_{y}^{1/2}\hat{\rho}_x^{1/2}}{\text{Tr}(\hat{\rho}_x \hat{\Pi}_{y})} &=
\frac{\hat{\Pi}_{y}^{1/2} \hat{\cal L}_x \hat{\rho}_x^{1/2}}{\text{Tr}(\hat{\rho}_x \hat{\Pi}_{y} \hat{\cal L}_x )},
\end{align}
where~$\hat{\cal L}_x$ is the symplectic logarithmic derivative (SLD) operator defined such that 
\begin{align}
\frac{\partial \hat{\rho}_x}{\partial x} = \frac{1}{2}\left( \hat{\rho}_x\hat{\cal L}_x +\hat{\cal L}_x \hat{\rho}_x\right).
\end{align}
The above two conditions are rather cryptic, however they can be fulfilled if a measurement setup with~$\{\hat{\Pi}_{y}\}$ is constructed by a set of projection operators over the eigenbasis of the SLD operator~$\hat{\cal L}_x$~\cite{Braunstein1994a,Paris2009}. This is a necessary and sufficient condition to reach the QCR bound for a full-rank state of~$\hat{\rho}_x$, for which the SLD operator is unique. On the other hand, for a rank-deficient state the optimal measurement satisfying the above two conditions is not unique~\cite{Braunstein1994a}.

The QFI,~$H$, of eq~\ref{eq:QuantumFisherInformation} can be written in terms of the SLD operator by
\begin{align}
H=\text{Tr}(\hat{\rho}_x \hat{\cal L}_x^{2}).
\label{eq:QFIbySLD}
\end{align}
For a parameter-encoded state with a given spectral decomposition, i.e.,~$\hat{\rho}_x =\sum_n p_n \ket{\psi_n}\bra{\psi_n}$ with~$\langle \psi_n\vert\psi_m\rangle=\delta_{n,m}$, the SLD operator can be written as~\cite{Braunstein1994a,Paris2009,Jiang2014}
\begin{align}
\hat{\cal L}_x=2\sum_{n,m}\frac{\bra{\psi_m}\partial_x \hat{\rho}_x \ket{\psi_n}}{p_n+p_m}\ket{\psi_m}\bra{\psi_n},
\end{align}
where the summation is taken over~$n,m$ for which~$p_n+p_m\neq 0$. 
When the state~$\hat{\rho}_x$ is pure, i.e.,~$\hat{\rho}_x=\ket{\psi_x}\bra{\psi_x}$ for some~$\ket{\psi_x}$, the SLD operator is given by~$\hat{\cal L}_x = 2\partial_x\hat{\rho}_x$, which simplifies eq~\ref{eq:QFIbySLD} to
\begin{align}
H = 4\left[\langle \partial_x\psi_x \vert \partial_x\psi_x\rangle + \langle \partial_x\psi_x \vert \psi_x\rangle^2 \right],
\label{eq:QFI_Pure}
\end{align}
where~$\vert \partial_x\psi_x\rangle \equiv \partial_x\vert \psi_x \rangle$. When the parameter is encoded through a unitary process, i.e.,~$\hat{\rho}_x=e^{ix\hat{G}}\hat{\rho}_0 e^{-ix\hat{G}}$, where~$\hat{\rho}_0$ is the initial probe state and~$\hat{G}$ is a generator of the parameter~$x$, one can show that~$H=4\langle (\Delta \hat{G})^2\rangle$, where~$\langle (\Delta \hat{G})^2\rangle=\langle \hat{G}^2 \rangle-\langle \hat{G} \rangle^2$ is the variance of the generator with respect to the probe state that undergoes parameter encoding. 

The evaluation of the QFI determines the lowest estimation uncertainty~$\Delta x_\text{est}$ (or the highest precision) for a given parameter encoding and probe state. As we shall show in section~\ref{sec:shot_noise_limited_sensing}, when a random feature obeying the Poisson distribution with a mean of~$N$ dominates the estimation uncertainty, the lower bound is called the SNL, where~$\Delta x_\text{est}$ scales with~$N^{-1/2}$. 

In interferometric sensing, the minimum estimation uncertainty achievable by a coherent probe state (the quantum state that represents light from a laser) is called the \textit{standard quantum limit} (SQL), where~$N$ represents the mean photon number of the state. We will show in section~\ref{sec:shot_noise_limited_sensing} that~$\Delta x_\text{est}$ scales with~$N^{-1/2}$~\cite{Giovannetti2011,Demkowicz2015,Degen2017}. The term SQL was introduced by Caves in his early papers in the sense that it is the limit of standard interferometers made of standard devices without quantum squeezing~\cite{Caves1980a, Caves1980b,Caves1981}. In the literature, the terms SNL and SQL have mostly been used in intensity sensing and phase sensing, respectively. However, they can sometimes be considered as synonymous because of the same fundamental process that causes them. 

On the other hand, when an optimal quantum probe state is used that maximizes the QFI for a given parameter encoding, we will show in section~\ref{sec:sub_shot_noise_phase_sensing} that the estimation uncertainty is reduced so that it scales with~$N^{-1}$~\cite{Giovannetti2011,Demkowicz2015,Degen2017}. The associated minimum is called the \textit{Heisenberg limit} (HL) or ultimate quantum limit, and scaling of $N^{-1}$ is often called \textit{Heisenberg scaling}.

An important note is due about the above mentioned scalings and their comparison. The size of the sample is assumed to be fixed in both the classical and quantum cases, i.e., $\nu$ corresponds to a constant number of repetitions, or probes, and therefore the integration time of the measurement is fixed. One could consider increasing $\nu$ for the classical case in order to achieve a similar scaling in precision to the quantum case. However, this may not be practical due to time constraints of the system being sensed, which is especially relevant for a dynamic biological or biochemical system. 

Furthermore, the scaling of the precision for the SNL is $N^{-1/2}$ using a classical probe state, but for a quantum probe state, for example in the interferometric setting, the precision is the HL with scaling $N^{-1}$. This means that one could obtain the same precision of a quantum probe state with mean photon number $N$ using a classical probe state with the mean photon number $N$ increased to $N^2$. However, this second approach to leveling the classical and quantum scaling may also not be feasible in a given sensor. First, the sensor may be at its physical limit in terms of the power being used to achieve a high precision, due to the type of materials the sensor is made from and its construction. Any higher power may cause structural changes and distort the response of the sensor, introducing additional sources of noise proportional to the intensity~\cite{Piliarik2009}. Second, and perhaps more importantly from the context of non-invasive sensing, is that the analyte itself may have a damage threshold for the power it can tolerate. This is particularly the case when monitoring small quantities of biological systems~\cite{Taylor2016}, for which plasmonic sensors are regularly used and a high precision is required with a limited optical power. 

The above considerations apply to any case where there is a gap between the scaling of the precision for the classical SNL and the quantum case, which may or may not have Heisenberg scaling. Thus, there is a clear benefit to using a quantum approach to improve the precision of a plasmonic sensor.

\subsubsection{Multiparameter estimation}
The formulation introduced above applies when a single parameter is estimated, but estimation of multiple parameters are often of interest in diverse areas of science and technology such as phase-contrast imaging~\cite{Preza1999} and gravitational-wave astronomy~\cite{Freise2009}. An extended theory for multiparameter estimation is thus required~\cite{Szczykulska2016, Liu2020}.  Consider the problem of estimating a set of multiple parameters~$\boldsymbol{x}=(x_1, x_2, \cdots, x_M)^\text{T}$ from the measurement results~$\boldsymbol{y}$ that have been drawn from a conditional probability density~$p(\boldsymbol{y}\vert \boldsymbol{x})$. The~$M\times M$ covariance matrix~$\text{Cov}(\boldsymbol{x}_\text{est})=\langle (\boldsymbol{x}_\text{est}-\langle \boldsymbol{x}_\text{est}\rangle)(\boldsymbol{x}_\text{est}-\langle \boldsymbol{x}_\text{est}\rangle)^\text{T} \rangle$ of any unbiased estimator~$\hat{\boldsymbol{x}}$ is found to be bounded by the so-called Fisher information matrix (FIM)~\cite{Helstrom1976,Paris2009}, written as
\begin{align}
\text{Cov}(\boldsymbol{x}_\text{est}) \ge \frac{\boldsymbol{F}^{-1}(\boldsymbol{x})}{\nu},
\label{eq:CRinequality_multi}
\end{align}
where the FIM,~$\boldsymbol{F}(\boldsymbol{x})$, is defined by
\begin{align}
[\boldsymbol{F}(\boldsymbol{x})]_{jk}
=\int d\boldsymbol{y} 
\frac{1}{p(\boldsymbol{y}\vert \boldsymbol{x})}\frac{\partial p(\boldsymbol{y}\vert \boldsymbol{x})}{\partial x_j}\frac{\partial p(\boldsymbol{y}\vert \boldsymbol{x})}{\partial x_k}.
\end{align}
The multiparameter CR inequality of eq~\ref{eq:CRinequality_multi} is satisfied when~$\text{Cov}(\boldsymbol{x}_\text{est}) - \boldsymbol{F}^{-1}(\boldsymbol{x})/\nu~$ is a positive semi-definite matrix~\cite{Helstrom1967, Helstrom1976, Paris2009}. The CR bound can always be saturated by a maximum likelihood method in the limit of large~$\nu$~\cite{Braunstein1992}.
As in single parameter estimation, the CR inequality of eq~\ref{eq:CRinequality_multi} for multiparameter estimation can be further reduced in the quantum regime, leading to the multiparameter QCR inequality written as~\cite{Helstrom1976,Paris2009}
\begin{align}
\text{Cov}(\boldsymbol{x}_\text{est}) \ge \frac{\boldsymbol{F}^{-1}(\boldsymbol{x})}{\nu}\ge \frac{\boldsymbol{H}^{-1}}{\nu},
\label{eq:QCRinequality_multi}
\end{align}
where the details of the quantum Fisher information matrix (QFIM),~$\boldsymbol{H}$, are given in Appendix~\ref{appendix:A}. The inequality in eq~\ref{eq:QCRinequality_multi} means that in the matrix sense
\begin{align}
\boldsymbol{n}^\text{T}\text{Cov}(\boldsymbol{x}_\text{est})\boldsymbol{n} \ge \frac{\boldsymbol{n}^\text{T}\boldsymbol{F}^{-1}(\boldsymbol{x})\boldsymbol{n}}{\nu}\ge \frac{\boldsymbol{n}^\text{T}\boldsymbol{H}^{-1}\boldsymbol{n}}{\nu},\label{eq:QCRinequality_multi_distributed}
\end{align}
for arbitrary~$M$-dimensional real vectors~$\boldsymbol{n}$~\cite{Pezze2017}. 
This can be exploited in the case when a global parameter~$\tilde{x}=\sum_{j}n_j x_j$, defined as a linear combination of multiple parameters is of interest. We then have~$\tilde{x}_{\rm est}=\boldsymbol{n}\cdot \boldsymbol{x}_\text{est}$ and~$(\Delta \tilde{x}_{\rm est})^2=\text{Var}(\tilde{x}_{\rm est})=\boldsymbol{n}^\text{T}\text{Cov}(\boldsymbol{x}_\text{est})\boldsymbol{n}$, which is of interest in distributed sensing and will be discussed further in section~\ref{sec:Multiple_phase_estimation}. Examples include relative phase estimation in a two-mode interferometer, where~$\boldsymbol{n}=(1,-1)$~\cite{Knott2016}, or the average phase estimation in an~$M$-mode interferometer, where~$\boldsymbol{n}=(1,\cdots,1)/M$~\cite{Guo2020,Oh2020}.

In the next sections, we use the introduced framework of the CR and QCR bound for single and multi-parameter estimation to give some basic examples of the SNL and SQL in optical probing schemes, distinguishing between intensity and phase sensing. The individual types of quantum sensor we discuss here will be connected to the corresponding types of plasmonic sensor already discussed in section~\ref{sec:Plasmonic_sensors}, eventually leading to the advances that will be discussed in section~\ref{sec:Quantum_plasmonic_sensors}.

\subsection{Shot-noise limited sensing}\label{sec:shot_noise_limited_sensing}
To understand where and how the shot noise appears, we focus on the two most canonical types of photonic sensing~\cite{Lee2003,Sabri2013,Santos2015}: Intensity sensing and phase sensing, as discussed for plasmonic sensors in section~\ref{sec:Intensity_vs_phase}. They are distinguished by the physical quantity that is being monitored and analyzed to identify the optical properties of an analyte. The goal of intensity sensing is to measure a change of the intensity of light when it passes through an analyte, whose properties under inspection are encoded in the change of intensity. On the other hand, the goal of phase sensing is to read out a change of the phase of the outgoing light from an analyte. These two kinds of sensing schemes lead to different sensitivities in plasmonic sensors, as described in section~\ref{sec:Intensity_vs_phase}. Here, we focus on how the estimation precision is determined according to the sensing type.

\subsubsection{Shot-noise limited intensity sensing}\label{sec:Intensity_quantum_sensing}
\begin{figure}[!t]
\centering
\includegraphics[width=0.5\textwidth]{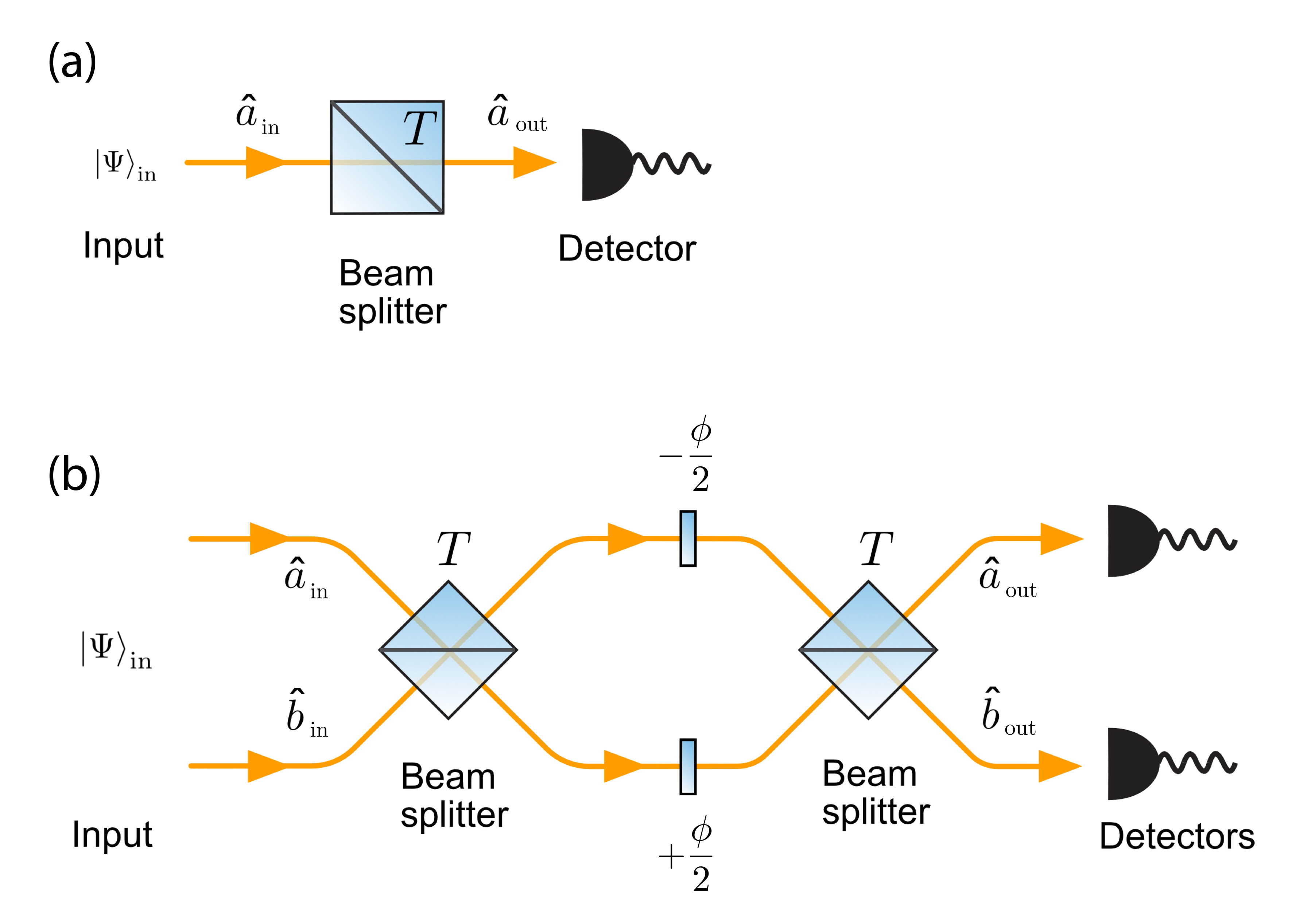}
\caption{
(a) Single-mode intensity sensing, where the transmittance of a BS models a transmissive object and is estimated by the ratio of the intensity of the transmitted light to that of the incident light. 
(b) Two-mode phase sensing, where a relative phase between two arms is estimated from a measurement analyzing the interference at the output ports.}
\label{Fig:Sec3:Basic_Scheme}
\end{figure}

Intensity sensing can be modeled as the estimation of the transmittance $T$ of a beam splitter~(BS), as shown in Figure~\ref{Fig:Sec3:Basic_Scheme}(a). A typical scheme to estimate the transmittance of the BS is to inject light with an intensity~$I_\text{i}$ and then to measure the intensity~$I_\text{t}$ of the transmitted light. The ratio of the incident intensity to the transmitted intensity can be used to estimate the transmittance, e.g., the estimated transmittance can be obtained by~$T_\text{est}=\bar{I}_\text{t}/I_\text{i}$, where~$\bar{I}_\text{t}=\sum_{j=1}^\nu I_{{\rm t},j}/\nu$ for~$\nu$ repetitions of the measurement, with each measurement yielding~$I_{\text{t},j}$. To take into account the effect of losses that further decreases the measured value~$I_{\text{t},j}$ on top of the transmission, one needs to replace the divisor~$I_\text{i}$ by~$\eta I_\text{i}$, with the transmission efficiency~$\eta\in [0,1]$, i.e.,~$T_\text{est}=\bar{I}_\text{t}/ \eta I_\text{i}$.
The estimation uncertainty is thus given by
\begin{align}
\Delta T_\text{est} = \frac{\Delta \bar{I}_\text{t}}{\eta I_i} = \frac{\Delta I_\text{t}}{\sqrt{\nu} \eta  I_i},
\label{eq:Delta_T_est}
\end{align}
where~$\Delta I_\text{t}$ denotes the standard deviation of the transmitted intensity based on the underlying distribution~\cite{Kay1993}, which depends on the type of light being injected. 

As mentioned, the classical source of light that has most widely been considered and employed in standard photonic sensing is modeled quantum mechanically by the coherent state~\cite{Loudon2000}. It is also used for setting the classical benchmark or SQL~\cite{Giovannetti2004,Giovannetti2011}. The coherent state is formally given as the displaced vacuum state~$\ket{\alpha}=\hat{D}(\alpha)\ket{0}$, where~$\hat{D}(\alpha)=\exp[\alpha \hat{a}^{\dagger}-\alpha^*\hat{a}]$, with a displacement parameter~$\alpha\in\mathbb{C}$ and the operators~$\hat{a}$ and~$\hat{a}^\dag$ representing the annihilation and creation operators for the optical mode~\cite{Loudon2000}. The displaced vacuum state can be projected into the Hilbert space spanned by the Fock states and is written as a superposition of photon number states,~$\ket{n}$, weighted by a Poisson distribution,~$p(n)$. When the coherent state is used as an input state in Figure~\ref{Fig:Sec3:Basic_Scheme}(a), i.e.,~$\ket{\Psi}_\text{in}=\ket{\alpha}$, and it passes through the BS with transmittance~$T$, whose value is unknown and is to be estimated, the outgoing state remains as a coherent state and is written as~$\ket{\Psi(T)}_\text{out}=\vert \sqrt{T\eta}\alpha\rangle$, with~$\eta$ characterizing the effect of loss~\cite{Loudon2000}. An intensity measurement is described quantum mechanically by the projectors onto the photon number state, i.e.,~$\ket{n}\bra{n}$, where we omit any geometric factors associated with photonic modes and detection for simplicity. The underlying conditional probability of finding the output state~$\ket{\Psi(T)}_\text{out}$ in the state~$\ket{n}$ is then~$p(n\vert T)\phantom{\,}=\phantom{\,}_{\rm out}\langle \Psi | n \rangle \langle n| \Psi \rangle_{\rm out}\phantom{\,}=e^{-T \eta N} (T\eta N)^n/ n!$, where~$N$ is the average photon number of the input probe state, i.e.,~$N=\langle \hat{n}\rangle_{\rm in} = I_{\rm i}=\bra{\alpha}\hat{n}\ket{\alpha}=\vert \alpha \vert^{2}=\sum_n n p(n)$, with~$\hat{n}=\hat{a}^\dag\hat{a}=\sum_{i=0}^\infty n \ket{n}\bra{n}$. This leads to the standard deviation of the transmitted intensity~$\Delta I_\text{t}=\langle(\Delta \hat{n})^2\rangle_\text{out}^{1/2}=(\langle\hat{n}^2\rangle_{\rm out}-\langle \hat{n}\rangle^2_{\rm out})^{1/2}=\sqrt{T \eta N}$, finally giving the estimation uncertainty of eq~\ref{eq:Delta_T_est} for the specific case of the coherent state input as
\begin{align}
\Delta T_\text{est}= \sqrt{\frac{T }{\nu \eta N}}.
\label{eq:min_Delta_T_est}
\end{align}
This equation is the result of using the sample mean ($\bar{I}_{\rm t}$) as an estimator. One can also show that the CR bound,~$\Delta T_\text{CR}$, defined in eq~\ref{eq:CRinequality} for the intensity measurement in the above scenario is the same as eq~\ref{eq:min_Delta_T_est}~\cite{Yoon2020}.
This implies that the sample mean chosen above is an optimal estimator when the intensity is measured. It is interesting to note that the QCR bound,~$\Delta T_\text{QCR}$, defined in eq~\ref{eq:QCRinequality} is also the same as eq~\ref{eq:min_Delta_T_est}, meaning that the measurement of the intensity is the optimal measurement in this scenario, even if quantum measurements were allowed. One more relevant observation is that the uncertainty analyzed by a linear error propagation model,~$\Delta T_\text{est} = \left\vert \frac{\partial \langle \hat{n}\rangle_\text{out}}{\partial T}\right\vert^{-1}\langle(\Delta \hat{n})^2\rangle_\text{out}^{1/2}$, turns out to be the same as eq~\ref{eq:min_Delta_T_est}. 

From eq~\ref{eq:min_Delta_T_est} it is clear that the estimation uncertainty scales with the inverse square root of the intensity, i.e.,~$N^{-1/2}$, where~$N$ is the mean of the underlying Poisson photon number distribution of the input coherent state. The right hand side of eq~\ref{eq:min_Delta_T_est} is the SNL and it can also be called the SQL in intensity sensing. Therefore, the estimation uncertainty or precision for intensity sensing using a coherent state of light is shot-noise limited.

\subsubsection{Shot-noise limited phase sensing}\label{sec:quantum_phase_sensing}
The paradigmatic scenario for phase sensing is in the form of a MZI, as shown in Figure~\ref{Fig:Sec3:Basic_Scheme}(b). Here, the goal is the estimation of the relative phase difference between the two paths. Consider that a coherent state~$\ket{\alpha}$ is fed into mode~$a$ of the first BS, and the vacuum is assumed to be in mode~$b$, i.e., the input state is~$\ket{\Psi}_\text{in}=\ket{\alpha}_\text{a}\ket{0}_\text{b}$. The BS operator can be written as~$\hat{B}(\tau,\theta)=\exp[\tau e^{i\theta}\hat{a}^{\dagger}\hat{b}-\tau e^{-i\theta} \hat{a}\hat{b}^{\dagger}]$ and transforms the operators~$\hat{a}$ and~$\hat{b}$ as~\cite{Campos1989, Fearn1987, Leonhardt1993}
\begin{align}
\hat{B}\hat{a}\hat{B}^{\dagger} &= \sqrt{T}\hat{a}-e^{i\theta}\sqrt{1-T}\hat{b},\\
\hat{B}\hat{b}\hat{B}^{\dagger} &= e^{-i\theta}\sqrt{1-T}\hat{a}+\sqrt{T}\hat{b},
\end{align}
where~$T=\cos^{2}\tau$ represents the transmittance of the BS and~$\theta$ is an associated phase shift which we choose to be~$\theta=\pi/2$ without loss of generality, so that the BS becomes symmetric between the two input ports. After the first BS with transmission~$T$, the outgoing state undergoes a relative phase shift that is induced by the operator~$\hat{U}(\phi)=\exp[i\frac{\phi}{2}(\hat{a}^{\dagger}\hat{a}-\hat{b}^{\dagger}\hat{b})]$, which transforms the operators as~$\hat{U}\hat{a}\hat{U}^\dag=e^{-i\phi/2}\hat{a}$ and~$\hat{U}\hat{b}\hat{U}^\dag=e^{i\phi/2}\hat{b}$. The state finally exits out of the second BS with transmission~$T$. For generality, we also include optical loss occurring between the two BSs, with the amount of loss characterized by channel transmission efficiencies~$\eta_\text{a}$ and~$\eta_\text{b}$ for the two modes. The output state is then written by~\cite{Loudon2000}
\begin{align}
\ket{\Psi(\phi)}_\text{out} 
=\ket{\alpha_\text{out}(\phi)}_\text{a} \ket{\beta_\text{out}(\phi)}_\text{b}, 
\end{align}
where~$\alpha_\text{out}(\phi)= T\sqrt{\eta_\text{a}} \alpha e^{i\frac{\phi}{2}}- (1-T) \sqrt{\eta_\text{b}} \alpha e^{-i\frac{\phi}{2}}$ and~$\beta_\text{out}(\phi)= i\sqrt{T(1-T)}\sqrt{\eta_\text{a}} \alpha e^{i\frac{\phi}{2}} + i\sqrt{T(1-T)} \sqrt{\eta_\text{b}} \alpha e^{-i\frac{\phi}{2}}$.
Suppose that we perform an intensity measurement at the two outputs and analyze the measurement results to estimate the relative phase~$\phi$. When the optimal estimator is chosen among many kinds of unbiased estimators, the estimation uncertainty is given by the CR bound, asymptotically saturable by the maximum likelihood method in the limit of a large sample size. It can be minimized by optimizing the transmittance~$T$ of the BSs, which can be shown to be~$T_\text{opt}=\sqrt{\eta_\text{b}}/(\sqrt{\eta_\text{a}}+\sqrt{\eta_\text{b}})$ in the vicinity of~$\phi=0$, for which the CR bound reads
\begin{align}
\Delta \phi_\text{CR} \approx \frac{1}{\sqrt{\nu}}\frac{\sqrt{\eta_\text{a}}+\sqrt{\eta_\text{b}}}{2\sqrt{\eta_\text{a}\eta_\text{b}}}\frac{1}{\sqrt{N}},
\label{eq:SIL}
\end{align}
where~$N=\vert \alpha \vert^2$ and~$\Delta \phi_\text{CR}$ is independent of the phase of the coherent state input. This is known as the standard interferometric limit (SIL) $\Delta\phi_\text{SIL}$~\cite{Demkowicz2009,Cooper2011}, which applies when an intensity measurement is used at the outputs of the second BS.
It is clear that~$\Delta \phi_\text{CR}$ of eq~\ref{eq:SIL} scales with~$N^{-1/2}$, also called the SQL. When~$\phi$ increases and approaches~$\pi/2$, the above optimal value of~$T_\text{opt}$ deviates, but the scaling with~$N$ is still kept. When~$\eta_\text{a}=\eta_\text{b}=\eta$,~$T_\text{opt}=1/2$, i.e., a 50:50 BS is the optimal choice in the above classical scenario, for which the CR bound reads~$\Delta \phi_\text{CR} = (\nu\eta N)^{-1/2}$, regardless of~$\phi$. The QCR bound of eq~\ref{eq:QCRinequality} can also be calculated using eq~\ref{eq:QFI_Pure} for the probe state which we take as the state just before the parameter encoding, i.e., the state present in between the first BS and the phase shifter in the MZI, which is written as~$\ket{\Psi}_\text{prob}=\ket{\alpha_\text{prob}}\ket{\beta_\text{prob}}$, with~$\alpha_\text{prob}=\sqrt{T \eta_\text{a}}\alpha$ and~$\beta_\text{prob}=i\sqrt{(1-T) \eta_\text{b}}\alpha$. The QFI is given by~$H = \vert \alpha_\text{prob}\vert^2 +\vert \beta_\text{prob}\vert^2= [T\eta_\text{a} +(1-T)\eta_\text{b}] N$, clearly showing that the QCR bound, $\Delta \phi_\text{QCR} \geq (\nu H)^{1/2}$,  is shot-noise-limited regardless of the value of~$T$ and type of measurement.

\subsection{Sub-shot-noise intensity sensing}\label{sec:sub_shot_noise_intensity_sensing}
Probing an analyte with a coherent state leads to the SQL (or SNL), where the estimation uncertainty scales with~$N^{-1/2}$. This is because a coherent state consists of discretized quantum particles (photons) that populate the energy levels of a spatial mode with a certain distribution, resulting in Poissonian photon number statistics on average over finite temporal intervals. As discussed in section~\ref{sec:parameter_estimation_theory}, the QCR bound on the estimation uncertainty is dependent on the probe state that becomes encoded with the information of a parameter to be estimated. An important question naturally arises: Can we further reduce the QCR bound below the SQL by optimizing the probe state? If this is possible, then we can also ask: What would be the optimal state to achieve the ultimate uncertainty bound? To address these questions, we now briefly show how intensity sensing and phase sensing can be further improved with the help of quantum states of light within the framework of the QCR bound. For more details, the latest review articles devoted to quantum metrology can be consulted~\cite{Demkowicz2015,Degen2017,Pirandola2018}. The techniques introduced here will then be used in section~\ref{sec:Quantum_plasmonic_sensors}, where we describe recent work on improving plasmonic sensors using quantum resources.

\subsubsection{Quantum-enhanced intensity sensing}\label{sec:Quantum_enhanced_intensity_sensing}
When a sample mean is used in intensity sensing as described above, the crucial quantity that affects the estimation uncertainty is~$\Delta I_\text{t}$. Is there a state that minimizes~$\Delta I_\text{t}$? To answer this, we need to look at photon number distributions. Individual photons undergo a Bernoulli process when passing through a BS, i.e., each photon is either transmitted with a probability~$T$ or reflected with a probability~$(1-T)$.
Given the incident photon number distribution~$p_\text{in}(m)$, the output photon distribution~$p_\text{out}(n\vert T)$ can be written as
\begin{align}
p_\text{out}(n\vert T)=\sum_{m=n}^{\infty}  \binom{m}{n}(\eta T)^n (1-\eta T)^{m-n} p_\text{in}(m).
\label{eq:pn_intensity_sensing_quantum}
\end{align}
This leads to the variance of the transmitted intensity given as 
\begin{align}
(\Delta I_\text{t})^2
&=\sum_{n=0}^{\infty} n^2 p_\text{out}(n\vert T) - (\sum_{m=0}^{\infty} n p_\text{out}(n\vert T))^2\nonumber\\
&= T\eta (1- T\eta) \langle \hat{n} \rangle + (T\eta)^{2} \langle (\Delta \hat{n})^2 \rangle, 
\end{align}
where~$\langle \hat{n} \rangle$ and~$\langle (\Delta \hat{n})^2 \rangle$ are the average and the variance of the photon number of the incident state, respectively. For a fixed average photon number~$\langle \hat{n} \rangle=N$, and given~$\eta$ and~$T$, we then have that the smaller the photon number variance~$\langle (\Delta \hat{n})^2 \rangle$ of the probe state, the smaller the variance~$(\Delta I_\text{t})^2$, thus reducing the estimation uncertainty. The state with the smallest photon number variance is the Fock state~$\ket{N}=(N!)^{-1/2}(\hat{a}^{\dagger})^{N}\ket{0}$, for which~$\langle (\Delta \hat{n})^2 \rangle=0$ and thus
\begin{align}
\Delta T_\text{est} =\sqrt{\frac{T(1-\eta T)}{\nu \eta N}}.
\label{eq:Delta_T_est_quantum}
\end{align}
This clearly shows a quantum enhancement by a factor of~$(1-\eta T)$ in comparison with eq~\ref{eq:min_Delta_T_est} obtained using a coherent state as a probe, although it still scales with~$N^{-1/2}$. Substituting the probability~$p_\text{out}(n\vert T)$ of eq~\ref{eq:pn_intensity_sensing_quantum} with~$p_\text{in}(m)=\delta_{m,N}$ into the discretized form for the FI of eq~\ref{eq:FisherInformation}, one can show that the CR bound is the same as eq~\ref{eq:Delta_T_est_quantum}.
For the outgoing state written as~$\hat{\rho}_{T}=\sum_n p_\text{out}(n\vert T) \ket{n}\bra{n}$, the QCR bound can also be calculated, consequently showing that the QCR bound is the same as the CR bound because~$\partial_{T} \ket{n}=0$~\cite{Paris2009}, as in the classical case discussed in section~\ref{sec:shot_noise_limited_sensing}. 
\begin{figure}[!t]
\centering
\includegraphics[width=0.47\textwidth]{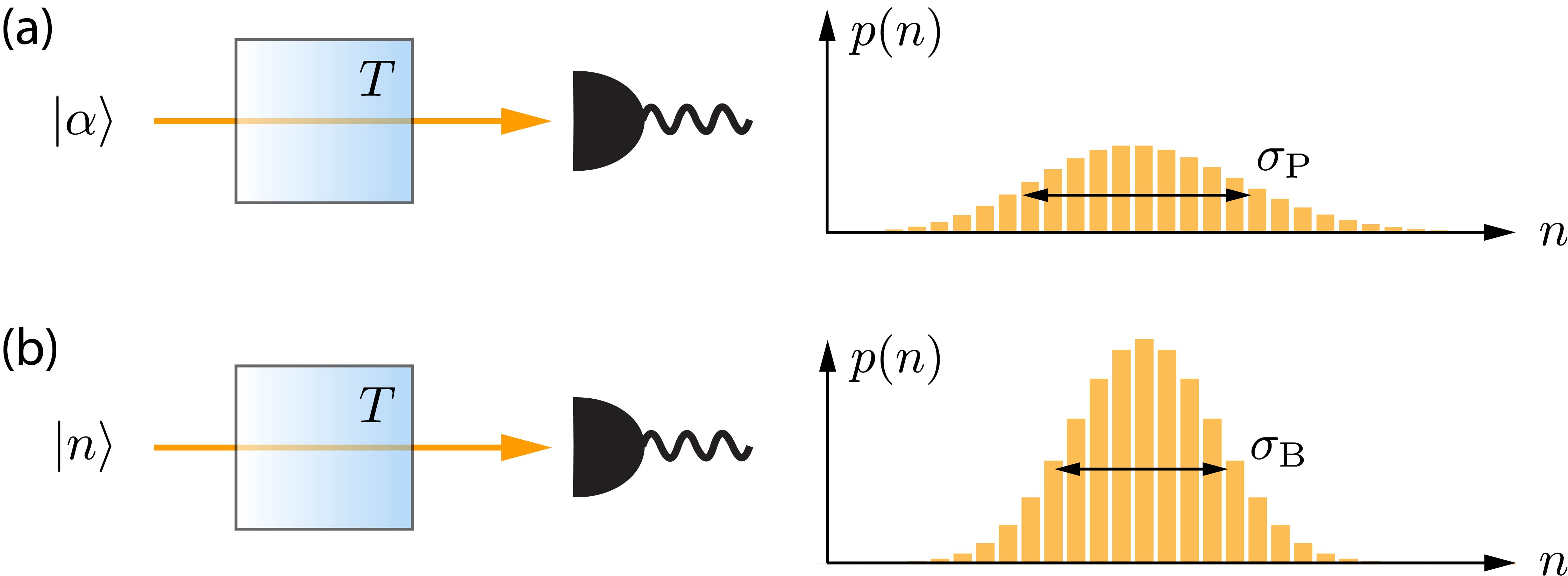}
\caption{Intensity sensing using classical and quantum states. (a) A coherent state is used to measure the transmission~$T$. The measured population distribution is Poissonian. (b) The photon number state is used to measure the transmission~$T$. The measured population distribution is binomial, with a smaller standard deviation,~$\sigma_B$, than the coherent state, $\sigma_P$.
}
\label{SNLintensity}
\end{figure}

The above quantum enhancement is achieved due to the fact that the photon number state has the least uncertainty, indeed no uncertainty, in photon number, whereas the coherent state has an uncertainty associated with the Poisson distribution in photon number. The probe states~$\ket{\alpha}$ and~$\ket{N}$ lead to Poisson and binomial photon number statistics in the transmitted light, respectively, as shown in Figs.~\ref{SNLintensity}(a) and (b). The variance of the binomial distribution,~$\sigma^{2}_\text{B}=\eta T(1-\eta T) N$, is smaller than that of the Poisson distribution,~$\sigma^{2}_\text{P}=\eta T N$, while their mean values are equal, i.e.,~$\mu_\text{B}=\mu_\text{P}=\eta T N$. 
As can be seen from the above simple formulation, it is known that the photon number state is the optimal state that minimizes the QCR bound, i.e., attaining the ultimate quantum limit in intensity sensing~\cite{Adesso2009, Monras2007}. 

\begin{figure*}[!t]
\centering
\includegraphics[width=0.7\textwidth]{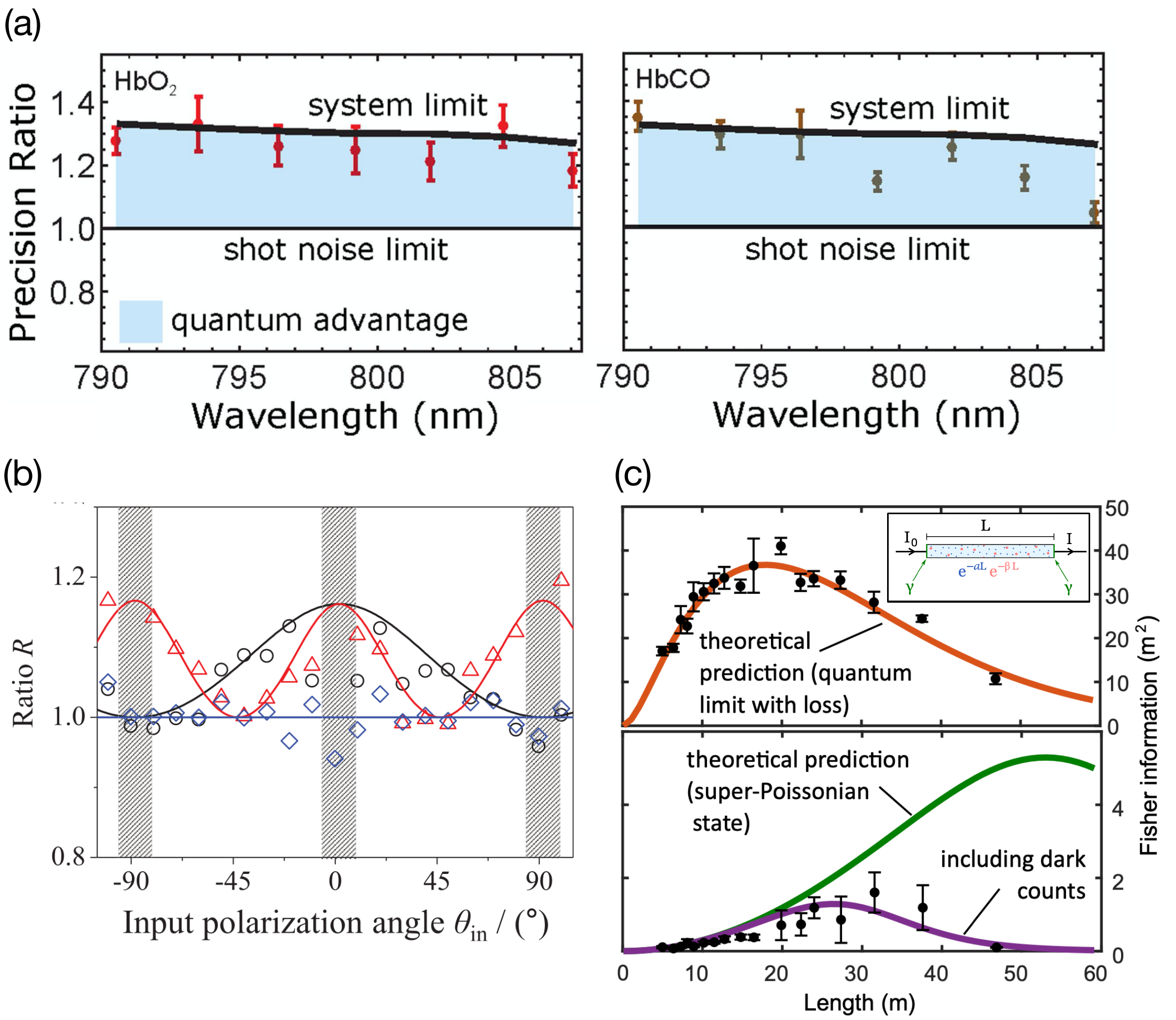}
\caption{
(a) Precision ratio of an absorption spectroscopy measurement using single photons ($\Delta T_{\rm est,q}$) to the SNL ($\Delta T_{\rm est,c}$) for different types of haemoglobin: (left) oxyhaemoglobin (HbO2) and (right) carboxyhaemoglobin (HbCO). The absorption parameter is experimentally estimated at individual wavelengths ranging from 790 to 808~nm. Adapted from ref~\citenum{Whittaker2017} (CC BY License).
(b) Precision ratio of a quantum polarimetry measurement using single photons to the SNL for sucrose solution. The optical activity of sucrose solution with concentration $C=$~0.5~g/ml is experimentally estimated at input polarization angles in a range from $-100^\circ$ to $100^\circ$ in steps of $10^\circ$. Reproduced with permission from ref~\citenum{Yoon2020}. Copyright 2020 IOP Publishing.
(c) (inset) The absorbance $a\in[0,\infty)$ of a medium is experimentally measured in the presence of other losses, such as surface losses $\gamma$ and copropagation loss $\beta$. The FI is analyzed, as a function of the length of a Pockels cell modulator, for an absorbance measurement with a two-mode squeezed vacuum state input and a coincidence counting scheme (upper). The latter is compared to the case using a single-mode thermal state of light with and without dark counts (lower). Adapted from ref~\citenum{Allen2019} (CC BY 4.0 License).
}
\label{fig:sensing_photons}
\end{figure*}
Intensity, or equivalently loss parameter sensing, has been considered with various probe states~\cite{Braun2018}. The optimal strategies of estimating a general one-parameter quantum process were studied in terms of the Kraus representation~\cite{Sarovar2006} and later the ultimate quantum bound on the estimation uncertainty in loss parameter sensing was derived~\cite{Monras2007}, which is eq~\ref{eq:Delta_T_est_quantum}. Gaussian probe states (a special class of quantum states whose Wigner function is a Gaussian function~\cite{Tan2019}) were considered in intensity parameter sensing~\cite{Monras2007}, and their sensing performances were shown to be improved by the use of a Kerr nonlinearity~\cite{Rossi2016}. It was shown that the ultimate bound is achievable only by the Fock state probe among all single-mode quantum probe states, including non-Gaussian probe states~\cite{Adesso2009}. The optimality of a Fock state probe in intensity sensing has been exploited in various types of sensing, such as in absorption spectroscopy to analyze the organic dye molecule dibenzanthanthrene~\cite{Rezus2012} or haemoglobin [see Figure~\ref{fig:sensing_photons}(a)]~\cite{Whittaker2017}. Other relevant studies include 
quantum polarimetry to measure the optical rotation occurring in chiral media [see Figure~\ref{fig:sensing_photons}(b)]~\cite{Yoon2020} and an experiment to measure the absorbance of a lossy medium [see Figure~\ref{fig:sensing_photons}(c)]~\cite{Allen2019}. Of particular interest to this review is that the Fock state probe has recently been used in plasmonic sensing~\cite{Lee2018}, whose details will be discussed in section~\ref{sec:Quantum_plasmonic_intensity_sensing}.

\subsubsection{Multiparameter or multimode intensity sensing}
\label{sec:Multi_Quantum_enhanced_intensity_sensing}
Single-mode intensity sensing can be improved by using ancillary modes~\cite{Monras2010}. 
It has been shown that entanglement can further improve the estimation of an unknown damping constant in an interferometric setup with specific probe states and measurements considered~\cite{Venzl2007}. Entangled Gaussian probes were shown to lead to a better strategy for discriminating lossy bosonic channels in the presence of a thermal environment~\cite{Invernizzi2011}. The estimation of a loss parameter can also be made simultaneously with the estimation of temperature~\cite{Monras2011} or phase~\cite{Crowley2014}. More generally, the estimation of multiple loss parameters needs to be considered with regards to several applications, such as in image sensing~\cite{Brida2010}, where the precise measurement of an image at individual pixels of an amplitude mask corresponds to the problem of the precise estimation of multiple loss parameters. 
Increasing the resolution of thermal electromagnetic sources can also be formulated as loss parameter estimation~\cite{Tsang2016, Nair2016, Lupo2016, Helstrom1973, Lu2018}. Other examples of importance are to probe the change of intensity parameters over different frequencies~\cite{Whittaker2017} or temporal modes~\cite{Taylor2016}. 

\begin{figure*}[!t]
\centering
\includegraphics[width=0.65\textwidth]{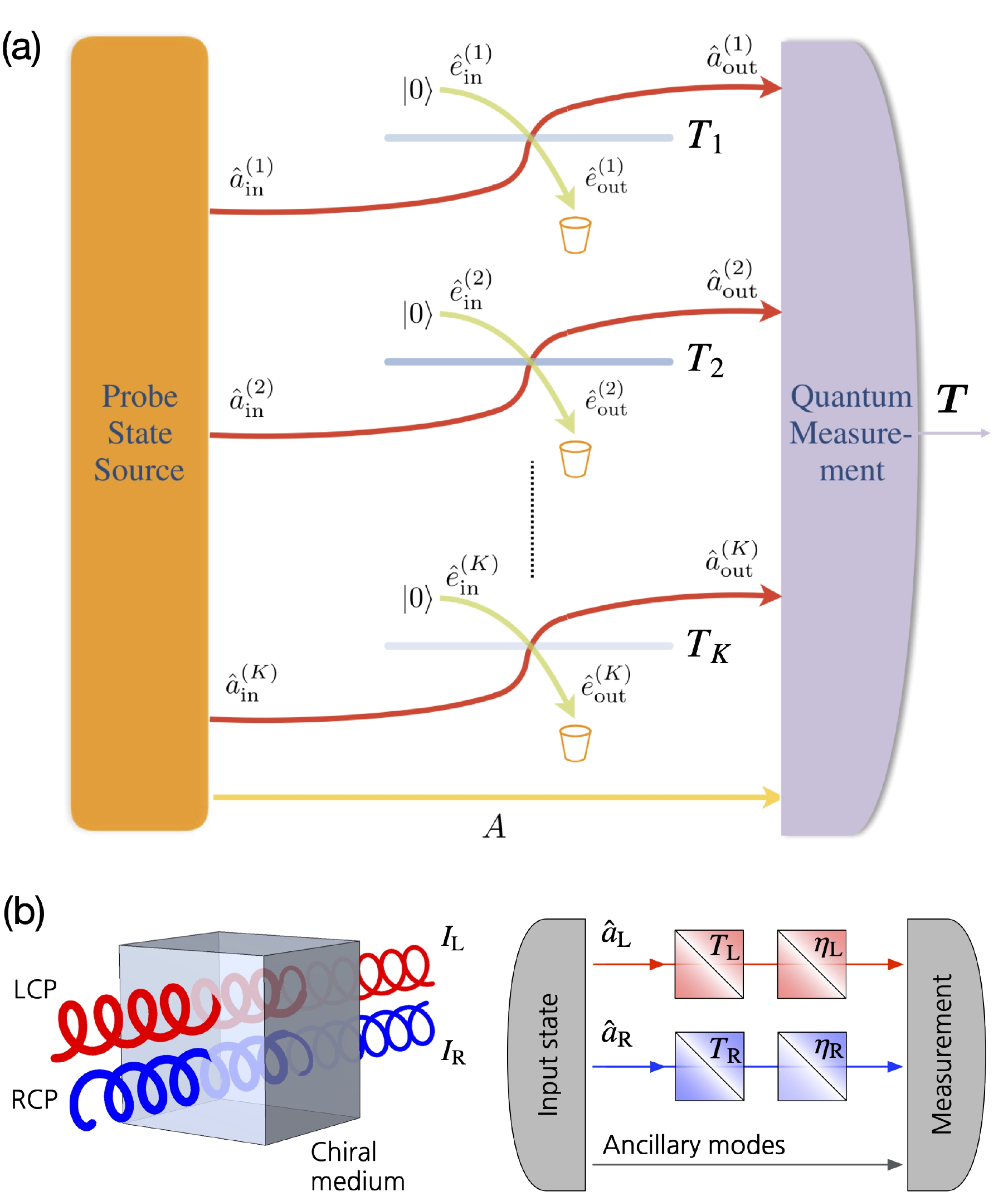}
\caption{Multiple intensity parameter sensing. (a) General ancilla-assisted strategy: Each signal mode (red) is sent to a fictitious BS with unknown transmittivity $T_j$ in the $j$th mode, while an ancilla (yellow) is kept lossless. The composite output state of the signal modes and the ancilla mode is measured to estimate the vector $\boldsymbol{T}$ consisting of all transmittivities. Virtual environment modes (green) are also considered and assumed to be in the vacuum state before entering the BSs. Adapted with permission from ref~\citenum{Nair2018}. Copyright 2018 the American Physical Society.  
(b) Left: Circular dichroism (CD) is experimentally observed by measuring the intensity-difference of the transmitted left-circularly polarized (LCP) light and right-circularly polarized (RCP) light upon propagation through a chiral medium. Right: General ancilla-assisted CD sensing scheme, similar to (a), but with additional losses $\eta_\text{L}$ and $\eta_\text{R}$ for the respective LCP and RCP modes. 
The transmittivities $T_\text{L}$ and $T_\text{R}$ represent the transmittivities of the LCP and RCP modes through a chiral medium, respectively. Reproduced from ref~\citenum{Ioannou2020}.
}
\label{fig:multi_intensity_sensing}
\end{figure*}
A recent study has derived the ultimate quantum bound to the estimation uncertainty of multiple intensity parameters and identified the optimal schemes reaching the ultimate quantum bound~\cite{Nair2018}. Consider the problem of the estimation of~$K$ transmittivities~$\{ T_k \}_{k=1}^{K}$ of~$K$ transmissive channels using what is called an `ancilla-assisted entangled parallel' strategy, as illustrated in Figure~\ref{fig:multi_intensity_sensing}(a). The probe states are sent through signal modes (S) representing the~$K$ transmissive channels modeled by~$K$ BSs, whereas ancillary modes (A) are used to transmit states entangled with the probe states and are kept lossless. To make the scenario more general, multimode features at individual channels are taken into account, so that~$M_k$ modes are assumed to be used for probing the~$k$th transmittance. The overall probe state can thus be written as a joint pure state~$\ket{\Psi}_\text{SA}$, with energy constraints on the individual channels, i.e.,~$\bra{\Psi}\hat{n}_{k}\ket{\Psi}_\text{SA}=N_k$ for~$k=1,\cdots,K$, where~$\hat{n}_k=\sum_{m=1}^{M_k}\hat{a}^{\dagger}_{k,m} \hat{a}_{k,m}$. It was shown that the maximum QFIM~$\boldsymbol{H}$ can be written as~\cite{Ioannou2020}
\begin{align}
\boldsymbol{H} = \text{diag} \left( \frac{\eta_1 N_1}{T_1 (1- \eta_1 T_1)}, \cdots, \frac{N_K}{\eta_K T_K (1-\eta_K T_K)} \right),
\label{eq:QFIM_ultimate_intensity}
\end{align}
where~$\eta_k$ is an additional but unwanted loss rate for the~$k$th channel. 
Such an ultimate quantum bound exhibits a quantum enhancement in comparison with the QFIM for a product coherent (PC) state probe written as
\begin{align}
\boldsymbol{H} = \text{diag} \left( \frac{\eta_1  N_1}{T_1}, \cdots, \frac{\eta_K N_K}{T_K} \right).
\end{align}
It has been shown that the above maximum QFIM of eq~\ref{eq:QFIM_ultimate_intensity} can be achieved by the product probe~$\ket{\Psi}=\bigotimes_{k=1}^{K} \ket{\text{NDS}}_k~$~\cite{Nair2018}, where~$\ket{\text{NDS}}_k$ is the so-called number-diagonal-signal (NDS) state with the energy constraint of~$N_k$ on the signal mode~\cite{Nair2011a,Nair2011b}, defined as
\begin{align}
\ket{\text{NDS}}_k = \sum_{\boldsymbol{n}_k\ge \boldsymbol{0}} \sqrt{p(\boldsymbol{n}_k)}\ket{\boldsymbol{n}_k}_\text{S}\ket{\phi_{\boldsymbol{n}_k}}_\text{A},
\end{align}
where~$\ket{\boldsymbol{n}_k}_\text{S}=\ket{n_1}_{k}\cdots \ket{n_{M_k}}_{k}$ is an~$M_k$-mode number state basis at the~$k$th signal channel,~$\{\ket{\phi_{\boldsymbol{n}_k}}_\text{A}\}$ is an orthonormal basis at the~$k$th ancillary channel, and~$p(\boldsymbol{n}_k)$ is the probability distribution of~$\boldsymbol{n}_k$. The signal energy constraint on~$k$th signal mode then reads as~$N_k=\sum_{m_k=0}^{\infty} m_{k} p(m_k)$ with~$p(m_k)=\sum_{n_1+\cdots+n_{M_k}=m_k} p(\boldsymbol{n}_k)$. 

According to the formulation introduced in section~\ref{sec:parameter_estimation_theory}, one can find the optimal measurement setting that reaches the ultimate multiparameter quantum bound. It has been shown that the optimal setting is the joint measurement with the Schmidt bases, i.e., the basis~$\ket{\phi_{\boldsymbol{n}_k}}$ on the ancillary modes and the number basis~$\ket{\boldsymbol{n}_k}_\text{S}$ on the signal modes~\cite{Nair2018}. This leads the FIM to be equal to the QFIM of eq~\ref{eq:QFIM_ultimate_intensity} for any~$\{ T_k \}$. Note that the multiparameter QCR bounds given by eq~\ref{eq:QFIM_ultimate_intensity} can be simultaneously achieved since the associated SLDs commute~\cite{Helstrom1976}.

The above formulation has recently been applied to one notable example, called circular dichrosim (CD) sensing, which is shown in Figure~\ref{fig:multi_intensity_sensing}(b). Typical CD sensing schemes aim to estimate the difference of the transmittivities between left- and right-circular polarization through a chiral medium composed of either chiral molecules~\cite{Greenfield2006, Micsonai2015} or chiral nanophotonic structures~\cite{Gansel2009, Passaseo2017, Collins2017}. By modeling CD sensing quantum mechanically [see Figure~\ref{fig:multi_intensity_sensing}(b)] and using eq~\ref{eq:QCRinequality_multi_distributed}, Ioannou \textit{et al.}~identified the ultimate quantum limit on the estimation uncertainty of an estimate $(T_\text{L}-T_\text{R})$ and investigated the optimality in terms of various quantum state inputs and quantum measurements~\cite{Ioannou2020}.

\subsubsection{Quantum noise reduction in intensity measurements}\label{sec:Quantum_noise_reduction_intensity_measurements}
The most widely exploited scheme for measuring intensity changes in both the classical and quantum regime is reference-assisted transmission or absorption spectroscopy. Here, the signal mode encodes a single intensity parameter of an object and a reference mode is kept unchanged. Examples include quantum imaging~\cite{Genovese2016}, quantum illumination~\cite{Lloyd2008, Lopaeva2013} and quantum sensing~\cite{Meda2017}. In particular, an intensity-difference measurement between the signal and reference output modes has often been performed in order to remove common excess noise between the two modes~\cite{Bachor2004}. This is a technique that has recently been used in classical plasmonic sensing to reach the SNL~\cite{Piliarik2009,Wang2011,Wu2004}.

The ultimate aim of using quantum probe states is to reduce the noise below the SNL. The noise reduction of the intensity-difference measurement of two modes~$a$ and~$b$ can be quantified by the so-called noise reduction factor (NRF)~\cite{Jedrkiewicz04,Bondani07,Blanchet08,Perina12}, defined by the ratio of the variance of the photon number difference between the signal and reference mode to that of coherent states with matching average photon number, written as
\begin{align}
\sigma=\frac{\langle [\Delta(\hat{n}_\text{b} - \hat{n}_\text{a})]^2 \rangle}{\langle \hat{n}_\text{a}\rangle+\langle \hat{n}_\text{b} \rangle}.
\label{eq:NRF}
\end{align}
It can be shown that~$\sigma\ge 1$ for all classical light, and thus the light is non-classical when~$\sigma < 1$ is measured. The NRF is minimized by light whose uncertainty in the photon number difference is minimal, i.e., the optimal probe state generally written in the form of the photon-number-correlated state~$\ket{\Psi}_\text{ab}=\sum_n w_n \ket{n,n}$ with any weights~$\{ w_n\}$, for which~$\sigma=0$. An example is the twin Fock (TF) state~$\ket{\text{TF}}=\ket{N,N}$ with~$\sigma^\text{(TF)}=0$. For a PC state~$\ket{\alpha,\beta}$, on the other hand,~$\sigma^\text{(PC)}=1$, which sets the SNL.

\begin{figure*}[!t]
\centering
\includegraphics[width=1\textwidth]{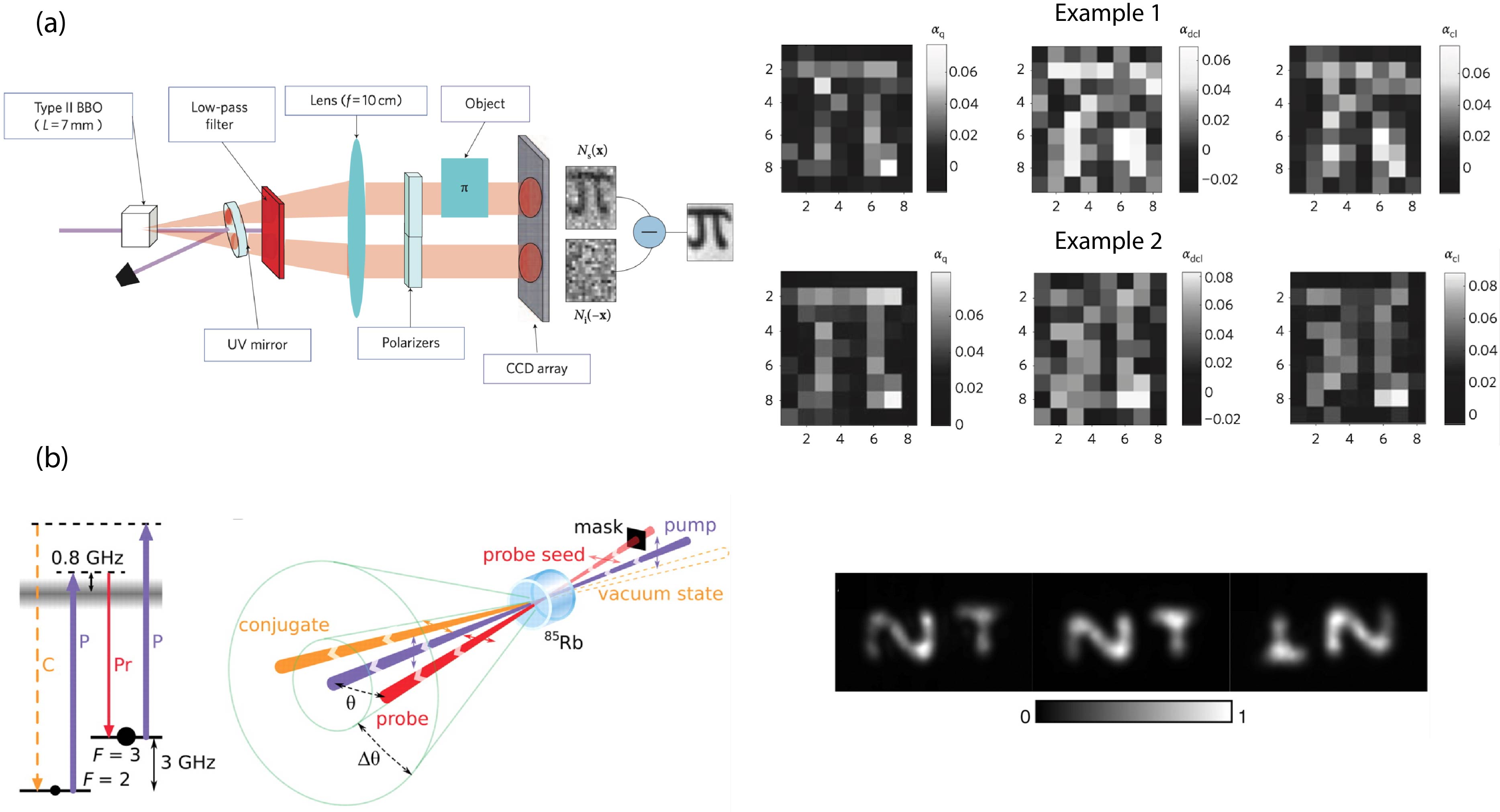}
\caption{
(a) (Left) Twin beam [i.e., the TMSV state of eq~\ref{eq:TMSV}] is generated via a SPDC process. One mode of the twin beam passes through a weakly absorbing object and arrives at a CCD camera array, while the other mode is directly sent to another area of the CCD array. A subtraction is performed between the two noisy images, realizing the intensity-difference measurement scheme, whereby an image of the object is constructed. (Right) For two example sets (upper and lower row) of the objects, three kinds of schemes are used: (left) differential-intensity measurement using a TMSV state, (middle) differential-intensity measurement using a PC state, and (right) direct-intensity measurement using single-mode coherent state. Adapted with permission from ref~\citenum{Brida2010}. Copyright 2010 Springer Nature. (b) (Bottom) The parametric process in the double-lambda scheme converts two pump (P) photons into one probe (Pr) photon and one conjugate (C) photon, generating a bright twin beam, i.e., the TMSD state of eq~\ref{eq:TMSD}. 
In the experiment, a bright pump beam couples with the probe and conjugate fields through a hot $^{85}$Rb vapor cell involving the double-lambda scheme. The spatially correlated bright twin beams are generated with cylindrical symmetry over a range of angles $\Delta \theta$ around the axis of the pump beam. When the spatial field profile of the probe seed is modified by an amplitude mask, an entangled image can be generated between the probe and conjugate outgoing field. (Left) Image of the probe seed entering a hot $^{85}$Rb vapor cell.
(Middle) Image of the outgoing probe beam. (Right) Image of the outgoing conjugate beam, clearly showing anti-spatial correlation with the image of the probe beam.
Reproduced with permission from ref~\citenum{Boyer2008}. Copyright 2008 AAAS.
}
\label{fig:Sub-Shot-Noise_Imaging}
\end{figure*}

Another useful photon-number-correlated state leading to~$\sigma=0$ is the two-mode squeezed vacuum (TMSV) state, often referred to as a twin beam. It has been widely used in experiments over the last few decades~\cite{Meda2017}. The TMSV state can be generated from a spontaneous parametric down conversion (SPDC) process~\cite{Burnham1970, Heidmann1987, Schumaker1985}. It is formally written as the outcome of a two-mode squeezing operation applied to a two-mode vacuum state and is written in the photon number basis as
\begin{align}
\ket{\text{TMSV}} = \hat{S}_{\text{2}}(\xi)\ket{0,0}=\sum_{n=0}^{\infty}c_{n}\ket{n,n},
\label{eq:TMSV}
\end{align}
where~$\hat{S}_{\text{2}}(\xi)=\exp[\xi^{*}\hat{a}\hat{b}-\xi \hat{a}^{\dagger}\hat{b}^{\dagger}]$ denotes the two-mode squeezing operator with~$\xi\in\mathbb{C}$ and~$c_{n}=(-e^{i\theta_\text{s}}\tanh r)^{n}/\cosh r$ for~$\xi=re^{i\theta_\text{s}}$. 
It is clear from eq~\ref{eq:TMSV} that the TMSV state exhibits a strong quantum correlation in the photon number between the two modes, leading to~$\sigma^\text{(TMSV)}=0$, whereas the individual modes follow thermal statistics, i.e.,~$\langle (\Delta \hat{n})^2 \rangle /\langle \hat{n} \rangle = N + 1$ with~$N=\sinh^2 r$. This strong photon number correlation has been used in various applications such as quantum ellipsometry~\cite{Abouraddy2001, Toussaint2004}, absorption/transmission measurements~\cite{Tapster1991, Jakeman1986, Hayat1999, Moreau2017, Losero2018}, quantum sensing~\cite{Zhang2015}, quantum illumination~\cite{Lopaeva2013}, quantum radiometry~\cite{Zwinkels2010, Polyakov2007}, and quantum gravity tests~\cite{Ruo-Berchera2013}. Figure~\ref{fig:Sub-Shot-Noise_Imaging}(a) presents an experimental demonstration of a quantum imaging technique using the nonclassical spatial correlation of TMSV states for a weakly absorbing object, in direct comparison with the experimental images obtained using classical light~\cite{Brida2010}. The TMSV state is normally regarded as a weak-field probe since the average photon number~$N$ is small, i.e., small~$r$, due to the weak nonlinearity of the material used for realizing the two-mode squeezing operation~\cite{Ruo-Berchera2009}. 

When sensors operating in the high-intensity regime are more preferable or favorable, one can employ a brighter entangled state at the expense of decreasing the quantum correlation. For example, the state produced by performing the two-mode squeezing operation on a vacuum state in one mode and coherent state in another (whose intensity can be high). This state is formally called a two-mode squeezed displaced (TMSD) state and has been investigated from various perspectives~\cite{Marino2009, Cai2015, Fang2015}. The TMSD state can be written as
\begin{align}
\ket{\text{TMSD}} = \hat{S}_{2}(\xi)\ket{\alpha}\ket{0},
\label{eq:TMSD}
\end{align}
for which~$\langle \hat{n}_\text{a} \rangle=\sinh^2 r + \vert \alpha \vert^2 \cosh^2 r$ and~$\langle \hat{n}_\text{b} \rangle= \sinh^2 r + \vert \alpha \vert^2 \sinh^2 r$ and 
\begin{align}
\sigma^\text{(TMSD)}=\frac{\vert \alpha \vert^2 }{ \vert \alpha \vert^2+2(1+\vert \alpha \vert^2)\sinh^2 r}.
\end{align} 
It is clear that for a given small~$r$, the intensity of the individual modes can be increased by cranking up the pump laser power (i.e.,~$\vert \alpha \vert^2$) incident to the nonlinear medium that produces the TMSD state. A large~$\vert \alpha \vert^2$ increases the NRF, but the NRF is always less than unity, implying that the quantum correlation of the TMSD state is always stronger than that of the classical case and can thus be exploited in quantum imaging or sensing. In an experiment, the TMSD state can be generated via a four-wave mixing (FWM) process in a double-lambda configuration provided by a~$^{85}\text{Rb}$ vapor cell~\cite{McCormick2008, Boyer2008a, Turnbull2013}, where a~$\text{NRF}\approx 87\%$, i.e., about~$-9~\text{dB}~[\approx 10\log_{10}(1-0.87)]$, has been measured. Such an intense quantum correlation with tens of~$\mu$W of optical power has been exploited in quantum sensing~\cite{Pooser2015, Clark2012}, imaging~\cite{Boyer2008, Lawrie2013b}, and quantum plasmonics~\cite{Lawrie2013a}. The application to quantum plasmonic sensing, in particular, will be described in more detail in section~\ref{sec:intmultimode}. Figure~\ref{fig:Sub-Shot-Noise_Imaging}(b) presents a demonstration of entangling spatial information of the probe beam with the conjugate beam by a FWM scheme that generates a TMSD state~\cite{Boyer2008}.

An interesting comparison can be made among the aforementioned quantum states of light in quantum imaging or intensity parameter sensing scenarios. Consider a setup, where the object with transmittance~$T$ is placed in the signal mode~$a$, and the signal and reference modes undergo inefficient transmission channels with factors~$\eta_\text{a}$ and~$\eta_\text{b}$, respectively, which include the detection efficiencies of the two detectors~\cite{Mandel1995}. 
The intensity-difference measurement operator~$\hat{n}_{-}=\hat{b}_\text{out}^{\dagger}\hat{b}_\text{out}-\hat{a}_\text{out}^{\dagger}\hat{a}_\text{out}$ can be written in terms of input operators using the Heisenberg picture~\cite{Loudon2000}. Here, states are static and operators contain the relevant dynamics. In this case, we can model the change of the transmittance and losses in the modes as fictitious BSs. This enables the output NRF to be calculated for the respective probe states~\cite{Loudon2000}. For example, the input-output relation given by 
\begin{align}
\hat{a}_{\text{out}}&=\sqrt{G\eta_\text{a} T}\hat{a}_{\text{in}}+\sqrt{(G-1)\eta_\text{a} T}\hat{b}^{\dagger}_{\text{in}}+\sqrt{(1-\eta_\text{a})T} \hat{a}_{\text{bath}}\nonumber\\
&\quad+\sqrt{(1-T)} \hat{a}_{\text{obj}},\label{aout}\\
\hat{b}_{\text{out}}&=\sqrt{G\eta_\text{b}}\hat{b}_{\text{in}}+\sqrt{(G-1)\eta_\text{b}}\hat{a}^{\dagger}_{\text{in}}+\sqrt{(1-\eta_\text{b})} \hat{b}_{\text{bath}}\label{bout}
\end{align}
can be applied to~$\hat{n}_{-}$ and the expectation value taken with respect to the initial vacuum state~$\ket{0}\ket{0}$ for the TMSV state probe, or the initial displaced state~$\ket{\alpha}\ket{0}$ for the TMSD state probe. 
Here,~$\hat{a}_{\text{bath}}~(\hat{b}_{\text{bath}})$ is the input operator of a fictitious BS describing a lossy channel of the signal (reference) mode~\cite{Loudon2000},~$\hat{a}_{\text{obj}}$ denotes the virtual input operator of the object, and~$G=\cosh^{2}r$ with~$r$ being the squeezing parameter and~$\theta_\text{s}=\pi$ assumed. The bath and object modes are taken to be in the vacuum initially.

\begin{figure*}[!t]
\centering
\includegraphics[width=0.7\textwidth]{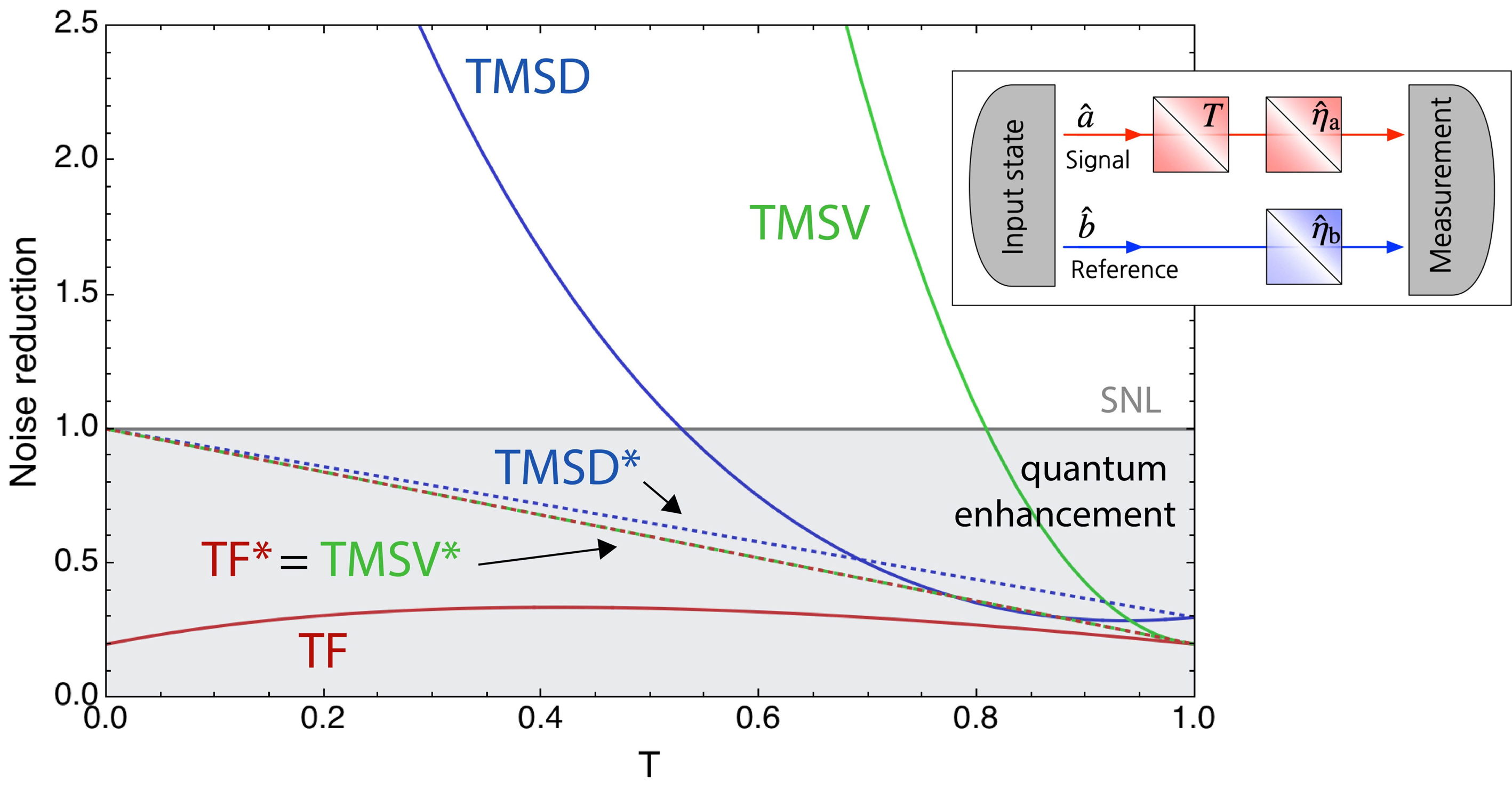}
\caption{Comparison of the noise reduction for different quantum states: TF, TMSV and TMSD. The parameters chosen are~$G=4.5$,~$|\alpha|^2=10$ and~$\eta_{\rm a}=\eta_{\rm b}=0.8$. The starred versions (TF$^*$, TMSV$^*$ and TMSD$^*$) correspond to the case when~$T\eta_{\rm a}=\eta_{\rm b}$.
Inset: General two-mode scheme using an input state that consists of the signal and reference mode to estimate the intensity parameter $T$ in the presence of losses $\eta_\text{a,b}$ for each mode. Here, the transmissive object with $T$ and linear losses are described by fictitious BSs. 
}
\label{NRFcompare}
\end{figure*}

The NRFs of the respective output states can thus be written as
\begin{align}
\sigma_\text{out}^\text{(TF)}
&=1- \frac{T^2 \eta_\text{a}^2+\eta_\text{b}^2}{T\eta_\text{a}+\eta_\text{b}}\label{TFratio},\\
\sigma_\text{out}^\text{(TMSV)}
&= 1+\frac{G|\alpha|^2(T \eta_\text{a}-\eta_\text{b})^{2}-(T^{2} \eta_\text{a}^{2}+\eta_\text{b}^{2})}{T \eta_\text{a}+\eta_\text{b}},\label{TMSVratio}\\
\sigma_\text{out}^\text{(TMSD)}
&\approx 1+\frac{2(G-1)\left[G(T \eta_\text{a}-\eta_\text{b})^{2}-\eta_\text{b}^{2}\right]}{GT \eta_\text{a}+(G-1)\eta_\text{b}},
\label{TMSDratio}
\end{align}
where the limit~$\vert \alpha\vert^{2}\gg1$ has been taken into account for the TMSD state probe~\cite{Jasperse2011}. An example of them is shown in Figure~\ref{NRFcompare}.

The quantum enhancement in noise reduction, i.e., a sub-shot-noise measurement, can be identified by the condition~$\sigma_\text{out}<1$. The latter holds for~$\sigma_\text{out}^\text{(TF)}$ at any value of~$T$, implying that the use of the TF state always achieves the sub-shot-noise measurement. The other two states, on the other hand, reach the noise level below the SNL when~$G(T\eta_\text{a}-\eta_\text{b})^2$ is less than~$T^2 \eta_\text{a}^2+\eta_\text{b}^2$ and~$\eta_\text{b}^2$, respectively. These conditions can be simply satisfied when~$T\eta_\text{a}$ is close to~$\eta_\text{b}$, motivating a consideration of the ideal case that~$T \eta_\text{a}=\eta_\text{b}=\eta$, for which the NRFs of eqs~\ref{TFratio} to \ref{TMSDratio} become
\begin{align}
\sigma_\text{out}^\text{(TF)}&= 1-\eta\label{TFratio2},\\
\sigma_\text{out}^\text{(TMSV)}&= 1-\eta\label{TMSVratio2},\\
\sigma_\text{out}^\text{(TMSD)}&\approx 1-\frac{2\eta(G-1)}{2G-1}.\label{TMSDratio2}
\end{align}
These NRFs are shown in Figure~\ref{NRFcompare} as the starred versions. As expected, a sub-shot-noise measurement can be achieved with all the considered states at any value of~$T$. This has been experimentally demonstrated in various studies~\cite{Boyer2008a, Pooser2016, McCormick2007}, including the imaging of a weakly absorbing object, for which~$\eta_\text{b}=\eta_\text{a}$ and~$T\approx 1$ are of interest~\cite{Brambilla2008, Brida2010}.
The behaviors of the TF and TMSV state exhibit the maximal noise reduction in intensity sensing, as predicted from the previous discussion where the NSD state has been shown to lead to the optimal results~\cite{Nair2018, Tame2019}.

Another useful and practical figure of merit that has often been used in the two-mode scheme is the SNR with respect to an intensity-difference measurement~\cite{Brida2010}, defined by 
\begin{align}
\text{SNR} = \frac{\vert \langle \hat{n}_\text{b}\rangle -\langle\hat{n}_\text{a} \rangle\vert}{\langle [\Delta(\hat{n}_\text{b} - \hat{n}_\text{a})]^2 \rangle^{1/2}}.
\end{align}
When symmetric probes are used (such as the TF, and TMSV states), i.e.,~$\langle \hat{n}_\text{a}\rangle_{\rm in}=\langle \hat{n}_\text{b}\rangle_{\rm in} =N$ and~$\langle (\Delta \hat{n}_\text{a})^2 \rangle_{\rm in}=\langle (\Delta \hat{n}_\text{b})^2 \rangle_{\rm in}$, the SNR can be rewritten as
\begin{align}
\text{SNR}=\frac{(\eta_\text{b}-\eta_\text{a} T)N}{\sqrt{N \left[ (T\eta_\text{a}-\eta_\text{b})^2 Q_\text{M} +2T\eta_\text{a}\eta_\text{b} (\sigma_\text{in}-1) +(T\eta_\text{a}+\eta_\text{b})\right]}},
\label{eq:SNR}
\end{align}
where~$Q_\text{M}=\langle (\Delta \hat{n})^2 \rangle_\text{in} /\langle \hat{n} \rangle_\text{in} -1$ is the Mandel Q-parameter~\cite{Mandel1979} and~$\sigma_\text{in}$ is the NRF for the input state~\cite{Jedrkiewicz04,Bondani07,Blanchet08,Perina12}. The Mandel Q-parameter is equal to or greater than zero for all classical light, so non-classical light can be identified by~$Q_\text{M}<0$. As mentioned previously, the NRF~$\sigma_\text{in}\ge 1$ for all classical light, and thus light is non-classical when~$\sigma_\text{in}<1$ is observed. It is clear to see that the SNR of eq~\ref{eq:SNR} increases with decreasing~$Q_\text{M}$ and~$\sigma_\text{in}$. One finds that the minimum values of~$Q_\text{M}$ and~$\sigma_\text{in}$ are~$-1$ and~$0$, respectively, which are obtained for the TF state~$\ket{\Psi_\text{in}}=\ket{N}_\text{a}\ket{N}_\text{b}$.

The quantum enhancement in the SNR can be quantified by the ratio of the SNR for the particular quantum state probe to that for the balanced PC probe state, written by
\begin{align}
R_\text{SNR}
&=\frac{\text{SNR}_\text{q}}{\text{SNR}_\text{c}}\nonumber\\
&=\sqrt{\frac{T\eta_\text{a} +\eta_\text{b}}
{(T\eta_\text{a}-\eta_\text{b})^2 Q_\text{M} +2T\eta_\text{a}\eta_\text{b} (\sigma_\text{in}-1) +(T\eta_\text{a}+\eta_\text{b})}}.
\label{eq:SNR_Ratio}
\end{align}
Note that quantum enhancement with~$R_\text{SNR}>1$ is observed when~$(T\eta_\text{a}-\eta_\text{b})^2 Q_\text{M} + 2T\eta_\text{a}\eta_\text{b}(\sigma_\text{in}-1)<0$, so a quantum probe with~$-1\le Q_\text{M}<0$ or~$0\le \sigma_\text{in}<1$ can be shown to be advantageous. 
The optimal state that maximizes the quantum enhancement is the TF state, for which~$Q_\text{M}=-1$ and~$\sigma_\text{in}=0$, which are the minimum bound for the respective parameters. In this optimal case,~$R_\text{SNR}>1$ regardless of~$T$ and~$\eta_\text{a,b}$.

The use of the TMSV state of eq~\ref{eq:TMSV}, for which~$Q_\text{M}=N$ and~$\sigma_\text{in}=0$, provides a quantum enhancement only when~$(T\eta_\text{a}-\eta_\text{b})^2 N < 2T\eta_\text{a}\eta_\text{b}$ with respect to the balanced PC state with~$Q_\text{M}=0$ and~$\sigma_\text{in}=1$. 
This indicates that probing with a TMSV state is more advantageous as~$N$ decreases. When~$\eta_\text{a}\approx \eta_\text{b}$ and~$T\approx1$, or~$\eta_\text{a} T \approx \eta_\text{b}$, the use of a TMSV state becomes more useful even with high~$N$. Such an enhancement has been observed in quantum imaging experiments~\cite{Brida2010, Meda2017}.
One can also find other useful states with~$\sigma_\text{in}=0$ and~$Q_\text{M}<0$, for example, pair coherent states~\cite{Agarwal1986} or finite-dimensional photon-number entangled states~\cite{Lee2012}. These states have all been considered in quantum plasmonic sensing, as we will discuss in detail in section~\ref{sec:Quantum_plasmonic_intensity_sensing}.

\subsection{Sub-shot-noise phase sensing}\label{sec:sub_shot_noise_phase_sensing}
\subsubsection{Phase sensing in Mach-Zehnder interferometers}
\label{losslesscohsqueeze}
The shot-noise limited interferometric phase sensor described in section~\ref{sec:shot_noise_limited_sensing} can be improved by making use of quantum resources. The original idea proposed by Caves, who largely contributed to the advent of the field of quantum metrology, is to inject a squeezed vacuum state~$\vert \xi \rangle$ into the input mode~$b$ of the first BS in the MZI shown in Figure~\ref{Fig:Sec3:Basic_Scheme}(b), where previously the input state in mode~$b$ was assumed to be the vacuum in the shot-noise limited classical sensor~\cite{Caves1981}. 
The squeezed vacuum state is defined as~$\vert \xi\rangle = \hat{S}_{1}(\xi)\vert 0\rangle$, where the single-mode squeezing operator is represented by~$\hat{S}_{1}(\xi)=\exp(\frac{1}{2}\xi^{*}\hat{a}^{2}-\frac{1}{2}\xi\hat{a}^{\dagger 2})$ and~$\xi=re^{i\theta_\text{s}}$, with~$r\ge0$ manifesting a squeezing magnitude in the variance of the corresponding quadrature variable,~$x_{\theta_\text{s}/2}=\langle \hat{x}_{\theta_\text{s}/2}\rangle$, with $\hat{x}_{\theta_\text{s}/2}=(e^{-i\theta_\text{s}/2}\hat{a}+e^{i\theta_\text{s}/2}\hat{a}^\dag)/\sqrt{2}$. Here,~$\theta_\text{s}/2$ represents the angle of the axis along which squeezing takes place~\cite{Scully1997}. The total input state is then~$\ket{\Psi}_\text{in}=\ket{\alpha}_\text{a}\ket{\xi}_\text{b}$, with the total average photon number of the two modes being~$N=|\alpha|^{2}+\sinh^{2}r$. In the absence of loss ($\eta_\text{a}=\eta_\text{b}=1$), this input state is transformed by the first BS into a non-separable state~$\ket{\Psi}_\text{prob}=\hat{B}(\pi/4,\pi/2)\ket{\Psi}_\text{in}$ that probes the phase information, resulting in~$\ket{\Psi(\phi)}=\hat{U}(\phi)\ket{\Psi}_\text{prob}$ before the measurement. For such a lossless case, Pezz{\'e} and Smerzi showed that the QCR bound to the estimation uncertainty of the relative phase~$\phi$ is written by~\cite{Pezze2008}
\begin{align}
\Delta \phi_\text{QCR}= \frac{1}{\sqrt{\nu}\sqrt{\vert\alpha\vert^{2}e^{2r}+\sinh^{2}r}},
\label{eq:QFI_phase_lossless_quantum}
\end{align}
where the optimal phase condition between the displacement and squeezing parameters is set, i.e.,~$\cos(2\theta_\text{c}-\theta_\text{s})=1$. Note that the QCR bound~$\Delta \phi_\text{QCR}$ is the same for any true value of~$\phi$, and it can reach Heisenberg scaling~$\Delta \phi_\text{QCR} \propto N^{-1}$ for~${N\gg1}$, when~$\vert\alpha\vert^{2}\simeq\sinh^{2}r\simeq N/2$. It can be shown that~$\Delta \phi_\text{QCR}\approx e^{-r}/\sqrt{\nu N}$ when~$\vert\alpha\vert^{2}\gg \sinh^{2}r$~\cite{Caves1981} and~$\Delta \phi_\text{QCR}\approx e^{-r}/\sqrt{\nu N (4\vert \alpha\vert^2+1)}$ when~$\vert\alpha\vert^{2}\ll \sinh^{2}r$~\cite{Pezze2008}. These cases show that too small or too large amounts of squeezing cannot achieve Heisenberg scaling, but they do provide sub-shot-noise sensing due to the additional factor that multiplies~$N^{-1/2}$.

Interestingly, the squeezed vacuum state considered above has been shown to be the optimal state that can be inserted into mode~$b$ of the lossless MZI when the coherent state is inserted into mode~$a$, as it minimizes the estimation uncertainty of the relative phase~$\phi$~\cite{Lang2013}. This indicates that the initial idea proposed by Caves turns out to be a good choice as a modification for better phase sensing in a MZI setup. The original goal here was to improve classical gravitational wave detectors, where a strong classical field is used in one mode of the interferometric setup~\cite{Ligo2011}. This has now led to the latest improvements to the LIGO detector~\cite{Tse2019} and the Virgo detector~\cite{Acernese2019}. Furthermore, the QCR bound of eq~\ref{eq:QFI_phase_lossless_quantum} is not only theoretical, but practically attainable by photon-number-counting detectors at the two output ports of the MZI~\cite{Pezze2008}, or a parity detection scheme at the output port~$a$ of the MZI~\cite{Seshadreesan2013}. 

When a coherent state does not need to be used in mode~$a$ of the MZI, the optimal input state for phase estimation using the MZI setup is two single-mode squeezed vacuum states with anti-phases, i.e.,~$\ket{\Psi}_\text{opt}=\hat{S}_\text{a}(-r)\hat{S}_\text{b}(r)\ket{0,0}$ with~$r \in \mathbb R$~\cite{Lang2014}, for which the QCR bound reads~$\Delta \phi_\text{QCR} = 1/\sqrt{\nu N(N+1)}$ with~$N=2\sinh^2 r$ being the total average photon number.

The MZI phase sensing setup has sometimes been considered with a single phase shift in mode~$a$, i.e.,~$\hat{U}(\phi)=e^{i\phi \hat{a}^{\dagger}\hat{a}}$~\cite{Sparaciari2015, Sparaciari2016}. However, in this setting particular care must be made in quantifying the extent to which the estimation uncertainty is reduced, since an optical phase can only be defined in a relative sense with respect to a reference phase~\cite{Jarzyna2012}. Hence, a sensing analysis with a single phase shift is valid only when a reference beam with a certain phase is assumed, whose resource also needs to be counted when determining the dependence of the uncertainty on~$N$.

In the case of loss in both arms of the MZI, the calculation of the QCR bound becomes complicated due to the probabilistic nature of photon loss that needs to be taken into account. In this case, it is more convenient to use the relation between the QFIM~$\boldsymbol{H}$ and the Bures distance~${\cal D}_\text{B}^2$ for the infinitesimally close states~$\hat{\rho}_{\boldsymbol{\phi}}$ and~$\hat{\rho}_{\boldsymbol{\phi}+\text{d}\boldsymbol{\phi}}$~\cite{Bures1969, Braunstein1994a,Facchi2010}. Using the relevant formulation introduced in Appendix~\ref{appendix:B}, one can thus derive the QCR bound for the case~$\eta_\text{a}=\eta_\text{b}=\eta$, written as~\cite{Gard2017, Ono2010}
\begin{align}
\Delta \phi_\text{QCR}= \frac{1}{\sqrt{\nu}\sqrt{\vert\alpha\vert^{2}\frac{\eta}{(1-\eta)+e^{-2r}\eta}+\eta \sinh^{2}r}}. 
\label{eq:QFI_phase_lossy_quantum}
\end{align}
It is clear that~$\Delta \phi_\text{QCR}$ increases with loss, i.e., as~$\eta$ decreases, whereas the lossless QCR bound of eq~\ref{eq:QFI_phase_lossless_quantum} is recovered when~$\eta=1$. 
It can also be shown that the Heisenberg scaling~$\Delta \phi_\text{QCR}\propto N^{-1}$ promised for the lossless cases starts to deteriorate as~$\eta$ decreases.

The QCR bound~$\Delta \phi_\text{QCR}$ of eq~\ref{eq:QFI_phase_lossy_quantum} 
has been shown to be achievable by a linear optical setup using a weak local oscillator field and photon counting~\cite{Ono2010}. Other types of measurement schemes have been investigated, for example, homodyne detection with a measurement operator~$\hat{M}=\hat{x}_{\phi_\text{HD}}$, a single-mode intensity measurement with~$\hat{M}=\hat{a}^{\dagger}\hat{a}$, an intensity-difference measurement with~$\hat{M}=\hat{b}^{\dagger}\hat{b}-\hat{a}^{\dagger}\hat{a}$, and a parity measurement with~$\hat{M}=(-1)^{\hat{a}^\dagger\hat{a}}$~\cite{Gard2017,Oh2017}. In the case of homodyne detection~(see Appendix~\ref{appendix:C}), it was shown that this offers a nearly optimal measurement scheme and approaches the QCR bound of eq~\ref{eq:QFI_phase_lossy_quantum} in the large power regime. Alternatively, the robustness of the TMSV state against photon loss enables one to achieve sub-shot-noise limited phase sensing for a wide range of the phase parameter space without pre- nor post-selection~\cite{You2020bb}.

\begin{figure*}[!t]
\centering
\includegraphics[width=1\textwidth]{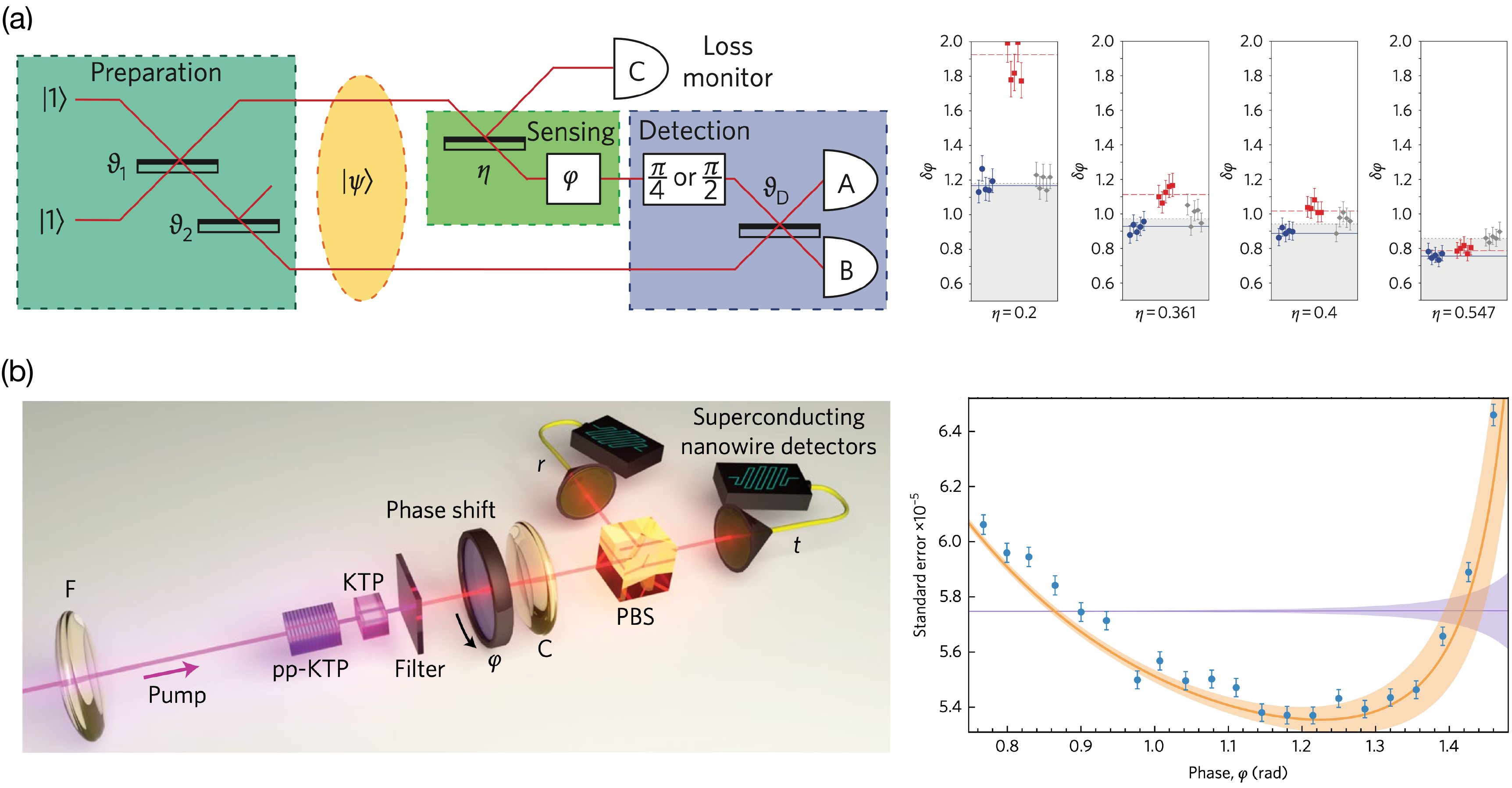}
\caption{
(a) (left) Linear optical network to demonstrate optimal phase estimation in the presence of loss. Initial BSs prepare the optimal probe state depending on the amount of loss measured via a loss monitor in a proof-of-principle demonstration. (right) Phase estimation uncertainties measured using two-photon optimal states (blue circles), NOON (red squares) states, and weak-coherent states to set the SIL regime (gray diamonds), for five phase values $\phi=0, \pm0.2, \pm0.4$~rad under several lossy conditions, given by different transmittivities $\eta$. Horizontal lines represent the theoretical CR bounds for individual input states in the presence of loss. Reproduced with permission from ref~\citenum{Kacprowicz2010}. Copyright 2010 Springer Nature. (b) (left) Two-photon NOON state undergoes a phase rotation in a collinear interferometer. The signal and idler photons are separated via a PBS and measured by superconducting nanowire detectors. The high efficiency and visibility in the experiment meant that post-selection was not needed, leading to the demonstration of unconditional quantum enhancement. (right) Experimentally measured standard deviation of the estimated phases with error bars determined via the standard bootstrapping technique. The theoretical SNL (purple line) and CR bound for phase sensing with NOON states (orange line) are shown with 95\% confidence regions. Reproduced with permission from ref~\citenum{Slussarenko2017}. Copyright 2017 Springer Nature.
}
\label{fig:Phase_sensing_definite_photons}
\end{figure*}

\subsubsection{Phase sensing with NOON states}\label{sec:Phase_sensing_with_NOON_states}
In addition to exploiting continuous variable states such as the squeezed vacuum state in phase sensing, discrete variable states or definite-photon-number states have also been considered as probe states for phase sensing~\cite{Polino2020}. The most famous state is the so-called NOON state, defined as~$\ket{\text{NOON}}=\frac{1}{\sqrt{2}} (\ket{N,0} + \ket{0,N})$, where~$N$ photons are found in either mode~$a$ or~$b$~\cite{Bollinger1996, Lee2002, Dowling2008}, exhibiting~$N$-particle maximal entanglement~\cite{Mermin1990}.  The generation of NOON states is very difficult with current technology, and they have been limited to~$N=5$ in photonic systems~\cite{Afek2010,Israel2012}. The NOON state possesses a very useful feature that the variance~$\langle (\Delta \hat{n}_{-})^2 \rangle~$ of the photon number difference between the two modes is maximal among the states with~$N$ photons in total, consequently maximizing the QFI written by~$H=4\langle (\Delta \hat{G})^2 \rangle$, where~$\hat{G}=\hat{n}_{-}/2$ is the generator of the relative phase shift. When the NOON state is employed as a probe state, the QCR bound is found to be~$\Delta\phi_\text{QCR}=1/\sqrt{\nu}N$, clearly manifesting a Heisenberg scaling enabled by the maximal variance of the photon number difference of the probe state. It is interesting to note that injecting a coherent state and a squeezed vacuum state into the two input ports of the first BS in a MZI generates an effective NOON state, as the output state has a large overlap with the NOON state, i.e.,~$\hat{B}(\pi/4,\pi/2)\ket{\alpha,\xi} \approx \ket{\text{NOON}}$ for~$ \sinh^2 r = \vert \alpha\vert^2=N/2$ and~$N \gg 1$~\cite{Pezze2008, Afek2010}. In the MZI setup, the coherent state and squeezed vacuum state thus exploit a large variance~$\langle (\Delta \hat{n}_{-})^2 \rangle$, similar to the NOON state probe, in order to achieve sub-shot-noise sensing. 

The QCR bound for the NOON state has been shown to be achievable by the measurement of the observable~$\hat{A}_N=\ket{0,N}\bra{N,0}+\ket{N,0}\bra{0,N}$~\cite{Mitchell04}, or parity detection~\cite{Bollinger1996}. One major obstacle in using the NOON state for phase sensing from a practical perspective is that it is extremely sensitive to photon loss, since the loss of a single or a few photons drastically changes the state's photon number distribution and its variance~$\langle (\Delta \hat{n}_{-})^2 \rangle$ quickly decreases. The estimation uncertainty associated with the NOON state thus becomes large even with a moderate amount of photon loss~\cite{Sahovar2006,Shaji2007,Gilbert2008,Huver2008}. Such an extreme sensitivity to photon loss can be alleviated by exploiting partial entanglement at the expense of a degraded performance, while still achieving sub-shot-noise limited behavior~\cite{Huelga1997}. Hence, adding other photon-number components (e.g.,~$\vert k, N-k\rangle$ for~$k<N$) into the NOON state helps phase sensing with NOON states become more robust to loss and provide sub-shot-noise sensing even in the presence of loss. In particular, in phase sensing with the definite-photon-number state~$\ket{\Psi}_\text{prob}=\sum_{k=0}^{N} c_k \ket{k,N-k}$ with~$N$ particles distributed over the two modes, one can find the optimal distribution~$\{c_k\}$ that maximizes the QFI and consequently enables one to beat the SQL or SNL at any loss level~\cite{Dorner2009, Demkowicz2009}. The latter has been experimentally demonstrated with optimally engineered definite-photon-number states~\cite{Kacprowicz2010} [see Figure~\ref{fig:Phase_sensing_definite_photons}(a)], and has been considered theoretically for quantum plasmonic sensing, as will be discussed in more detail in section~\ref{sec:phasedesc}. The unconditional quantum enhancement in phase estimation precision with NOON state has also been experimentally demonstrated in a nearly lossless scenario with significantly increased efficiency and visibility~\cite{Slussarenko2017} [see Figure~\ref{fig:Phase_sensing_definite_photons}(b)].

\begin{figure*}[!t]
\centering
\includegraphics[width=0.65\textwidth]{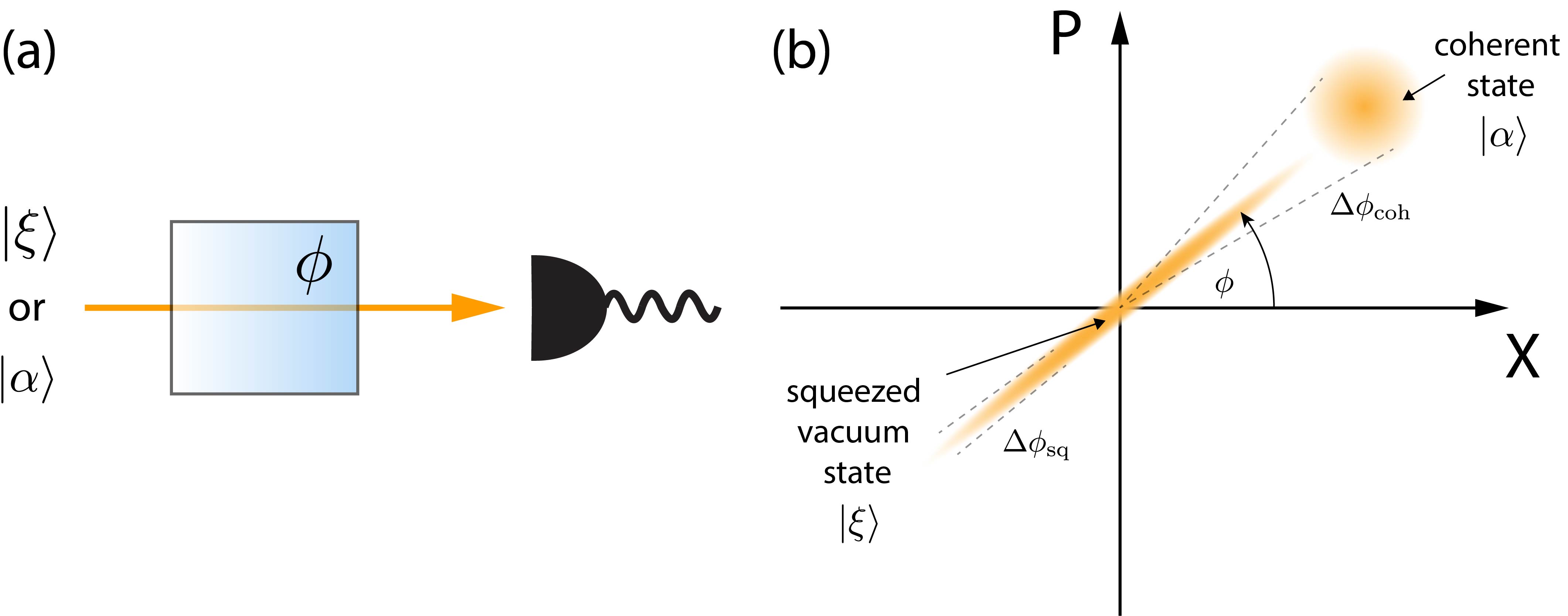}
\caption{Single-mode phase sensing. (a) A coherent state or a squeezed vacuum state are used to measure the phase~$\phi$. (b) The phase space representation for the states. The coherent state has an estimation precision of~$\Delta \phi_{\rm coh}$ for the phase~$\phi$, whereas the squeezed vacuum state has a smaller estimation precision of~$\Delta \phi_{\rm sq}$. The initial phase of the squeezed state, $\ket{\xi}$, is $\theta_{\rm s}=\pi$ and the phase affected squeezed state $\ket{\xi e^{2 i\phi}}$ rotates in phase space counter-clockwise by the angle $\phi$.
}
\label{Singlephase}
\end{figure*}

The most commonly used detection scheme for phase sensing with NOON states is the~$N$-fold coincidence detection scheme that counts the number of events when~$N$ photons are detected simultaneously~\cite{Wolfgramm2013}. This allows one to investigate the~$N$-fold detection modulation with the phase~$\phi$, leading to the probability of a coincidence detection given by
\begin{align}
p_\text{coin}(\phi) = \frac{ f_N\left[1+ {\cal V}_N \cos(N\phi) \right]}{2},
\label{eq:prob_coincidence}
\end{align}
where~$f_N$ is the proportion of the input state that causes an~$N$-fold coincidence detection event, and~${\cal V}_N$ is the~$N$-photon visibility. The CR bound for this scheme can be calculated with the underlying probabilities~$p_\text{coin}(\phi)$ for the detection event and~$1-p_\text{coin}(\phi)$ for the no detection event, and is written as
\begin{align}
\Delta \phi_\text{coin} =\frac{2\sqrt{p_\text{coin}(\phi)\left[1-p_\text{coin}(\phi)\right]}}{f_N {\cal V}_N N \vert \sin(N\phi)\vert} \le  \frac{1}{f_N {\cal V}_N N \vert \sin(N\phi)\vert},
\label{eq:CRbound_coincidence}
\end{align}
where the upper bound takes into account the worst case of the probability~$p_\text{coin}(\phi)=1/2$.
For comparison with the classical benchmark, the CR bound~$\Delta \phi_\text{coin}$ in eq~\ref{eq:CRbound_coincidence} can be compared with~$\Delta \phi_\text{SIL}$ of eq~\ref{eq:SIL}, leading to an inequality~$\Delta \phi_\text{coin} < \Delta \phi_\text{SIL}$ for a sensor to be so-called super-sensitive~\cite{Resch2007, Thomas-Peter2011}. The inequality can be specified, when~$\vert \sin(N \phi)\vert =1$, as
\begin{align}
1<\frac{f_N^2 {\cal V}^2_N N}{\tilde{\eta}},
\label{eq:Inequality_super-sensitivity}
\end{align}
where~$\tilde{\eta}= \left[ 2\sqrt{\eta_\text{a} \eta_\text{b}}/(\sqrt{\eta_\text{a}}+\sqrt{\eta_\text{b}})\right]^2$. For the threshold visibility defined as~${\cal V}_N^{\text{(th)}}=\sqrt{\tilde{\eta}/f_N^2 N}$, the above inequality can be written as~${\cal V}_N > {\cal V}_N^{\text{(th)}}$, i.e., only a measured visibility higher than the threshold visibility demonstrates `super-sensitivity'~\cite{Resch2007}. The latter can be considered as a genuine criterion for the quantum enhancement of the precision in sensing experiments with NOON states instead of `super-resolution', which has sometimes been misinterpreted, since it can also be produced by only classical light and projective measurements~\cite{Resch2007}. The above theory will be used in section~\ref{sec:qplasphasesense} when describing quantum plasmonic sensors that have considered the use of NOON states.

A simple example of NOON state phase sensing uses the two-photon NOON state~$\frac{1}{\sqrt{2}}(\ket{2,0}+\ket{0,2})$. It can be readily created by exploiting Hong-Ou-Mandel (HOM) interference~\cite{HOM1987}. The HOM interference occurs when two single photons are injected into a lossless 50/50 BS, i.e., the output state is written as~$\hat{B}(\pi/4,\pi/2)\ket{1,1}=(\ket{2,0}+\ket{0,2})/\sqrt{2}$ up to a global phase, where the component~$\ket{1,1}$ is absent due to destructive interference caused by the indistinguishable paths leading to the same output state~$\ket{1,1}$. When the two-photon NOON state undergoes a relative phase shift described by the operator~$\hat{U}(\phi)$, the outgoing state can be written up to a global phase as~$(\ket{2,0}+e^{-2i\phi}\ket{0,2})/\sqrt{2}$ before the measurement. The super-resolution with~$\phi$-modulation is observed due to the multiplied constant factor of~$N=2$ in the exponential factor~\cite{Wolfgramm2013}. For two-photon NOON states, one way to realize the required two-fold modulation measurement is by sending the two modes of the NOON state (now with the phase encoded) into a lossless 50/50 BS and detecting the coincidence of single photons at the output using single-photon detectors. The visibility of the two-photon measurement signal (as~$\phi$ is modulated) is then~${\cal V}_2=2\eta_\text{a}\eta_\text{b}/(\eta_\text{a}^2+\eta_\text{b}^2)$, which enables the super-sensitivity inequality to be reduced to approximately~$(\eta_\text{a}+\eta_\text{b})/2>0.8$, i.e., super-sensitive phase sensing with the two-photon NOON state can be achieved when the total loss is less than~$20\%$ on average~\cite{Chen2018}. This is an important point when using a plasmonic sensor with NOON states, as will be discussed in more detail in section~\ref{sec:qplasphasesense}.

Interferometric phase sensing has been studied with various quantum states and measurements that beat the SQL and achieve the HL, including the twin-Fock state~$\ket{N}\ket{N}$ incident into a MZI~\cite{Holland1993}, the photon-number correlated state~\cite{Yurke1986a, Yurke1986b, Shapiro1993, Dowling1998, Pezze2006, Huver2008, Tae-Woo_Lee2009}, the TMSV state with parity detection~\cite{Anisimov2010}, Schr{\" o}dinger's cat states~\cite{Joo2011}, and multi-headed coherent states~\cite{Lee2015a,Lee2020a}. More details of these states and their performance can be found in ref~\citenum{Polino2020}.

\subsubsection{Single-mode phase sensing}
Instead of two-mode phase sensing, single-mode phase sensing has also been investigated for the sake of simplicity. Here, the implicit assumption is that the phase of a reference mode is set. The latter is enabled by employing
phase-sensitive detection schemes, e.g., homodyne detection~\cite{Yuen1983,Schumaker1984} or general-dyne measurement~\cite{Genoni2014}. Thus, the phase information in single-mode phase sensing cannot be extracted from a probe state by phase-insensitive detection schemes, such as photon-number counting, i.e.,~$\partial_\phi p(n\vert \phi)=0$.

Consider that a single-mode state of light~$\vert \Psi\rangle_\text{in}$ undergoes a phase shift described by the operator~$\hat{U}(\phi)=e^{i\phi\hat{a}^{\dagger}\hat{a}}$, as shown in Figure~\ref{Singlephase}(a). As in the intensity sensing case, one may consider states having a small uncertainty in their phase for single-mode phase sensing. One typical quantum state whose phase uncertainty can be reduced below the SNL is the squeezed vacuum state,~$\vert \xi\rangle$. The phase shift operator~$\hat{U}(\phi)$ rotates a state in phase space about the origin and the output state is~$\vert \Psi\rangle_\text{out}=\hat{U}(\phi)\vert \xi\rangle =\vert \xi e^{2i \phi}\rangle$, for which the QCR bound is obtained as~\cite{Monras2006,Pinel2013}
\begin{align}
\Delta \phi_\text{QCR}=\frac{1}{\sqrt{\nu}\sqrt{8(N+N^2)}},
\label{eq:QCRB_sm_sq}
\end{align}
where~$N=\sinh^{2}r$ represents the average photon number of the initial squeezed vacuum state. It is clear that the QCR bound of eq~\ref{eq:QCRB_sm_sq} scales with~$N^{-1}$, not only reaching Heisenberg scaling~\cite{Monras2006, Pinel2013} but also beating the SQL of~$\Delta\phi=1/\sqrt{\nu 4 N}$ that is obtained for a single-mode coherent state input. Such a quantum enhancement in phase sensing is enabled by the smaller phase uncertainty of the squeezed vacuum state as compared to the coherent state with the same energy, as shown in Figure~\ref{Singlephase}(b). One also finds that the photon number state that is the optimal state for single-mode intensity sensing cannot encode any phase information since the photon number state exhibits a full phase uncertainty~\cite{Loudon2000}. 

Generally speaking, to measure the change of a particular physical quantity being induced by a parameter~$x$ with better precision or lower estimation uncertainty, the probe state showing the least uncertainty in that particular parameter is the most useful. A more analytical understanding can be made through the relation between the QFI and the fidelity ${\cal F}$ between two infinitesimally close states~$\hat{\rho}(x)$ and~$\hat{\rho}(x+\text{d}x)$ for a given parameter~$x$~\cite{Oh2019b,Banchi2015}, i.e.,~$H(x)=\lim_{\text{d}x\rightarrow 0} 8\{1-{\cal F}[\hat{\rho}(x),\hat{\rho}(x+\text{d}x)]\}/(\text{d}x)^2$ (see also Appendix~\ref{appendix:B}). That is, the estimation capability is related to the ability to distinguish two infinitesimally close states~$\hat{\rho}(x)$ and~$\hat{\rho}(x+\text{d}x)$ and a better distinction can be made with the least uncertainty in~$x$ of the probe state.

One can show that the QCR bound of eq~\ref{eq:QCRB_sm_sq} is reached by homodyne detection, which measures the quadrature variable of~$\hat{x}_{\theta_\text{HD}}=(e^{-i\theta_\text{HD}}\hat{a}+e^{i\theta_\text{HD}}\hat{a}^{\dagger})/\sqrt{2}$~\cite{Monras2006,Olivares2009,Oh2019a} by setting the optimal homodyne angle~$\theta_\text{HD}$ depending on the value of~$r, \theta_\text{s}$ and $\phi$,~\cite{Oh2019a}. The optimality of homodyne detection holds in the absence of loss, i.e., probing with a pure squeezed vacuum state, whereas a realistic squeezed state of light involves an inevitable thermal photon contribution~\cite{Vahlbruch2016}, for which the QCR bound can only be obtained by performing an exotic measurement with projectors over the eigenstates of the SLD operator~\cite{Oh2019b,Oh2019a}.

\subsubsection{Multiple-phase sensing}\label{sec:Multiple_phase_estimation}
Single-phase estimation can be extended to estimating multiple phases,~$\boldsymbol{\theta}=(\theta_{1},\theta_{2},\cdots,\theta_d)^\text{T}$. This is relevant to applications such as phase imaging~\cite{Preza1999}, which measure phase contrast and interference or gravitational wave detectors~\cite{Freise2009}, which measure multiple parameters. It is also relevant to quantum plasmonic imaging, which will be discussed in more detail in section~\ref{sec:intmultimode}. The estimation uncertainty of the total of all phases is governed by the covariance matrix~$\text{Cov}(\boldsymbol{\theta})$ and lower bounded by the QCR bound, as given in eq~\ref{eq:QCRinequality_multi}. The uncertainty of the total of all phases estimated can be quantified by the sum of the individual uncertainties:~$\vert \Delta\boldsymbol{\theta}\vert^2 =\sum_j (\Delta \theta_j)^2 =\text{Tr}[\text{Cov}(\boldsymbol{\theta})]$. This quantity is often compared using three typical cases: (i) a scheme using optimal classical state inputs, setting the classical benchmark or SQL, (ii) a scheme that estimates the phases individually by using optimal separable quantum state inputs, which is called `individual estimation' or a `local strategy', and (iii) a scheme that estimates all the phases simultaneously by using optimal entangled state inputs and a collective measurement (when necessary), which is called `simultaneous estimation' or a `global strategy'. The comparison of these three cases aims to address the following questions: Can schemes using quantum resources beat the SQL in multi-parameter estimation? Is a simultaneous estimation approach beneficial as compared to an optimal (maybe quantum) individual estimation approach? Relevant studies attempting to answer these questions have recently been started from various points of view, arising from the fact that multiparameter estimation is non-trivial, and depends on the kind of parameters estimated, as well as type of sensing scenario~\cite{Liu2020}.

\begin{figure*}[!t]
\centering
\includegraphics[width=0.95\textwidth]{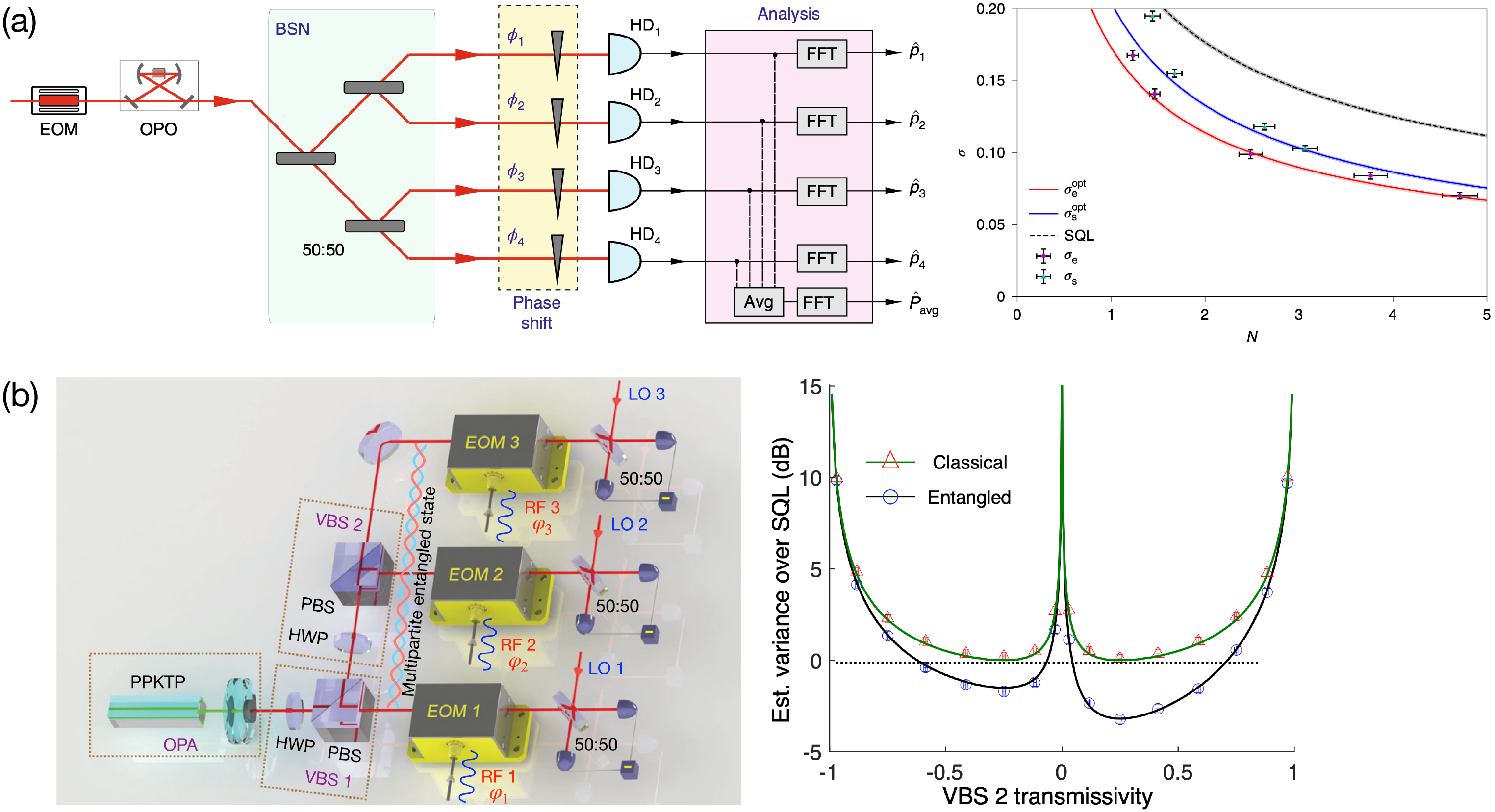}
\caption{(a) (left) An experimental set-up for distributed phase sensing with $M = 4$. A displaced squeezed state input is split into four identical and entangled probes through a BS network (BSN). The multiple-phase-encoded probe state is measured with homodyne detection (HD) set-ups, from which the average phase of multiple phases is estimated. 
(right) The precision of the estimated average phase for different average numbers $N$ of photons per sample are compared between the entangled scheme ($\sigma_\text{e}$) and the separable scheme ($\sigma_\text{s}$). Their theoretical precision (solid lines) as well as the SQL (dashed line) are also presented. Reproduced with permission from~\cite{Guo2020}. Copyright 2020 Springer Nature.
(b) (left) A phase-squeezed state injected into two variable BSs configures the continuous-variable multipartite entangled probe to estimate the average displacement. An individual displacement on the squeezed phase quadrature variable is induced by an electro-optic modulator (EOM) driven by radio-frequency (RF) fields. The output state is measured by homodyne detection, leading to the estimate of the average displacement. (right) The variance of the estimated values as a function of the transmittivity of the second BS (VBS2) is compared among different cases: entangled sensors (circles and black curves for experiment and theory), classical separable sensors (triangles and green curves for experiment and theory) and the SQL (black orizontal dotted line). Reproduced and adapted with permission from ref~\citenum{Xia2020}. Copyright 2020 the American Physical Society.
}
\label{fig:Distributed_sensing}
\end{figure*}

In one of the earliest works on this topic, Humphreys \textit{et al.}~showed that the quantum enhancement in the precision of simultaneous estimation of multiple phases can be obtained by a coherent superposition of~$N$ photons among~$d$ modes, in which individual phases~$\theta_{j=1,\cdots,d}$ are encoded~\cite{Humphreys2013}. It has been shown that the considered entangled state is optimal among the class of definite-photon-number states, with~$N$ being the total number of photons, and the advantage over individual estimation scaling with~${\cal O}(d)$. The particular optimal measurements identified in ref~\citenum{Humphreys2013} have recently been generalized in ref~\citenum{Pezze2017}, which derived the necessary and sufficient conditions for projective measurements to saturate the multiparameter QCR bound in the case of pure probe states. 

Among~$N$-particle photonic states, Holland-Burnett states~\cite{Holland1993} and NOON states have been shown to be the optimal states for simultaneous multiple-phase estimation when~$d=2$ in the absence of loss and decoherence~\cite{Liu2017}. Such non-Gaussian states achieve a factor of~$d$ improvement over individual estimation, as shown in ref~\citenum{Humphreys2013}, but the use of Gaussian states offers no more than a factor of 2 improvement~\cite{Gagatsos2016}. Furthermore, when both definite-photon-number states and indefinite-photon-number states are all considered together, the precision given by the QCR bound for simultaneous estimation can be obtained or even further enhanced through an individual estimation strategy using proper mode-separable input states~\cite{Knott2016}. It was found that a large particle-number variance within each mode plays a crucial role in improving the QCR bound of multiple-phase estimation. This comparable or even better sensing performance provided by individual schemes suggests experimentally more favorable settings for multiple-phase estimation. Moreover, the ultimate quantum limit for simultaneous multiple-phase estimation is only achievable by an optimal state that manipulates entanglement among both particles and modes~\cite{Gessner2018}. A more conclusive mathematical proof has been provided by Proctor \textit{et al.}, showing that entanglement in simultaneous estimation leads to no fundamental precision enhancement over individual estimation when the generators of the multiple parameters commute and yields no more than a factor of 2 enhancement even when the generators do not commute~\cite{Proctor2018}. 

Another interesting but more challenging scenario is to estimate multiple phases that are governed by non-commuting generators, e.g., the three components of a magnetic field in terms of the spatial coordinates~\cite{Jones2009}. For this scenario, a three-fold improvement over individual estimation strategies has been shown to be achievable by simultaneous estimation schemes using permutationally invariant quantum states~\cite{Baumgratz2016}, including the superposition of three Greenberger-Horne-Zeilinger-type states, each of which is known to attain the QCR bound of estimating a magnetic field aligned along one of the specific axes~\cite{Giovannetti2006}. The work in ref~\citenum{Baumgratz2016} also showed that too much entanglement is detrimental for achieving a Heisenberg scaling in terms of the total number of particles~$N$. In general, for non-commuting generators, there exits a tradeoff for the precisions among the individual estimators, i.e., more precise estimation of one parameter leads to less precise estimation of the others~\cite{Vidrighin2014, Crowley2014, Yuan2006}. Hou \textit{et al.} have recently derived the minimal tradeoff for the precision of simultaneous estimation of a three-dimensional magnetic field, finally leading to the identification of the ultimate quantum limit~\cite{Hou2020}.

Entanglement appears to lie at the heart of multiparameter estimation and has been studied from several perspectives, but here again, in the estimation of multiple phases, entanglement does not always guarantee a quantum enhancement~\cite{Knott2016, Gessner2018}. However, entanglement becomes a more significant factor in estimating a global parameter that is composed of multiple parameters to be encoded across multiple modes or locations, e.g., distributed sensing or networked sensing. The global phase parameter that is often considered is a linear combination of multiple phases, written as~$\tilde{\theta}=\sum_j n_j \theta_j$, with positive~$n_j$ and normalization~$\sum_j n_j=1$. For this, eq~\ref{eq:QCRinequality_multi_distributed} needs to be used. The effect of quantum entanglement in the uncertainty~$\Delta \tilde{\theta}$ of distributed sensing has sparked some interest. Recent work suggests useful schemes using quantum entanglement to gain an enhancement over schemes that measure the~$\{ \theta_j\}$ individually and then compute a global parameter~$\tilde{\theta}$ via classical communication.

Along the lines mentioned above, Proctor \textit{et al.} have shown that entanglement can significantly enhance the precision only when global parameters are of interest in estimation~\cite{Proctor2018}. Ge \textit{et al.}~considered a linear optical network and separable quantum inputs to show that injecting separable quantum states into only a few modes out of all the modes achieves a quantum enhancement~\cite{Ge2018}, resulting in the Heisenberg scaling~$1/Nd$ in terms of both the constrained photon number~$N$ and the number of parameters~$d$. Particularly, TF states have been shown to be useful quantum states achieving the Heisenberg scaling~$\propto1/Nd$, showing a quantum improvement over an individual estimation approach with the precision scaled as~$1/N\sqrt{d}$. Zhuang \textit{et al.}~have proposed theoretical schemes that use a squeezed vacuum input being injected into a BS array to estimate a linear combination of displacement parameters, or that of multiple phases~\cite{Zhuang2018}. The proposed schemes were shown to achieve Heisenberg scaling in the absence of loss, and have recently been experimentally demonstrated for phase parameters~\cite{Guo2020} [see Figure~\ref{fig:Distributed_sensing}(a)]
and for displacement parameters~\cite{Xia2020} [see Figure~\ref{fig:Distributed_sensing}(b)]. These proposed schemes can be enhanced by continuous-variable error correction to reinstate the Heisenberg scaling, at least up to moderate values of~$d$, even in the presence of loss or decoherence~\cite{Zhuang2020}. The schemes have all used homodyne detectors, but single-photon detectors can also be used with an anti-squeezing operation that transforms the initial squeezed vacuum state into the vacuum, which achieves Heisenberg scaling in distributed phase sensing~\cite{Gatto2019}. 

In the special case that Gaussian states are employed as an input into an array of BSs to estimate the average of independent phase shifts, Oh \textit{et al.}~have identified the optimal scheme that exploits partially entangled Gaussian probe states, as maximally entangled probe states are rather detrimental~\cite{Oh2020}. 

As is evident from the above-mentioned recent work, many studies appear to show that entanglement may not be the only quantity that determines the characteristic behavior of the multiple-phase quantum sensors under investigation. In such a sense, a new operational concept called multiparameter squeezing has recently been suggested to identify metrologically useful states and optimal estimation strategies~\cite{Gessner2020}. This can be seen as a counterpart to the NRF that characterizes quantum enhancement in quantum noise reduction in intensity measurements. This is highly relevant to quantum plasmonic imaging.

\subsubsection{Quantum sensing with SU(1,1) interferometers}\label{sec:nonlinear_interferometer}
\begin{figure*}[!t]
\centering
\includegraphics[width=0.95\textwidth]{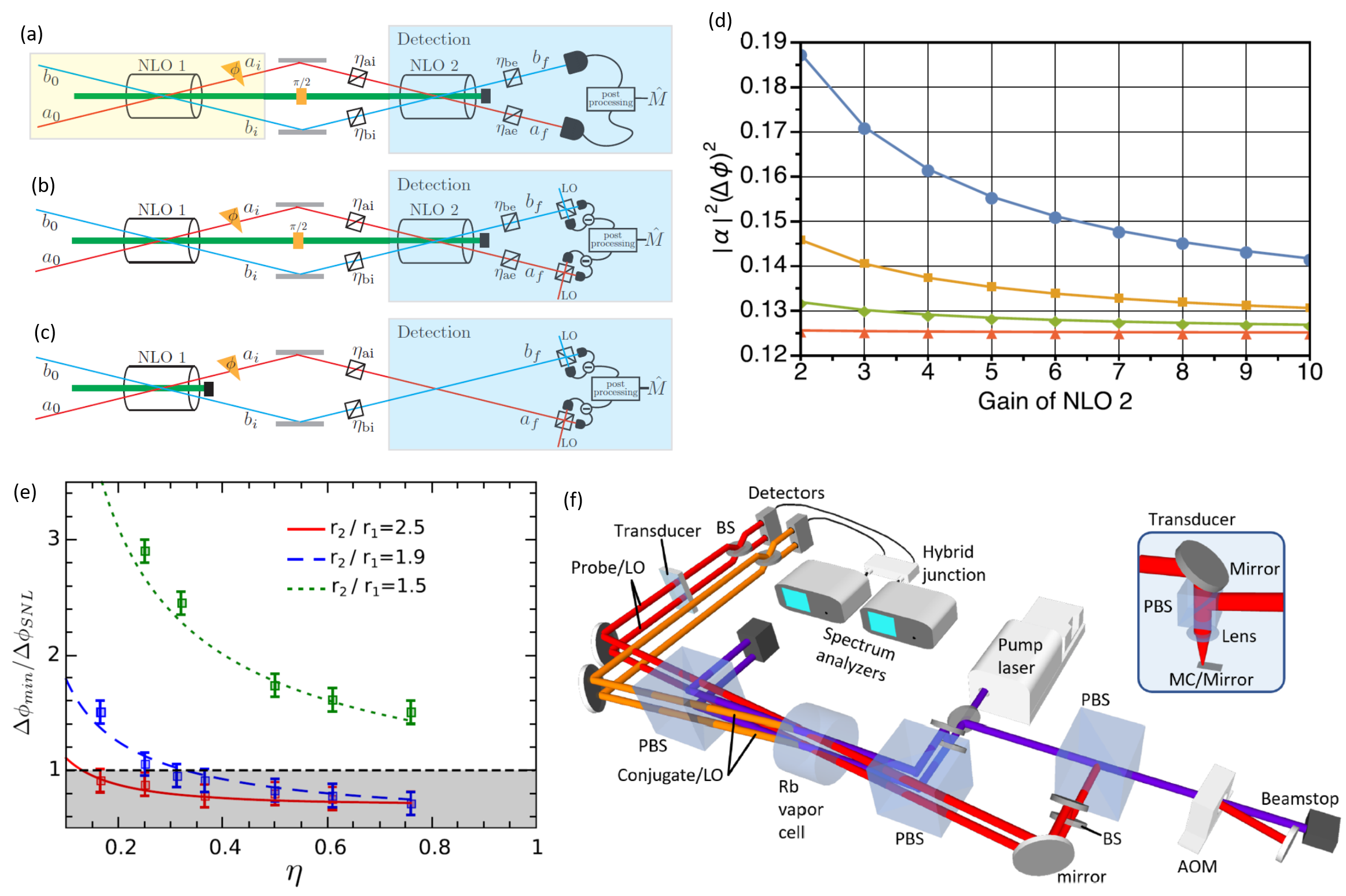}
\caption{
(a-c): SU(1,1) nonlinear interferometers with direct intensity readout, with dual homodyne readout, and with a truncated layout including dual homodyne readout after a single nonlinear amplifier. A pump laser is illustrated in green, probe and conjugate fields are represented in red and blue, and BSs are illustrated to represent loss in the interferometer, with transmission~$\eta$. A transducer is shown inside the interferometer that imparts the sensors phase $\phi$ onto the probe. Reproduced with permission from ref~\citenum{anderson2017optimal}. Copyright 2017 the American Physical Society. (d): Precision (normalized here by the mean photon number~$\vert\alpha\vert^2$, and presented as the variance rather than the standard deviation of $\phi$) as a function of the gain of the second nonlinear amplifier in a nonlinear interferometer with direct intensity readout. The gain of the first nonlinear amplifier is set to G=2, and the internal transmissions are set to~$\eta_{\rm ai}=\eta_{\rm bi}=\eta_\text{i}=1$.  The blue, yellow, green, and orange lines represent~$\eta_\text{ae}=\eta_\text{be}=\eta_\text{e}$=0.5, 0.75, 0.9, and 0.99 respectively. Reproduced and adapted with permission from ref~\citenum{anderson2017optimal}. Copyright 2017 the American Physical Society. (e): Optimal phase precision normalized to the SNL as a function of~$\eta_\text{e}$  for a nonlinear interferometer with asymmetric gain between the two nonlinear amplifiers.  Increased gain~$r_2$ in the second nonlinear amplifier reduces the sensitivity to external losses~$1-\eta_\text{e}$. Reproduced with permission from ref~\citenum{manceau2017detection}. Copyright 2017 the American Physical Society. (f): Example of a sensor based on a truncated nonlinear interferometer. The Rb vapor cell serves as a nonlinear amplifier and the transducer imparts the sensor's phase onto the probe or the probe's local oscillator. Reproduced with permission from ref~\citenum{pooser2020truncated}. Copyright 2020 the American Physical Society.}
\label{Fig:Sec3:NLIs}
\end{figure*}

Conventional MZIs like those illustrated in Figure~\ref{Fig:Sec3:Basic_Scheme}(b) consist of two inputs and two outputs, and can be described in a group-theoretical framework as SU(2)~\cite{Yurke1986b}, the special unitary group of degree 2. The SU(2) group, which is a simple Lie group, is equivalent to the rotation group in three dimensions, which has the nice feature that it allows one to visualize the operations of BSs
and phase shifters as rotations in three-dimensional space. This is true for both classical and quantum optical readout fields.  When the BSs in linear MZIs, or other interferometers like Fabry-Perot interferometers, are replaced by nonlinear amplifiers like four-wave mixers or optical parametric amplifiers, as shown in Figs.~\ref{Fig:Sec3:NLIs}(a) and (b), the resulting nonlinear interferometer can be described by the SU(1,1) group~\cite{Yurke1986b}, another type of simple Lie-group where the operations can be visualized as rotations and Lorentz boosts. Similarly, truncated nonlinear interferometers composed of one nonlinear amplifier followed by dual homodyne detectors, as shown in Figs.~\ref{Fig:Sec3:NLIs}(c) and (f), can also be described by the SU(1,1) group~\cite{anderson2017phase,gupta2018optimized,pooser2020truncated}. Describing interferometric sensors with such a group-theoretical approach allows for a straightforward analysis of the quantum states that optimize an interferometericic estimation precision.  Both SU(2) and SU(1,1) interferometers can achieve a phase precision with a scaling approaching~$1/N$ for~$N$ photons, but SU(2) interferometers require a squeezed vacuum state to be injected into the second channel of the first BS in order to achieve the HL when the first channel is fed by a coherent state of light, as discussed in section ~\ref{losslesscohsqueeze}.  In contrast, SU(1,1) interferometers can be operated with vacuum, coherent state, or combined coherent state and squeezed vacuum state inputs, and they offer an enhancement in precision that scales with the gain of the nonlinear amplifier~\cite{li2014phase,Yurke1986b,plick2010coherent}. However, because at least one of the nonlinear amplifiers in a conventional SU(1,1) nonlinear interferometer relies on phase-sensitive amplification, it is essential to maintain a near-zero phase difference within the interferometer to maintain the precision advantages~\cite{marino2012effect}.

As discussed in section~\ref{losslesscohsqueeze}, the main obstacle to practical quantum phase sensing with squeezed states of light is protecting the vulnerable quantum advantage against losses, as described in eq~\ref{eq:QFI_phase_lossy_quantum}.  Indeed, in the presence of losses, SU(2) interferometers with a squeezed vacuum input asymptotically approach the sensitivity of classical SU(2) interferometers with the same mean photon number~\cite{aspachs2009phase}. As a result, the development of detectors with near-unity detection efficiency has been a critical research topic for years. While some progress has been made with Bayesian parameter estimation as a tool for mitigating the effect of losses on squeezed SU(2) interferometers~\cite{Pezze2008}, SU(1,1) interferometers offer an alternative approach that is more robust against losses in some cases. Their dependence on loss can be broken down into dependencies on internal and external losses, represented here by~$(1-\eta_\text{i})$ and~$(1-\eta_\text{e})$, respectively. Here,~$\eta_\text{i}$ and~$\eta_\text{e}$ are assumed to be balanced for both arms of the interferometer, i.e. $\eta_\text{ai}=\eta_\text{bi}=\eta_\text{i}$ and $\eta_\text{ae}=\eta_\text{be}=\eta_\text{e}$, as shown in Figures~\ref{Fig:Sec3:NLIs}(a)-(c).  For a SU(1,1) interferometer with direct intensity readout, as illustrated in Figure~\ref{Fig:Sec3:NLIs}(a), external losses, or losses after the second nonlinear amplifier largely represent losses due to filtering before detection and losses due to imperfect detector efficiency. Internal losses include losses within the nonlinear amplifiers as well as losses from optical interactions with a sensor in the interferometer. In the limit of~$\eta_\text{i}=1$, and when the interferometer is operated in a balanced mode where the two nonlinear amplifiers have identical gain and loss, with a coherent state used in each input port, the external loss modifies the precision~$\Delta \phi$ of the SU(1,1) interferometer as~\cite{marino2012effect} 
\begin{align}
\Delta \phi_\text{e}=\eta_\text{e}^{-1/2}\Delta \phi,
\label{eq:NLI_eta_e}
\end{align}
In other words, unlike SU(2) interferometers, external loss in a balanced SU(1,1) interferometer does not change the functional description of the precision except for the addition of a prefactor because of the second nonlinear amplifier~\cite{kim2005precision}. In fact, because the second nonlinear amplifier is a phase-sensitive amplifier that can exhibit noiseless amplification, increasing its gain can compensate for external loss. The effect of external loss in an unbalanced SU(1,1) interferometer is shown in Figure~\ref{Fig:Sec3:NLIs}(d)~\cite{anderson2017optimal}. Here, the gain of the first nonlinear amplifier is set to 2, internal losses are neglected, and the gain of the second nonlinear amplifier is varied.  In the case of no external loss, the precision is almost independent of gain. On the other hand, in the limit of arbitrarily large gain, the interferometeric precision becomes insensitive to external loss~\cite{anderson2017optimal}. Figure~\ref{Fig:Sec3:NLIs}(e) expresses this insensitivity to external loss more directly by plotting the interferometeric precision, normalized by the SNL, as a function of $\eta_{\rm e}$ for three different gain ratios, $r_2/r_1$.  As the gain of the second amplifier, $r_2$, is increased relatively to that of the first, $r_1$, the precision becomes increasingly insensitive to external loss.

However, for the purposes of plasmonic sensing, internal losses that occur on a plasmonic sensing element within one arm of the interferometer must be considered.  When~$\eta_\text{e}$ is set equal to unity and the internal transmission~$\eta_\text{i}$ is varied (setting $\eta_{\rm bi}=\eta_{\rm ai}=\eta_\text{i}$ such that any additional loss from the sensor is matched), for bright coherent inputs~$\vert\alpha\rangle$ and~$\vert\beta\rangle$ seeding the first nonlinear amplifier and~$N_\text{i}$ photons inside the balanced interferometer, the precision $\Delta \phi$ of the SU(1,1) interferometer becomes~\cite{marino2012effect}
\begin{align}
\Delta \phi_\text{i}=\left[1+\frac{1-\eta_\text{i}}{\eta_\text{i}} \frac{N_\text{i}}{\lvert \alpha \rvert ^2 + \lvert \beta \rvert ^2}\right]^{1/2} \Delta \phi,
\label{eq:NLI_eta_i}
\end{align}
This suggests that balanced SU(1,1) interferometers are robust against small levels of loss, but as with conventional squeezed interferometers, increasing loss will push the precision to the SNL.

Truncated nonlinear interferometers that include only a single nonlinear amplifier followed by dual homodyne detection, as illustrated in Figs.~\ref{Fig:Sec3:NLIs}(c) and (f), provide the same phase precision as conventional nonlinear interferometers, but they offer some substantial advantages. First, they offer a simpler experimental design by removing the phase-sensitive nonlinear amplifier from the nonlinear interferometer.  Second, they typically exhibit reduced loss (and thus a greater quantum enhancement) because the nonlinear amplifier itself exhibits a tradeoff between gain and loss that always introduces some loss into the measurement.  Finally, truncated nonlinear interferometers can use arbitrarily high power local oscillators.  This is critical because the SNL for such a measurement is defined by the sum of the powers of the two-mode squeezed state output from the interferometer and the local oscillators. Thus, it is possible to arbitrarily increase the power of the local oscillators, thereby increasing the SNR of the measurement, without introducing excess power to the sensor, and while taking advantage of the quantum noise reduction in a two-mode squeezed state. No truncated nonlinear interferometric plasmonic sensor has been demonstrated in the literature to date, but a growing literature has described the precision of this approach for photosensitive sensing applications~\cite{anderson2017phase,gupta2018optimized,pooser2020truncated}.

\subsection{Quantum sensing beyond the Cram{\'e}r-Rao bound}
For phase estimation using a pure probe state in a single-mode setup with~$\hat{U}(\phi)=e^{i\phi\hat{a}^{\dagger}\hat{a}}$ or in a two-mode setup with~$\hat{U}(\phi)=e^{i\frac{\phi}{2}(\hat{a}^{\dagger}\hat{a}-\hat{b}^{\dagger}\hat{b})}$, the QFIs of eq~\ref{eq:QFI_Pure} are given as~$H=4\langle (\Delta\hat{a}^{\dagger}\hat{a})^{2}\rangle$ and~$H=\langle [\Delta(\hat{a}^{\dagger}\hat{a}-\hat{b}^{\dagger}\hat{b})]^{2}\rangle$, respectively, where~$\langle ..\rangle$ denotes the average over the probe state. This indicates that phase estimation becomes more precise by increasing the photon number variance of the probe state in a single-mode setup and the variance of the photon number difference of the two-mode probe state in a two-mode setup, respectively. A natural question is then, how large can these variances be increased by? 

Rivas and Luis~\cite{Rivas2012} have suggested an estimation strategy in a single-mode setup using a superposition of a vacuum and a squeezed vacuum state. The considered probe state has been shown to achieve an arbitrary scaling in precision, i.e.,~$\Delta\phi\propto N^{-k}$, where~$N$ is the mean photon number and~$k$ is an arbitrary value with~$k>1$. Such an alluring analysis sparked an intensive debate in the literature because it seemed to beat the Heisenberg scaling ($\Delta\phi\propto N^{-1}$) that is regarded as the fundamental ultimate limit that can never be beaten by any other physical strategy~\cite{Giovannetti2012b}. The state studied by Rivas and Luis is not the only example and states with various photon number distributions have been considered as candidates for increasing the QFI, for example, the SSW state~\cite{Shapiro1989}, the SS state~\cite{Shapiro1991}, Dowling's model~\cite{Dowling1991}, and the small-peak model~\cite{Braunstein1994b}. When the maximal photon number of a probe state is upper bounded, the so-called ON state---a superposition of vacuum and a Fock state---is known to lead to the maximal photon number variance~\cite{Lee2019}. One can show that the use of the ON state in single-mode phase estimation can indefinitely increase the QFI, while keeping the average photon number fixed. The ON state is also known to be useful in quantum computation~\cite{Sabapathy2018}, and a version with 18~excitations has been realized in the harmonic motion of a single trapped ion~\cite{McCormick2019}. 

On the other hand, when the maximal photon number of a probe state is unbounded~\cite{Johnson2005}, more bizarre photon number statistics can be found, as discussed in ref~\citenum{Lee2019}. For example, phase estimation using a probe state with a Borel photon number distribution~\cite{Borel1942, Tanner1961} can lead to sub-Heisenberg scaling~$\Delta\phi\propto N^{-3/2}$, calculated by the QCR bound~\cite{Lee2019}. Some heavy-tailed and sub-exponential distributions exhibit a diverging or even an infinite variance~\cite{Foss2013}. Particular interest has been paid to the Riemann-Zeta distribution as an example showing an infinite QFI, leading to completely precise phase estimation without uncertainty in a two-mode scheme~\cite{Zhang2013}. This last example is rather mysterious. Together with the other sub-Heisenberg strategies mentioned above, these examples need to be justified in order to certify if their precisions are achievable in practice, putting aside the question of how one might generate those photon number distributions in the first place. 

Apart from exploiting exotic photon number statistics, sub-Heisenberg-limited precision can also be achieved by using nonlinear effects in many-body systems~\cite{Boixo2007, Boixo2008, Choi2008, Roy2008, Woolley2008, Napolitano2010, Rams2018}. The nonlinear effects in atomic ensemble systems have experimentally demonstrated sub-Heisenberg-scaling~\cite{Napolitano2011}. However, a careful examination of the total resources is needed, as this determines how the precision scales. 

Over the last decade, the debate has been devoted to address one simple question: Can the HL or scaling be beaten?~\cite{Bollinger1996, Yurke1986b, Zwierz2010, Luis2013a, Luis2013b, Anisimov2010}
Fortunately, a conclusive answer has finally been proved~\cite{Zwierz2010, Tsang2012, Giovannetti2012a, Giovannetti2012b, Berry2012, Hall2012a, Hall2012b, Hall2012c, Jarzyna2015}, and shows that the overall scaling should properly include the amount of resources required to obtain a priori probability distribution of the unknown parameter and the number of measurements repeated to achieve the asymptotic QCR bound. With such an accounting of the total resources, one can show that the aforementioned scenarios all turn out to be Heisenberg scaling-limited.

In particular, when the likelihood function being used for the QCR bounds is highly non-Gaussian and the sample size is small, Ziv-Zakai (ZZ) bounds are known to be more appropriate~\cite{Ziv1969, Seidman1970, Chazan1975, Bellini1974, Weinstein1988}. The quantum version of the ZZ bound has been derived, showing that the MSE would be higher and thus tighter than a corresponding QCR bound~\cite{Tsang2012}. Using ZZ bounds, Giovannetti and Maccone have proved that sub-Heisenberg strategies are ineffective~\cite{Giovannetti2012a}. When a small amount of prior information is given, no sub-Heisenberg scaling is achievable, i.e., sub-Heisenberg scaling requires a large amount of prior information. When a large amount of prior information is given, however, one can just guess a random value based on the prior distribution without performing any measurement. A random guess with a large amount of prior information has been proven to achieve a comparable precision to the corresponding ZZ bound. Considering a finite amount of prior information, G{\'o}recki \textit{et al.}~have recently showed that the HL needs to be corrected by an additional constant factor of~$\pi$~\cite{Gorecki2020}.

To investigate the practical achievable precision with a finite amount of prior information and a limited number of measurements, which often nullifies QCR bounds in practice, Bayesian approaches have been considered in various situations~\cite{Jarzyna2015, Friis2017, Rubio2018, Rubio2019, Rubio2020b}. In this respect, it has been shown that quantum sensors can be enhanced by machine learning~\cite{Lumio2018} or calibrated by neural networks~\cite{Cimini2019}. One can also find a review article that discusses quantum multi-parameter estimation in terms of the Holevo CR bound, the quantum local asymptotic normality approach and Bayesian methods~\cite{Demkowicz-Dobrzanski2020}.
\begin{figure}[!b]
\centering
\includegraphics[width=0.45\textwidth]{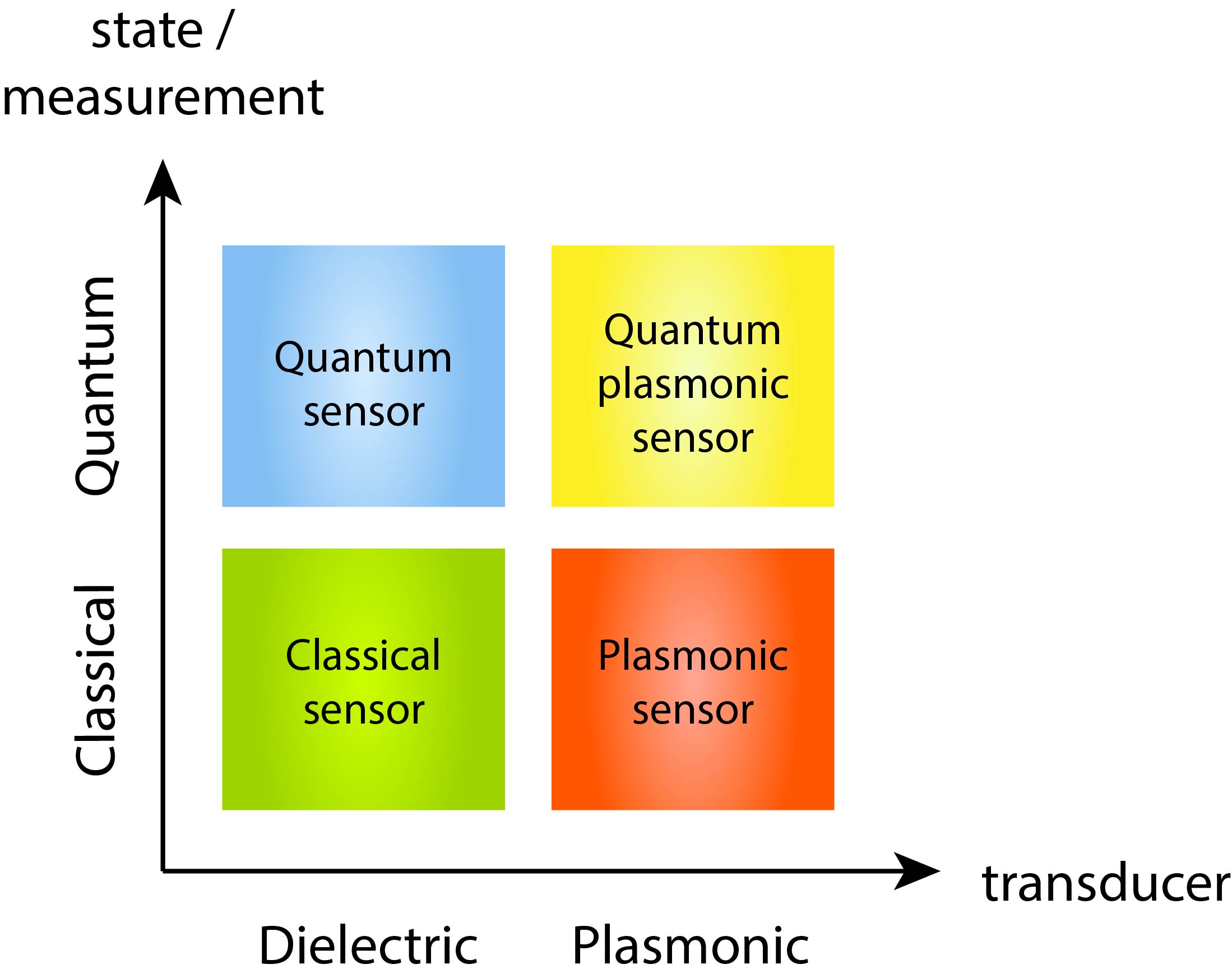}
\caption{The four different types of photonic sensors, classified depending on the type of physical system used for the probe state, measurement scheme, and transducer encoding the parameter to be sensed. The default type is the classical sensor (bottom left), which can be upgraded by moving in either the horizontal or vertical direction to become a plasmonic sensor (bottom right) or a quantum sensor (top left). Such an upgrade involves the individual benefits of either type, however, the benefits of both types can be achieved by an ultimate type of photonic sensor, called a quantum plasmonic sensor (top right). }
\label{sensortype}
\end{figure}

\section{Quantum-enhanced plasmonic sensors}\label{sec:Quantum_plasmonic_sensors}

As seen in section~\ref{sec:Plasmonic_sensors}, plasmonic structures provide sub-diffraction sensing with high sensitivity~\cite{Homola1999a,Homola1999b,Homola2006,Homola2008,Lal2007,Stewart2008,Shalabney2011,Li2015}, 
whereas from section~\ref{sec:Quantum_sensors} it is clear that quantum resources provide the ability to reduce the noise and estimation uncertainty below the SNL or SQL~\cite{Giovannetti2004,Paris2009,Giovannetti2011,Toth2014,Demkowicz2015,Taylor2016,Degen2017,Pirandola2018,Braun2018,Tan2019}. Individual sensing techniques have been developed for plasmonic sensing and quantum sensing independently from each other over the last few decades, which has led to the establishment of separate scientific fields in academia and industry. Research is now being devoted to combining the techniques of plasmonic sensing and quantum sensing with the aim of providing a new breed of plasmonic sensors with high sensitivity and high precision at scales below the diffraction limit. These new types of sensors are called `quantum plasmonic sensors' and they provide a sensing performance that cannot be obtained solely by either classical plasmonic sensors or conventional quantum optical sensors. In this section, we review recent studies that have exploited quantum resources to improve the sensing performance of plasmonic sensors. 

The decomposition of a general sensing procedure discussed in section~\ref{sec:parameter_estimation_theory} (see also Figure~\ref{estimationprocess}) allows us to classify four different kinds of sensor~\cite{Lee2016}: First, a sensor is called a `classical sensor' if a classical state input, an ordinary optical transducer, and a classical measurement scheme are used. Second, when the classical input and measurement are replaced by a quantum state input and a quantum measurement, this introduces quantum observables, and the sensor is called a `quantum sensor', enabling the potential for sub-shot-noise sensing. Third, when the ordinary optical transducer in a classical sensor is replaced by a plasmonic transducer, but the state and measurement remain classical, the sensor is called a `plasmonic sensor', enabling sub-diffraction sensing with high sensitivity, but ultimately shot-noise limited. Finally, a `quantum plasmonic sensor' is formed when a plasmonic structure is employed for the transducer, and a quantum state input and a quantum measurement are used. In this case, the plasmonic structure provides a high sensitivity via the sub-diffraction confinement of light~\cite{Takahara97,Takahara09,Gramotnev2010} , while the quantum input state and measurement provide the potential for sub-shot-noise sensing. The four kinds of sensors are illustrated in Figure~\ref{sensortype}, with the hope that quantum plasmonic sensors are expected to achieve combined benefits that have only been obtained individually by plasmonic sensors or quantum sensors.

More generally, the quantum properties of a plasmonic system can also play a role in sensing, e.g., electron tunneling~\cite{Zuloaga2009, Savage2012}, quantum size effects~\cite{Halperin1986,Halas2011,Scholl2012} or quantum surface response~\cite{Goncalves2020a}. In this case, the quantum plasmonic sensors do not necessarily use external quantum resources, such as input states or measurements, but exploit inherent quantum effects stemming from the plasmonic systems themselves. A careful analysis of whether the quantum effects give rise to sub-shot-noise sensing or simply improve the sensitivity must be performed. While the methods presented in this review are invaluable for such analyses, in the following sections we first focus on recent studies that fit into the above introduced classification of quantum plasmonic sensors, where the plasmonic structure provides a high sensitivity and quantum input states and measurements provide the capability for sub-shot-noise sensing, thus improving the LOD. We then briefly review other approaches to quantum plasmonic sensors at the end of the section that do not fit into this classification.

\begin{figure*}[!t]
\centering
\includegraphics[width=0.8\textwidth]{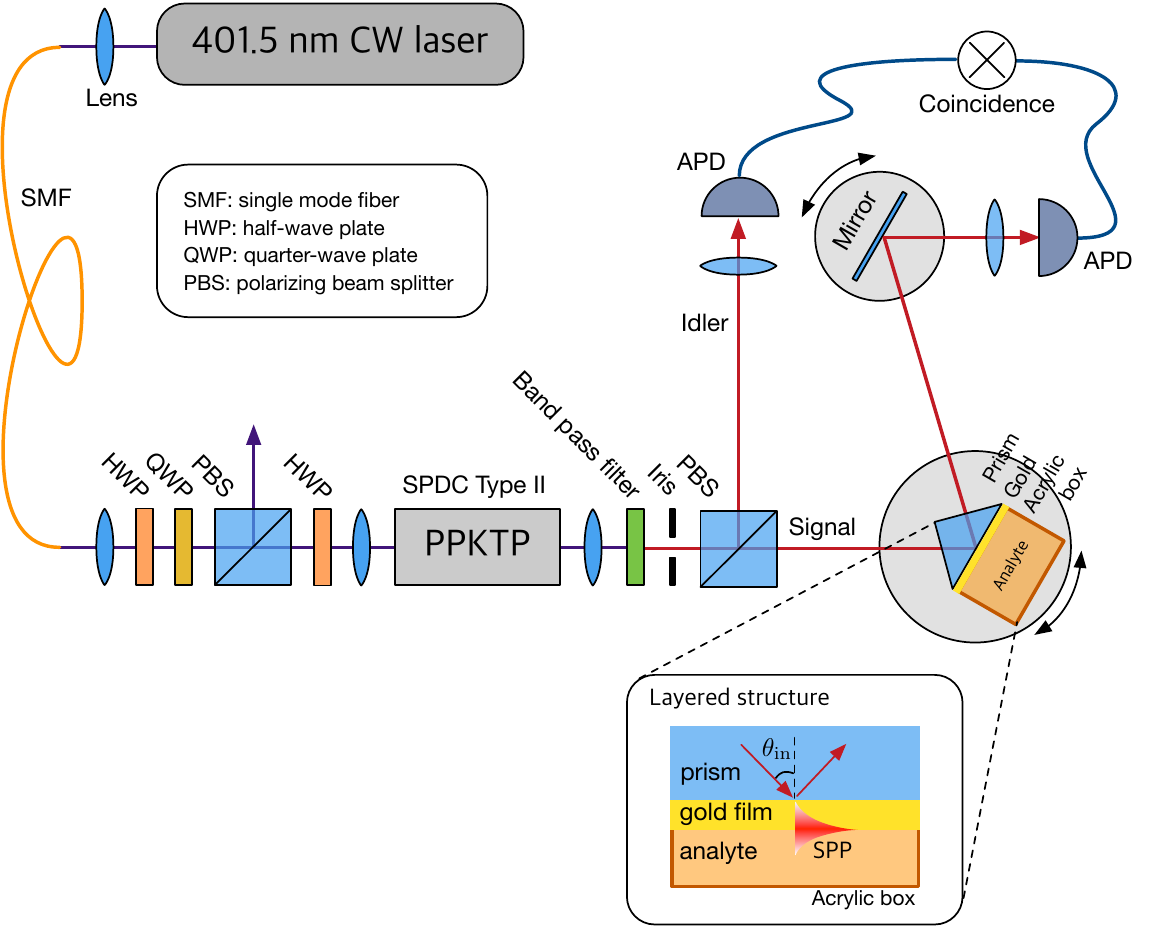}
\caption{Schematic of the prism setup probing the concentration of BSA with heralded single photons. The photons of a photon pair generated by a PPKTP nonlinear crystal are sent to the ATR prism setup and to a detector of the idler channel, respectively. The detection of the idler photon heralds that the signal photon has been injected into the ATR prism setup. The number of transmitted signal photons are then counted for a given analyte and an incident angle that can vary. Reproduced with permission from ref~\citenum{Lee2018}. Copyright 2018 The Optical Society.
}
\label{OELee1}
\end{figure*}

\subsection{Quantum plasmonic intensity sensing}\label{sec:Quantum_plasmonic_intensity_sensing}
The first method we describe is the use of a plasmonic transducer that encodes parameter information into light where the intensity of the light is dependent on the parameter to be estimated. Below, we review `intensity-sensitive' plasmonic sensors that use quantum states of light and quantum measurements.

\subsubsection{Intensity sensing with discrete variable states}
As introduced in section~\ref{sec:SPR_Sensing}, the ATR prism setup can be used in such a way that the intensity of the reflected light from the prism is modulated depending on the refractive index of an optical sample under characterization. To use the theory of intensity parameter sensing discussed in section~\ref{sec:Intensity_quantum_sensing}, we describe the reflection from the prism as the transmission through the prism setup, i.e.,~$\vert r_\text{spp}\vert^2=T$, where~$r_\text{spp}$ is given by eq~\ref{eq:rspp}.

A practical and useful intensity-parameter SPR sensing scheme is the two-mode scheme shown in the inset of Figure~\ref{NRFcompare}, but where the BS with transmittance $T$ is replaced by a prism setup. A relevant quantum theory of the ATR prism setup~\cite{Tame2008, Ballester2009} can be used. Within the ATR prism setup, the two-mode scheme can be considered in terms of a signal mode that passes through the prism setup ($a$) and an idler mode that is kept as a reference ($b$). In this regard, Lee \textit{et al.}~theoretically studied the two-mode scheme with a particular type of two-mode state~\cite{Lee2017} called a twin-mode (TM) state~$\ket{\Psi_\text{twin}}=\sum c_{n,m}\ket{n}_a\ket{m}_b$ with~$\vert{c_{n,m}\vert}=\vert{c_{m,n}\vert}$. Such a state has the following symmetric properties:~$\langle \hat{a}^{\dagger}\hat{a}\rangle =\langle \hat{b}^{\dagger}\hat{b}\rangle~$ and~$\langle (\Delta \hat{a}^{\dagger}\hat{a})^{2}\rangle =\langle (\Delta \hat{b}^{\dagger}\hat{b})^{2}\rangle$, which covers path-symmetric states with~$c_{n,m}=c_{m,n}^{*}e^{-2i\gamma}$~\cite{Hofmann2009,Seshadreesan2013}. 
In this two-mode plasmonic sensing scheme, an intensity-difference measurement of the observable~$\hat{n}_{-}=\hat{b}^{\dagger}_\text{out}\hat{b}_\text{out}-\hat{a}^{\dagger}_\text{out}\hat{a}_\text{out}$ is considered, i.e., the intensity of the transmitted light of the signal mode (reflected from the prism) is compared with the intensity of the reference mode. The measured intensity-difference can be inserted into an appropriate estimator together with eq~\ref{eq:rspp} in order to estimate the refractive index of an analyte. 

As a classical benchmark, the balanced PC state input~$\vert \alpha\rangle_\text{a} \vert \alpha\rangle_\text{b}$ can be considered. When the average photon number of the individual modes is restricted by~$N$ for both the TM state input and the PC state input, i.e.,~$\langle \hat{a}^{\dagger}_\text{in}\hat{a}_\text{in}\rangle=\langle \hat{b}^{\dagger}_\text{in}\hat{b}_\text{in}\rangle=N$, the intensity-difference signal for both states is the same, i.e.,~$\langle \hat{n}_{-}\rangle=N(\eta_\text{b}-\eta_\text{a}T)$ with non-ideal channel transmittance~$\eta_\text{a,b}$, which includes the detection efficiency, as modeled in Figure~\ref{NRFcompare}. However, the associated measurement noise~$\langle (\Delta \hat{n}_{-})^{2} \rangle$ is different, leading to different SNRs. By comparing the SNR of the TM state with that of the PC state, one finds that their SNR ratio~$R_\text{SNR}=\text{SNR}_\text{TM}/\text{SNR}_\text{PC}$ can be written as eq~\ref{eq:SNR_Ratio}, where the Mandel-Q parameter~$Q_\text{M}$ and the NRF~$\sigma$ of eq~\ref{eq:NRF} play important roles in intensity-sensitive SPR sensing. 
As discussed in section~\ref{sec:Quantum_noise_reduction_intensity_measurements}, a quantum probe with~$-1\le Q_\text{M}<0$ or~$0\le \sigma_\text{in}<1$ leads to a quantum enhancement with~$R_\text{SNR}>1$. The optimal state that maximizes the ratio~$R_\text{SNR}$ is the TF state, which exhibits an unconditional quantum enhancement regardless of the values of~$N$,~$T$ and~$\eta_\text{a,b}$. The TMSV state of eq~\ref{eq:TMSV}, on the other hand, provides a conditional quantum enhancement that depends on the values of~$N$,~$T$ and~$\eta_\text{a,b}$. In particular, when~$\eta_\text{a}\approx \eta_\text{b}$ and~$T\approx1$, or~$\eta_\text{a} T \approx \eta_\text{b}$, the ratio~$R_\text{SNR}$ for the TMSV state case becomes almost the same as the case of using a TF state input.

\begin{figure*}[!t]
\centering
\includegraphics[width=0.85\textwidth]{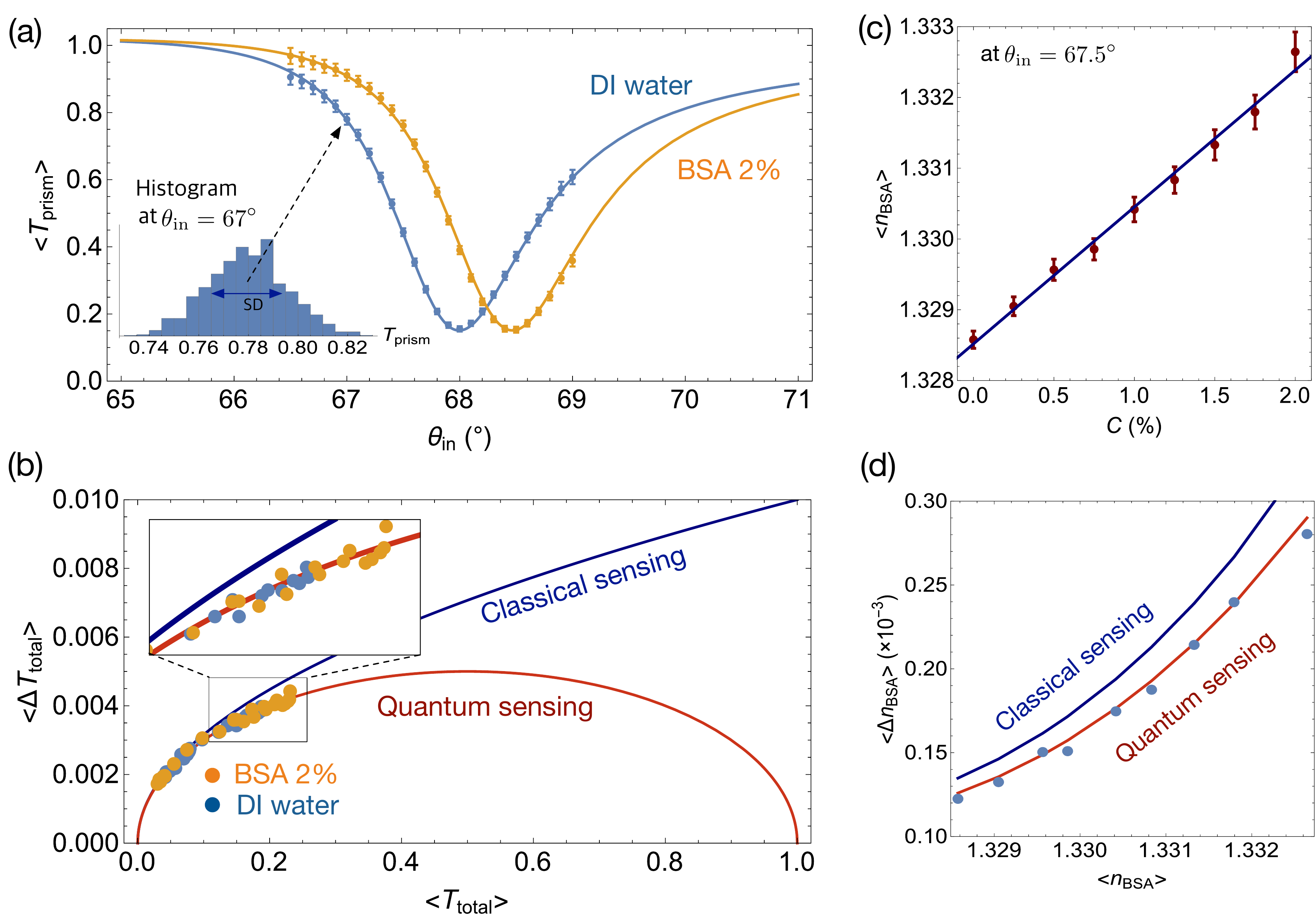}
\caption{(a) The effective transmittance through the ATR prism setup is measured as the incidence angle $\theta_\text{in}$ is varied, and normalized by the data measured with air. Two samples are used: deionized water and BSA with~$C=2$~\%. 
(b) The standard deviation of the total transmittance histograms [the same data as (a)] is presented as a function of the total transmittance, in comparison with the theoretical expectation of classical sensing (upper solid curve) and quantum sensing (lower solid curve). 
(c) At a particular incident angle~$\theta_\text{in}=67.5^{\circ}$, where the change of intensity is expected to be maximal (the inflection point), the refractive index of the BSA solution is estimated for samples with concentration varying from~$0~\%$ to~$2~\%$ in~$0.25\%$ steps. 
(d) The estimation uncertainty of the refractive index, given as the error bar at each point in (c), corresponding to the theoretical expectation for the classical and quantum scenarios, demonstrating a $10\sim20$\% quantum enhancement. Reproduced with permission from ref~\citenum{Lee2018}. Copyright 2018 The Optical Society.
}
\label{OELee2}
\end{figure*}

As mentioned in section~\ref{sec:Quantum_sensors}, the Fock state input is known to be the optimal state reaching the ultimate quantum limit in the estimation uncertainty in both single parameter (section \ref{sec:Quantum_enhanced_intensity_sensing}) and multiparameter (sections \ref{sec:Multi_Quantum_enhanced_intensity_sensing} and \ref{sec:Quantum_noise_reduction_intensity_measurements}) estimation scenarios that monitor the change of the intensity of the transmitted light through an optical sample~\cite{Adesso2009, Nair2018, Ioannou2020}. However, Fock states with high photon number~$N\gg 1$ cannot be readily realized in experiments with current technology~\cite{Varcoe2000, Bertet2002, Varcoe2004, Waks2006}. A few novel schemes have been suggested for the generation of large Fock states, but the fidelity is limited~\cite{Clausen2001, Sanaka2005a, Ou2006, Deleglise2008, Wang2008, Hofheinz2008, Zhou2012, Uria2020, Cosacchi2020, Groiseau2020, Zapletal2020}. Alternatively, one can use~$N$ single photons to achieve the same limit that would be obtained by the Fock state with~$\vert N\rangle$. This can be seen by the fact that the individual single photons in the Fock state~$\ket{N}$ undergo an independent Bernoulli sampling. Such an equivalence allows the use of single photons for various quantum sensors~\cite{Rezus2012,Whittaker2017, Allen2019, Yoon2020}, including a plasmonic sensor~\cite{Lee2018} which we now introduce. 

The experimental quantum plasmonic sensing setup considered in ref~\citenum{Lee2018} is shown in Figure~\ref{OELee1}. Single-photon pairs are generated via SPDC in a nonlinear periodically poled potassium-titanyl-phosphate (PPKTP) crystal, where the detection of the idler photon heralds that the signal photon has been injected into the prism setup. The detection of~$N$ idler photons thus corresponds to the use of~$N$ signal photons in the signal channel, i.e., equivalent to the use of the Fock state~$\vert N\rangle$ in the sense of estimation uncertainty. 
The experimental estimation uncertainty is measured by the repetition of an identical experimental observation a thousand times, which yields a distribution of the estimated total transmittance through the whole setup,~$T_\text{total}=N_{\text{t}}/N$, where~$N_\text{t}$ is the number of transmitted and detected signal photons out of~$N$ injected heralded single photons for each repetition. The effective transmittance through the prism setup can be obtained by normalizing the total transmittance by a normalization factor~${\cal N}$, i.e.,~$T_\text{prism}=T_\text{total}/{\cal N}$. In the experiment, the normalization factor is defined as the average transmittance measured with an air analyte at an individual incident angle~$\theta_\text{in}$, i.e.,~${\cal N}(\theta_\text{in})=\langle T_\text{air}(\theta_\text{in}) \rangle$ which leads to~$T_\text{prism}(\theta_\text{in})=T_\text{total}(\theta_\text{in})/\langle T_\text{air}(\theta_\text{in})\rangle$. The use of air is because the light entering the prism is off-resonant from the plasmonic excitation across the entire range of incident angles considered. This approach enables the elimination of the effect of an incident angle-dependent misalignment when scanning through a wide range of incident angles. 

\begin{figure*}[!t]
\centering
\includegraphics[width=0.85\textwidth]{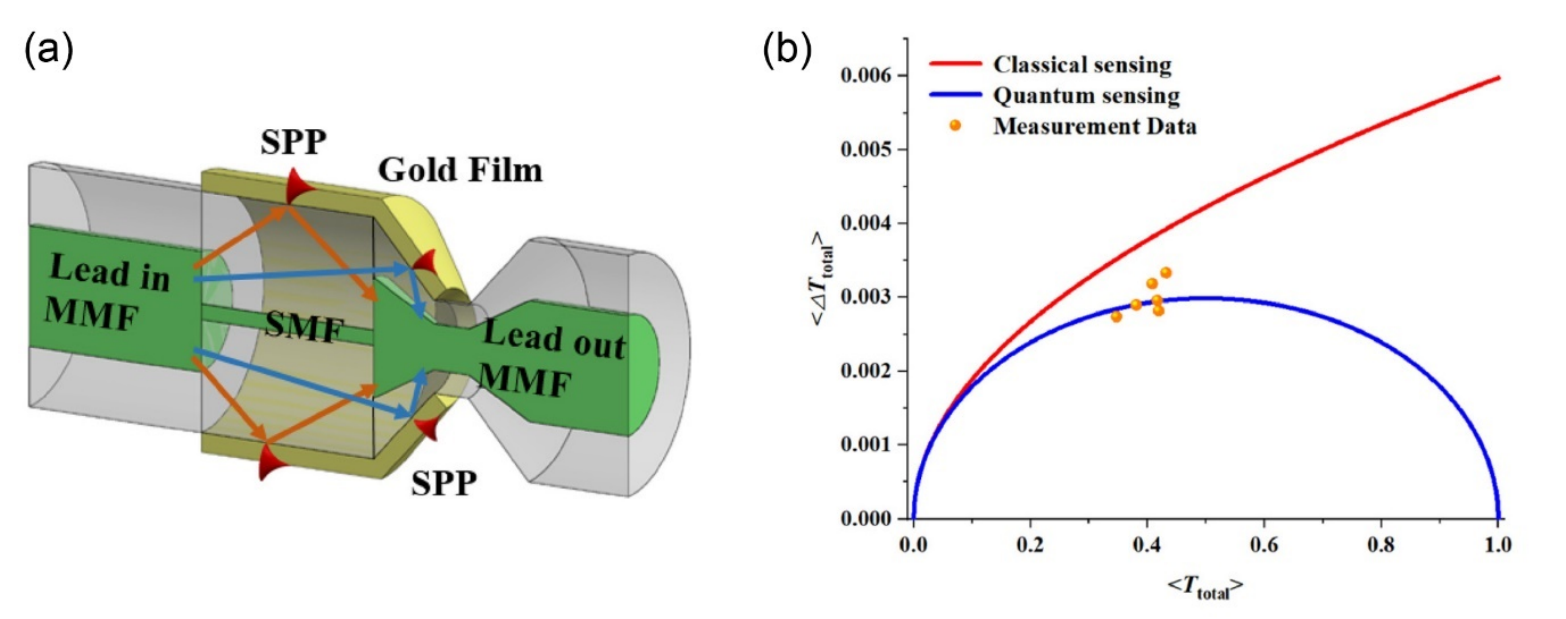}
\caption{(a) Schematic of the tapered hetero-core fiber quantum sensor. (b) Experimentally obtained variance in the total transmittance of the system shown together with the theoretical values for classical (red) and quantum (blue) cases. Reproduced with permission from ref~\citenum{Zhao2020}. Copyright 2020 Elsevier.
}
\label{fig:Fiber_sensor}
\end{figure*}

The above quantum plasmonic sensor aims to estimate the refractive index of the blood protein of bovine serum albumin (BSA) in aqueous solution whose concentration~$C$ is controlled in this proof-of-principle demonstration. From the measurement of the average transmittance~$\langle T_\text{prism} \rangle$, the refractive index of the analyte is estimated according to the theory model of the reflectance [see eq~\ref{eq:rspp}] by setting~$\langle T_\text{prism}\rangle =R_\text{spp}$. The analyte is placed on the opposite side of the gold film (see the inset in Figure~\ref{OELee1}) and its refractive index affects the resonant condition of SPPs, changing the resonance angle satisfying eq~\ref{eq:SPP_dispersion} or equivalently the intensity of the transmitted light for a fixed incident angle, i.e.,~$\langle T_\text{prism}\rangle$.

The overall behavior of the measurements are presented in Figs.~\ref{OELee2}(a) and (b). The average transmittance~$\langle T_\text{prism}(\theta_\text{in})\rangle$ measured in the experiment when varying the incident angle $\theta_\text{in}$ is shown in Figure~\ref{OELee2}(a) for deionized water, i.e.,~$C=0\%$, and BSA with~$C=2\%$. At each incident angle, the measurement is repeated a number of times, which yields a histogram [see the inset of Figure~\ref{OELee2}(a)] whose standard deviation quantifies the error bar in Figure~\ref{OELee2}(a). The measured errors depend on the incident angle, which modulates the transmittance of the setup and thereby the output photon number statistics. Note that the size of the error bars represents the estimation uncertainty of the effective transmittance for~$N$ single-photon inputs. The corresponding classical benchmark is defined by the case of using a coherent state with the same input power as~$N$ single photons. The experimentally measured quantum noise is thus compared with the classical benchmark, as shown in Figure~\ref{OELee2}(b). Here, the measured uncertainty is interpreted in terms of the total transmittance~$\langle T_{\text{total}}\rangle$ that consists of the effective transmittance through the prism setup and loss including the detection efficiency. Such a comparison clearly exhibits a quantum enhancement in the experiment and is in good agreement with the ultimate quantum limit introduced in section~\ref{sec:Quantum_enhanced_intensity_sensing}. 

For a specific estimation of the refractive index, the incidence angle is set to~$\theta_\text{in}=67.5^{\circ}$ as an example and the measurement is repeated for each concentration in a range from~$0\%$ to~$2\%$ in~$0.25\%$ steps. From the repeated measurements, the refractive indices are estimated using eq~\ref{eq:rspp} (i.e., the reflectance $R_\text{spp}$). The relation between the refractive index and the concentration of BSA solution can be determined by the slope of the linear fitting function, yielding~$\text{d}\langle n_\text{BSA}\rangle/\text{d}C=(1.933\pm 0.107) \times 10^{-3}$ [see Figure~\ref{OELee2}(c)], which is in good agreement with the previously measured value of~$1.82\times10^{-3}$~\cite{Barer1954}. The quantum property of light plays no role in this relation of the mean values, which can be thought of as more of a calibration, but it plays an important role in reducing the estimation uncertainty of the refractive index. The latter is demonstrated in the experiment by repeating the estimation, which produces a distribution of the estimated refractive indices. The estimation uncertainties with varying~$n_\text{BSA}$ are shown in Figure~\ref{OELee2}(d), which evidently demonstrates a~$10\sim20\%$ quantum enhancement in comparison with the classical benchmark. As described in section~\ref{sec:Limit_of_detection}, this enhancement directly affects the LOD of the sensor and improves its sensing performance.

\begin{figure*}[!t]
\begin{center}
\includegraphics[width=0.85\textwidth]{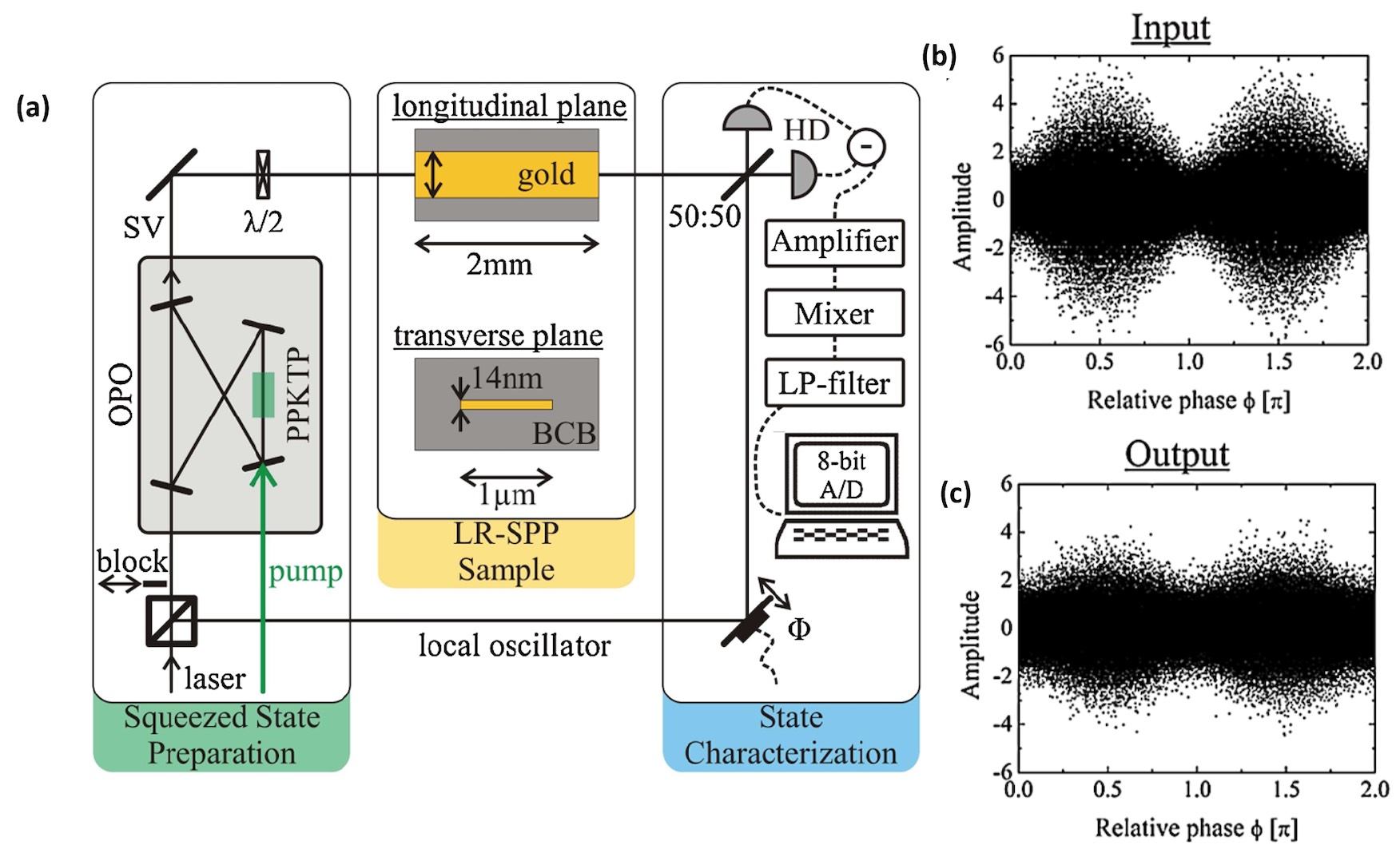}
\end{center}
\caption{(a) The experimental setup utilized by Huck \textit{et al.} generated squeezed light in an optical parametric oscillator (OPO), excited plasmon polaritons in a gold waveguide, and characterized the squeezing in the photons emitted from the end of the gold waveguide. PPKTP, periodically poled potassium titanyl phosphate crystal; SV, squeezed vacuum; $\lambda/2$, half-wave plate; $\Phi$, piezo actuated mirror for phase variation; 50:50, symmetric beam splitter; and HD, homodyne detection scheme. The measured squeezed vacuum state generated prior to injection into the gold waveguide (b) and the measured quantum state after excitation of SPPs in the gold waveguide (c) demonstrate that long range SPPs are capable of maintaining squeezing. Reproduced with permission from ref~\citenum{PhysRevLett.102.246802}. Copyright 2009 the American Physical Society.}
\label{Hucksqueeze}
\end{figure*}

As discussed in section~\ref{sec:Quantum_enhanced_intensity_sensing} and shown in Figure~\ref{OELee2}(b), the quantum enhancement in the intensity-sensitive sensor depends on the total transmittance of the whole sensing setup. Therefore, the quantum enhancement can be further increased by improving the total transmittivity of the setup by minimizing channel loss and maximizing the detection efficiency. 

A tapered fiber based quantum SPR sensing scheme has also been demonstrated recently for measuring BSA concentration~\cite{Peng2020} and salinity~\cite{Zhao2020}. In these works, a tapered hetero-core structure was fabricated, which is composed of two multi-mode fibers and a single-mode fiber, coated with a 50~nm thick gold film [see Figure~\ref{fig:Fiber_sensor}(a)]. Heralded single photons were used as inputs, prepared by SPDC using a 2 mm thick type-II metaborate crystal ({$\beta$-$\text{BaB}_2\text{O}_4$, BBO}) and by applying a coincident detection scheme. Based on the long evanescent field decay of the single-mode fiber and by reducing the diameter of the fiber, the authors showed that it was possible to reach the theoretically predicted estimation uncertainty below the SNL, as shown in Figure~\ref{fig:Fiber_sensor}(b).

In the given examples and in many other quantum sensing scenarios, heralded single photons have been used to successfully beat the classical limit in the estimation uncertainty. Here, it is important to point out a couple of limitations of heralded single photons for sensing purposes. First, heralded single photons require a high total detection efficiency to truly achieve a quantum enhancement. One main goal of quantum sensing is to achieve a high detection resolution with limited photon intensity at the sample position to avoid any possible damage. Therefore, all the photons impinging on the sample should be considered, not just the post-selected ones, for a fair comparison with the classical counterpart~\cite{Slussarenko2017}. Second, heralded photons are not true single-photon states. While the statistics of photon pairs produced by SPDC follows a thermal distribution, the post-selected photons can reveal highly anti-bunched nature at a low flux regime, however, this degrades further at a higher flux~\cite{Razavi2009,Bashkansky2014}. 

In another approach, many researchers have been pursuing the development of a true, or so-called `on-demand' single-photon source, which is an essential ingredient in many quantum optical applications, including quantum plasmonic sensing. These `deterministic' single photons are known to be one of the optimal states for low-noise intensity measurements~\cite{Lee2017}. An ideal single-photon source requires several important conditions: a high purity (strong anti-bunching), a high level of indistinguishability depending on the purpose, a high flux, and a high fidelity~\cite{Lounis2005,Aharonovich2016}. Current state-of-art nanophotonic methodologies utilizing many different types of single emitters have demonstrated improved quality of single-photon sources~\cite{Berchera2019,Chu2017,
Grosso2017,Wang2019c}, however, no single one has yet satisfied all of the requirements at the same time. In particular, the demonstrated photon fluxes are still too low (<$10^8$ cps) for practical use.

\begin{figure*}[!t]
\begin{center}
\includegraphics[width=5.5in]{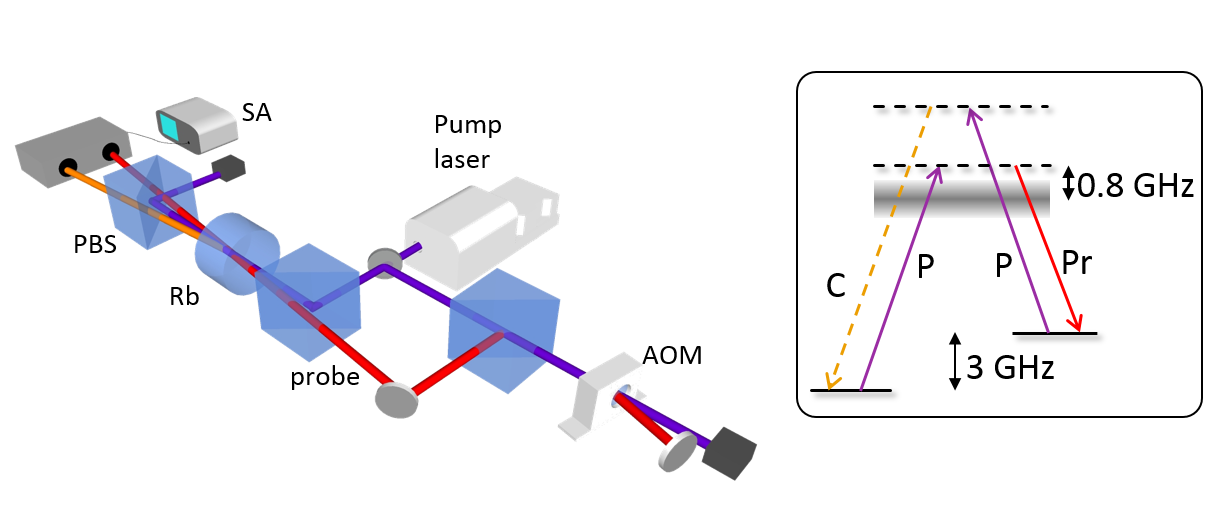}
\end{center}
\caption{FWM in rubidium vapor, simplified schematic for quantum noise reduction. A pump laser (P) is used to derive a probe (Pr) field by frequency-shifting the pump in an acousto-optic modulator (AOM). The double-pass configuration ensures the probe is not displaced relative to the pump as the AOM frequency is selected. The probe is offset from the pump by 3 GHz, or approximately equal to the hyperfine ground state splitting at the D1 line (795~nm). The pump and probe are jointly detuned from resonance by approximately 0.8~GHz. They overlap at a small angle (0.3$^\circ$) inside the Rb vapor cell. The resulting probe and conjugate fields are separated from the pump by a polarizing beam splitter (PBS) and produces a TMSD state. When incident on a balanced detector, the intensity-difference between the probe and conjugate (C) shows quantum noise reduction, visible on a spectrum analyzer (SA). The inset shows the energy levels in Rb associated with a double~$\Lambda$ system. The excited state hyperfine splittings are much smaller than the ground state, and are blurred with respect to each other in the Doppler-broadened vapor due to heating to approximately 120$^\circ$~C. As two pump photons are absorbed, coherence between the hyperfine levels ensures that when the probe field stimulates emission, a third field must be emitted from the virtual level illustrated by the upper dashed line in order to maintain energy conservation.}
\label{FWMsqueeze}
\end{figure*}

\subsubsection{Intensity sensing with continuous variable states}
\label{sec:intmultimode}
Sensors that exploit quantum noise reduction, or squeezed light~\cite{Caves1981}, have seen renewed interest in recent years with applications ranging from gravitational wave detection to ultra-trace plasmonic sensing at the nanoscale. The implementations of these sensors are increasingly limited by their ultimate limits of detection as defined by the Heisenberg uncertainty principle. At this limit, the noise is dominated by back action and the quantum statistics of light, leading to the HL, or the SNL when a coherent state of light is used. Simultaneously, many devices, including plasmonic sensors, have reached tolerance thresholds in which power in the readout field can no longer be increased without increasing the noise due to back action or thermal effects. Beyond these limits, squeezed light can be used to further improve the precision in these platforms. In recent years a growing number of sensors based on quantum noise reduction have been demonstrated~\cite{schnabel2010,schnabel2013,Taylor2013,hoff2013,treps2003,treps2002surpassing,Brida2010,Wolfgramm2010,Otterstrom2014,Pooser2015,Lawrie2013b,Lawrie2013a,Fan2015,Pooser2016}.

As discussed in section~\ref{sec:Quantum_noise_reduction_intensity_measurements}, intensity-difference noise reduction allows one to replace shot-noise limited sensors that operate in a differential configuration in order to reject common-mode noise and go below the SNL by exploiting quantum correlations present in two-mode squeezed states. The first observation of squeezing in plasmonic media characterized the generation, propagation, and subsequent re-emission of squeezed long-range SPPs in a gold waveguide~\cite{PhysRevLett.102.246802}, as shown in Figure \ref{Hucksqueeze}. This effort generated a squeezed vacuum state in an optical parametric oscillator, excited long-range SPPs in a gold strip waveguide with that squeezed vacuum state, and then characterized the quantum state of the photons emitted from the end of the waveguide. The authors concluded that squeezing can be efficiently and coherently transduced from an optical field to a SPP, and back to an optical field with no loss of quantum information except for that due to the introduction of vacuum noise. This result led to a flurry of subsequent demonstrations that (a) squeezing is maintained in localized surface plasmons also~\cite{Lawrie2013a,Holtfrerich2016}, and (b) this coherent interface between squeezed states of light and plasmonic media can be used to enable ultratrace plasmonic sensors with noise floors below the SNL~\cite{Fan2015,Pooser2016,Dowran2018}. The results of work on this type of quantum plasmonic sensor are outlined here.

This new type of quantum plasmonic sensor generates intensity-difference squeezing via FWM~\cite{McCormick2008, Boyer2008a, Turnbull2013}, which is based on a third-order optical nonlinearity. An example setup is shown in Figure~\ref{FWMsqueeze}. The noise in this amplifier can be derived in the interaction frame of the Heisenberg picture. In this frame, the interaction Hamiltonian for the case with degenerate pumping fields is 
\begin{equation}
{\cal H} = i\hbar\chi^{(3)} \hat{a}_\text{pr} \hat{a}_\text{c} \hat{a}^\dagger_\text{p} \hat{a}^\dagger_\text{p} + \text{H.C.}, \label{H1}
\end{equation}
where~$\chi^{(3)}$ is the nonlinear coefficient,~$\hat{a}_\text{p}$ is the pump field operator,~$\hat{a}_\text{pr}$ is the input probe field's operator, and~$\hat{a}_\text{c}$ is a third `conjugate' field which is parametrically amplified from the vacuum. The energy level diagram describing this system is shown in Figure~\ref{FWMsqueeze}.

\begin{figure}[!t]
\begin{center}
\includegraphics[width=3in]{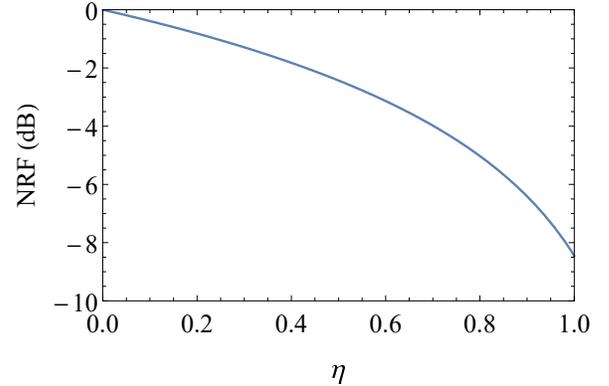}
\end{center}
\caption{Quantum noise reduction as a function of transmission $\eta$ on both the probe and conjugate fields ($\eta_{\rm a}=\eta_{\rm b}=\eta$) for nonlinear gain~$G=4$. This accounts for imperfect detection efficiency of the two fields on identical detectors. Here, the noise reduction, or `noise power' is given by $10\log_{10}(1-P/P_0)$~dB, where $P_0$ is the shot-noise, $P$ is the noise from the TMSD state, and $P/P_0$ is the NRF $\sigma$.}\label{sqzg=4equalloss}
\end{figure}

In many experiments the pump field is powerful relative to the probe and is undepleted. In this simplified scenario the field operators take the form of eqs~\ref{aout} and~\ref{bout} in section~\ref{sec:Quantum_noise_reduction_intensity_measurements}. This leads to a TMSD state and intensity-difference noise reduction as given in eq~\ref{TMSDratio2}. The important point about eq~\ref{TMSDratio2} is that the noise is less than one shot-noise unit for~$\eta>0$ and~$G>1$. Thus, using two-mode squeezing in lieu of a traditional reference subtraction on a balanced detector will yield a SNR increased by~$2\eta(G-1)/(2G-1)$ over the shot-noise limited measurement~\cite{McCormick2007,McCormick2008,Boyer2008}.
The NRF, or total `noise power' contained within a power spectral measurement, as a function of transmission on both the probe mode (a) and conjugate mode (b), with $\eta_{\rm a}=\eta_{\rm b}=\eta$ and $G=4$ is shown in Figure~\ref{sqzg=4equalloss}. 
While squeezing is typically known for its fragility in the presence of loss, and for decades this has limited its usage in sensors, Figure~\ref{sqzg=4equalloss} shows that if one starts with a pure TMSD state and if losses are symmetric and moderate, the noise reduction over classical detection still remains substantial. Plasmonic sensors can be configured to fulfill precisely these conditions. Despite losses, they can operate with large enhancements to their precision over their classical counterparts, and they represent state-of-the-art of quantum sensing with two-mode squeezed states. Here we outline several sensors that exploit quantum noise reduction and enable practical quantum sensing below the SNL and that beat state-of-the-art classical sensor in some cases.

The first transmission of multi-mode squeezed states through an extraordinary optical transmission (EOT) medium was demonstrated in ref~\citenum{Lawrie2013a}. Using FWM to provide the squeezed light, the transmission of multiple optical modes through the plasmonic medium in the form of quantum images was also demonstrated by exploiting properties of LSPs.

\begin{figure}[!t]
\begin{center}
\includegraphics[width=0.47\textwidth]{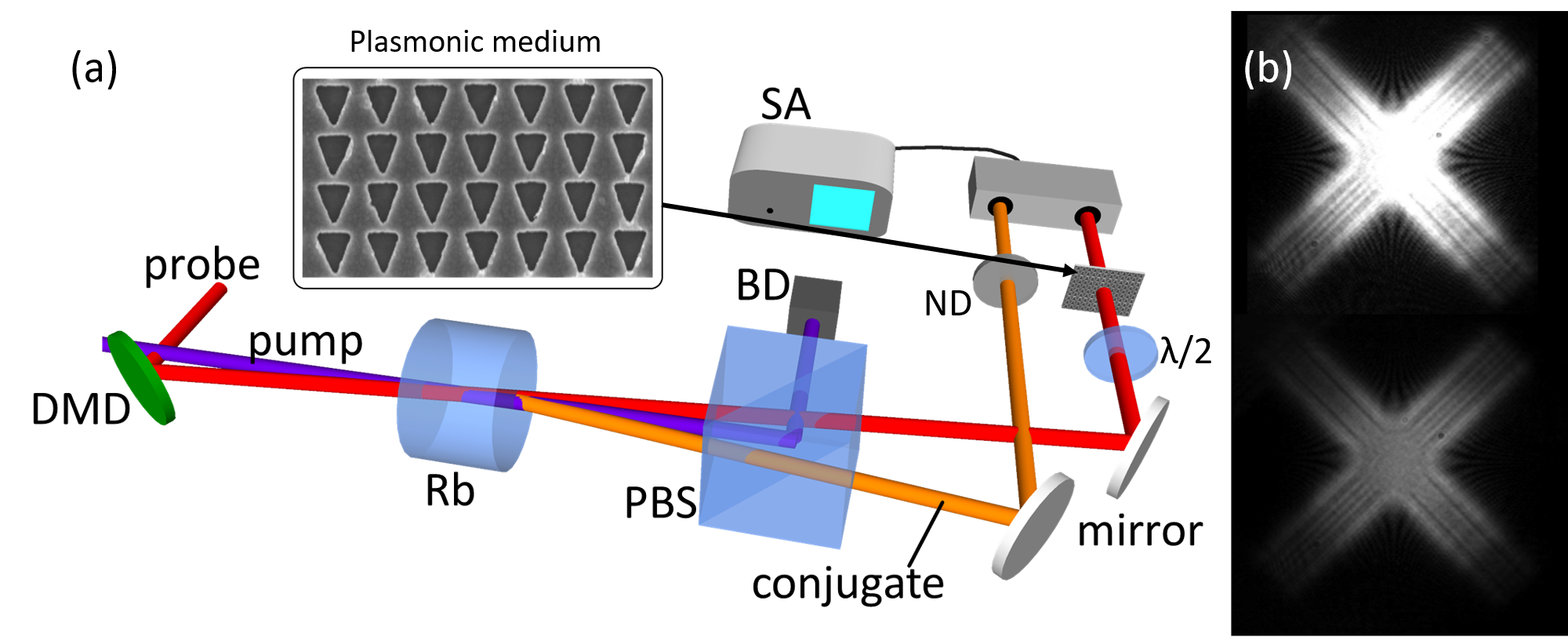}
\caption{(a) Optical setup showing the FWM experiment with abbreviations DMD: digital micromirror device, PBS: polarizing beam splitter, BD: beam dump, SA: spectrum analyzer, Rb: rubidium,~$\lambda/2$: half wave plate and ND: neutral density. One beam passes through the EOT medium while the other is attenuated with the neutral density filter in order to balance losses and maximize the noise reduction, or squeezing. Inset: an SEM image of a zoomed in subarray of the nanostructures. (b) Upper right: the image of the probe beam imprinted by the DMD before the plasmonic medium. Lower right: the transmitted image after EOT. Only the intensity has been attenuated while the number of spatial modes present remains the same.}\label{EOTfig}
\end{center}
\end{figure}

Several works have shown transmission of classical light by LSPs using triangles~\cite{Rodrigo2010} and crosses~\cite{Lin2011}, among others. 
The inset of Figure~\ref{EOTfig}(a) shows an SEM image of isosceles triangular hole nanostructures. The transmission spectrum of this plasmonic medium has a broad resonance centered at approximately 900~nm, with ample transmission at the wavelength of common squeezed light sources, 795~nm.

The experimental setup for EOT of squeezed light is shown in Figure~\ref {EOTfig}(a). A digital micromirror device is used as a spatial light modulator in order to imprint an image onto the probe  field. The image is amplified in the FWM medium due to its multi-spatial-mode nature~\cite {Boyer2008}, resulting in two twin images in the probe and conjugate fields that exhibit quantum noise reduction.
The plasmonic medium transmits the probe image while the conjugate image is attenuated by a neutral density filter. The LSPs confined to the edges of the triangular holes transmit all of the spatial information contained in the image, unlike the case for SPPs. Figure~\ref{EOTfig}(b) shows the incident probe image imprinted by the DMD (top) and the transmitted probe image after the EOT (bottom).

\begin{figure*}[!t]
\begin{center}
\includegraphics[width=0.85\textwidth]{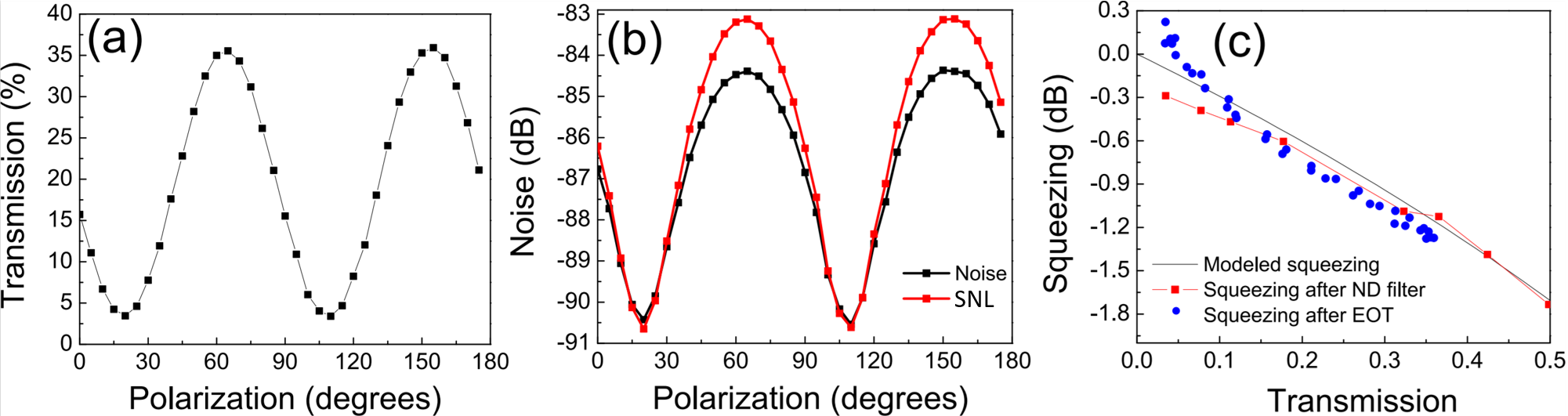}
\end{center}
\caption[EOTsqz]{\label{fig:data} Extraordinary optical transmission of squeezed light. (a) Single color transmission for the hole array for a polarization of~$0-175^{\circ}$, and (b) The relative noise intensity and SNL for the rebalanced probe and conjugate after the probe has passed through the EOT medium. (c) The measured squeezing as a function of transmission through the hole array and through a variable neutral density filter.  The noise based on the model in eq~(\ref{TMSDratio2}) is shown in black.  The error bars in (a) and some error bars in (b) are smaller than the symbols. Reproduced from with permission from ref~\citenum{Lawrie2013a}. Copyright 2013 the American Physical Society.
}\label{EOTsqz}
\end{figure*}

The transmission is strongly dependent on polarization due to the shape of the nanostructures, such that when the polarization overlaps with an edge of the triangles, a LSP is excited, resulting in transmission of multiple spatial modes.    
Figure~\ref{EOTsqz}(a) and (b) show the transmission and noise power as a function of polarization, respectively. As the polarization becomes incident with the base edge of the triangles, the noise power is well below the SNL, resulting in a quantum noise reduction. Figure~\ref{EOTsqz}(c) shows the quantum noise reduction (squeezing) as a function of transmission. The effects of losses in SPP waveguides~\cite{Ballester2009} and scattering in metal nanoparticle arrays~\cite{Lee2012} have been previously treated as effective BSs. A BS model also matches the experimental data, as the theoretical noise reduction as a function of BS transmission in Figure~\ref{EOTsqz}(c) shows. This experimentally demonstrates the capability for LSP-mediated EOT to transmit quantum images  while conserving macroscopic quantum information, such as quantum noise reduction.   

Exploiting the plasmons' propensity to conserve quantum information means that they can serve not only as good nanoscale quantum information platforms, but also as quantum sensors which make use of squeezing. SPPs have been used extensively to detect trace biochemical compounds in the Kretschmann configuration~\cite{Kretschmann1968}. State-of-the-art classical SPR sensors utilize differential detection with a reference field that does not interact with the SPP in order to eliminate noise present in the probe laser~\cite{Blanchard-Dionne2011,Wang2011,Wu2004}. Many of these sensors are now only limited by the SNL\cite{Piliarik2009,Wang2011,Wu2004}, and quantum sensors are required for further improvements in precision.

\begin{figure}[!t]
\begin{center}
\includegraphics[width=0.47\textwidth]{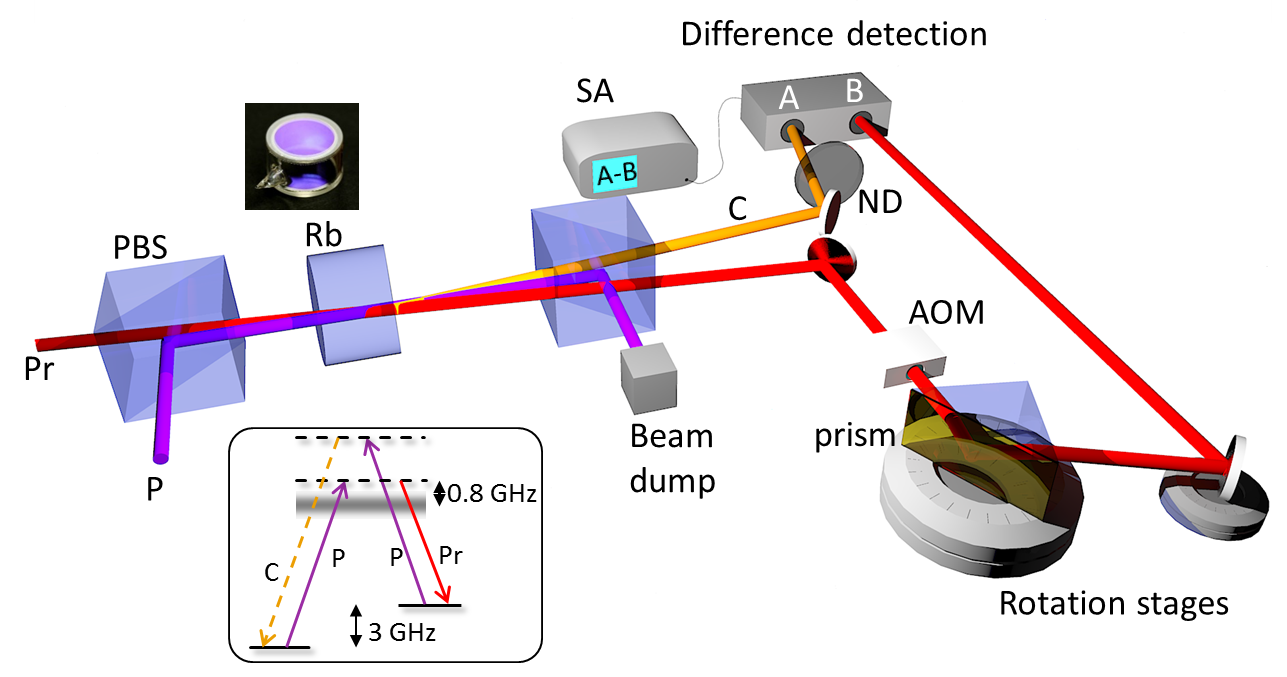}
\caption{Experimental setup for plasmonic sensing in the Kretschmann configuration using FWM to provide the signal probe and reference beams, with abbreviations: SA, spectrum analyzer; PBS, polarizing beam splitter; ND, neutral density filter; P, pump beam; Pr, probe beam; C, conjugate beam; AOM, acousto-optic modulator. The probe passes through a prism coated with a plasmonic thin film on the hypoteneuse, while the conjugate beam serves as a reference after balancing its intensity with that of the transmitted probe. Differential detection (A-B) reveals a signal with transmission-dependent NRF. Reproduced with permission from ref~\citenum{Pooser2016}. Copyright 2016 American Chemical Society.}
\label{sqzplas1}
\end{center}
\end{figure}

\begin{figure*}[!t]
\begin{center}
\includegraphics[width=0.754\textwidth]{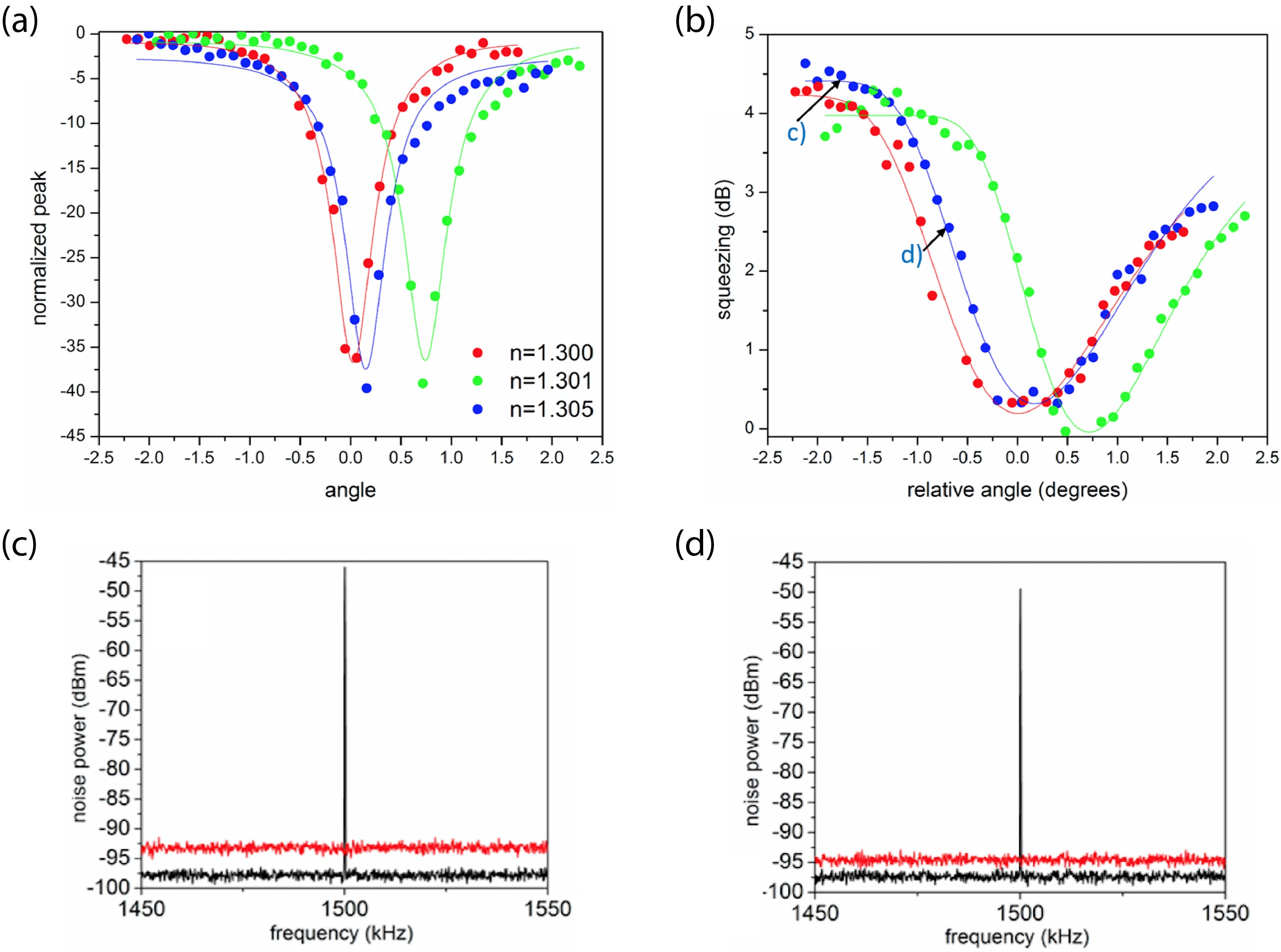}
\caption{Resonance curves and noise reduction for plasmonic sensing in the Kretschmann configuration with squeezed light. (a) Plasmonic resonance as a function of relative angle of incidence ($0^\circ$ corresponds to on-resonance for~$n=1.3$) and index of refraction (green: n=1.305; blue: n=1.301; red: n=1.300). The numerical curve fits are intended as a guide to the eye. (b) Quantum noise reduction (squeezing) as a function of angle of incidence and index of refraction (green: n=1.305; blue: n=1.301; red: n=1.300). (c) \& (d) The raw data for two points along the curve. The red line indicates the SNL, while the peak is associated with the transmitted probe power. The sidebands that indicate the minimum noise (or noise floor) associated with each data point fall below the SNL - squeezing (noise reduction) is equal to the difference between the black and red sidebands. Squeezing decreases as transmission decreases, but virtually all data points are squeezed. A maximum of 4.6~dB below the SNL is observed in the left shoulder of the peaks in (a). A minimum of 0.3~dB is observed in the n=1.300 and n=1.301 datasets, while the n=1.305 dataset minimum is at the SNL. Reproduced with permission from ref~\citenum{Pooser2016}. Copyright 2016 American Chemical Society.}
\label{sqzplas2}
\end{center}
\end{figure*}

A plasmonic SPP sensor in the Kretschmann configuration can use quantum noise reduction to supercede the precision of state-of-the-art classical device. Figure~\ref{sqzplas1} shows an experimental setup using FWM to produce a TMSD state and a Kretschmann sensor to obtain a precision below the SNL.
In the experiment, a 43.5~nm thick gold film was deposited on a prism, and index matching oils were deposited on the film in order to measure a shift in the plasmonic resonance as a function of the refractive index. The intensity-difference is measured at the two output channels A and B, with a ND filter placed on the conjugate mode (A) to equal the transmittivity of mode B and reduce the noise in the measurement signal to a minimum, as discussed in section~\ref{sec:Quantum_noise_reduction_intensity_measurements}. An acousto-optic modulator (AOM) was used to place a modulation on the probe mode (B). This leads to a peak in the measurement signal at the driving frequency 1.5~MHz when viewed on a spectrum analyzer, as shown in Figure~\ref{sqzplas2}(c) and (d) for two example cases. The height of the peak is proportional to the magnitude of the amplitude modulation on the probe field and due to the conjugate field being unmodulated the peak is also proportional to the magnitude of the intensity-difference signal. On the other hand, the sidebands of the signal (either side of the peak) correspond to the noise floor (the variance) of the measurement signal, as there is minimal amplitude modulation of the probe (and thus of the intensity-difference measurement signal) at frequencies away from the AOM resonance. Using a spectrum analyzer to give a power spectrum of the measurement signal in this way allows the intensity-difference (proportional to the peak at resonance) and the intensity-difference noise (the side-bands) to be obtained from the same plot. Thus, both the signal amplitude and noise can be read off simultaneously, allowing an immediate characterization of the SNR of the system. An additional feature is that technical noise at low frequencies (such as laser or vibration fluctuations) can effectively be eliminated in a manner similar to lock-in detection, allowing the sensor's precision to be brought close to the theoretical limit. 

\begin{figure}[t]
    \centering
    \includegraphics[width=0.41\textwidth]{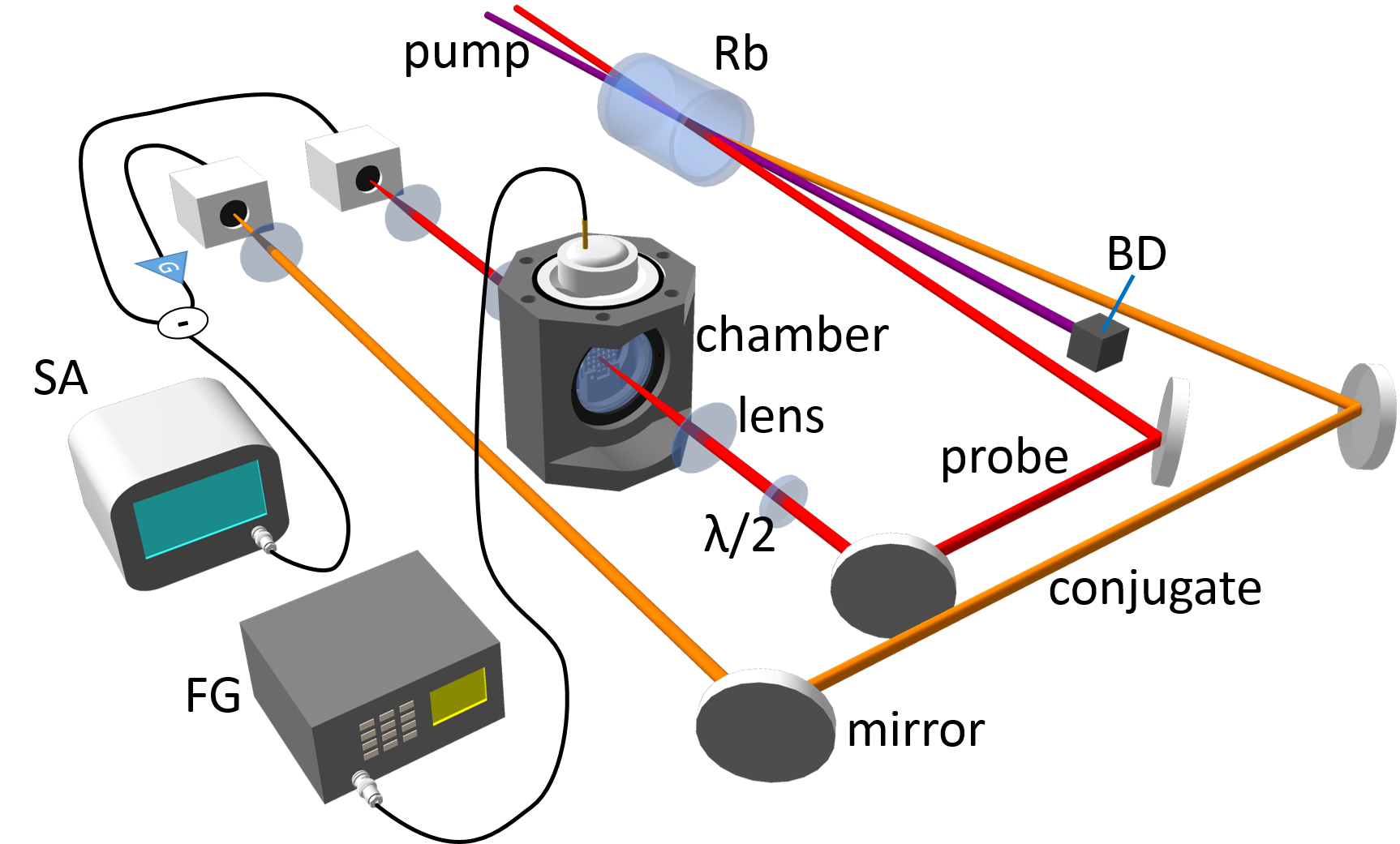}
    \caption{Ultrasonic pressure wave sensing. A FWM configuration is used to generate entangled probe and conjugate beams in a TMSD state. The probe beam serves as a probe for the plasmonic sensor while the conjugate beam serves as a quantum-correlated reference. SA: spectrum analyzer; BD: beam dump; FG: function generator. Adapted with permission from ref~\citenum{Dowran2018}. Copyright 2018 The Optical Society.}
    \label{fig:sqz_pressure}
\end{figure}
\begin{figure}[t]
    \centering
    \includegraphics[width=0.3\textwidth]{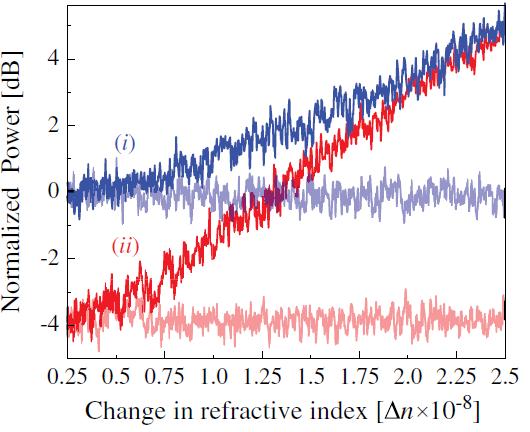}
    \caption{Measured signal while linearly ramping the driving voltage of the ultrasonic transducer, thus increasing the change of refractive index of air ($\Delta n$), when probing with coherent states, trace (i), and with twin beams (as a TMSD state), trace (ii). The lighter-weight lines represent the shot-noise (i) and squeezed noise floors (ii). Adapted with permission from ref~\citenum{Dowran2018}. Copyright 2018 The Optical Society.}
    \label{fig:my_label}
\end{figure}

Figure~\ref{sqzplas2}(a) shows the resulting resonance curve of the signal as a function of the angle of incidence and the refractive index. Every data point corresponds to a quantum noise reduction, as shown in Figure~\ref{sqzplas2}(b). These points are obtained by comparing the sidebands of the signal on the spectrum analyzer, as shown in Figure~\ref{sqzplas2}(c) and (d) for two example cases, where the peak occurs at 1.5 MHz due to the frequency set by the AOM. A maximum noise reduction of 4.6~dB is obtained on the left-hand shoulder of Figure~\ref{sqzplas2}(b). It is notable that a large amount of squeezing (noise reduction) is observed near the inflection points in Figure~\ref{sqzplas2}(b), meaning that one can choose to operate this SPR sensor at a single position\cite{Wang2011}, where a large quantum effect is still observed. At all points on a transmission vs index curve, a higher SNR would be achieved than is possible with the classical version of this sensor. Comparing the quantum sensor with a classical sensor in which the incident power may be turned up indefinitely, the quantum sensor still compares favorably. The quantum light source used here can be operated at equivalent optical powers, and theoretically the amount of quantum noise reduction does not depend on the incident optical power [see eq~(\ref{TMSDratio2}) in section~\ref{sec:Quantum_noise_reduction_intensity_measurements}]. 
On the other hand, the classical sensor cannot be used at powers beyond the point of thermal modulation\cite{Kaya2013} of the plasmon or the damage threshold of photo-sensitive ligands\cite{Robinson2014,Bartczak2011}. These thresholds are easily within the power capabilities of the squeezed light source used here and many others. Notably, it is also possible to use this configuration without amplitude modulation to obtain a quantum-modulated signal at DC~\cite{Fan2015}.

Similar sensing configurations have been used to detect ultrasonic pressure waves~\cite{Dowran2018} and establish long range quantum plasmonic networks~\cite{Holtfrerich2016}. In the case of ultrasonic measurements, the index of refraction shift is caused by local variations in air pressure above the plasmonic nanosurface at a frequency of 199 kHz. Using LSPs and EOT, a change in transmission serves as a transduction of index of refraction shifts onto optical intensity modulation. The magnitude of the intensity modulation (representing the magnitude of the refractive index modulation $\Delta n$) can then be detected with noise below the SNL using a TMSD probe state. Figure~\ref{fig:sqz_pressure} shows the experimental setup. A chamber with an ultrasonic transducer is used to provide pressure waves at the surface of a plasmonic thin film, which consists of triangular hole arrays. One of the two twin beams passes through the EOT sensor while the other continues in free space and acts as a reference. The resulting difference measurement contains the modulation signal, provided by the function generator to the ultrasonic transducer, on top of a reduced noise background.

Figure~\ref{fig:my_label} shows the measured signal (and noise floor) as the magnitude of the change in refractive index $\Delta n$ increases. The classical signal (coherent state probing) is shown in blue (i) and the TMSD state signal is shown in red (ii). The signals are normalized by the average shot-noise of the classical case, shown as a light blue line. Clearly the TMSD state has reduced noise compared to the classical case: the minimum noise levels show a difference of 4~dB. The SNR for the TMSD state is higher than the classical case as $\Delta n$ decreases. For very small index shifts, the TMSD version of the sensor is therefore able to discern index shifts that are smaller than the classical sensor, enabling a detectable shift~$\Delta n$ of~$O(10^{-9})$ RIU. The minimum resolution of the refractive index modulation can be determined at a 99\% confidence level in the SNR. In the quantum case this gives $\Delta n_{\rm min} = 5.5 \times 10^{-9}$ RIU with the TMSD state and $\Delta n_{\rm min} = 9.6 \times 10^{-9}$ RIU with classical coherent states. This demonstrates a quantum enhancement of 56\% in the precision of the sensor. 

\begin{figure*}[!t]
    \centering
    \includegraphics[width=5in]{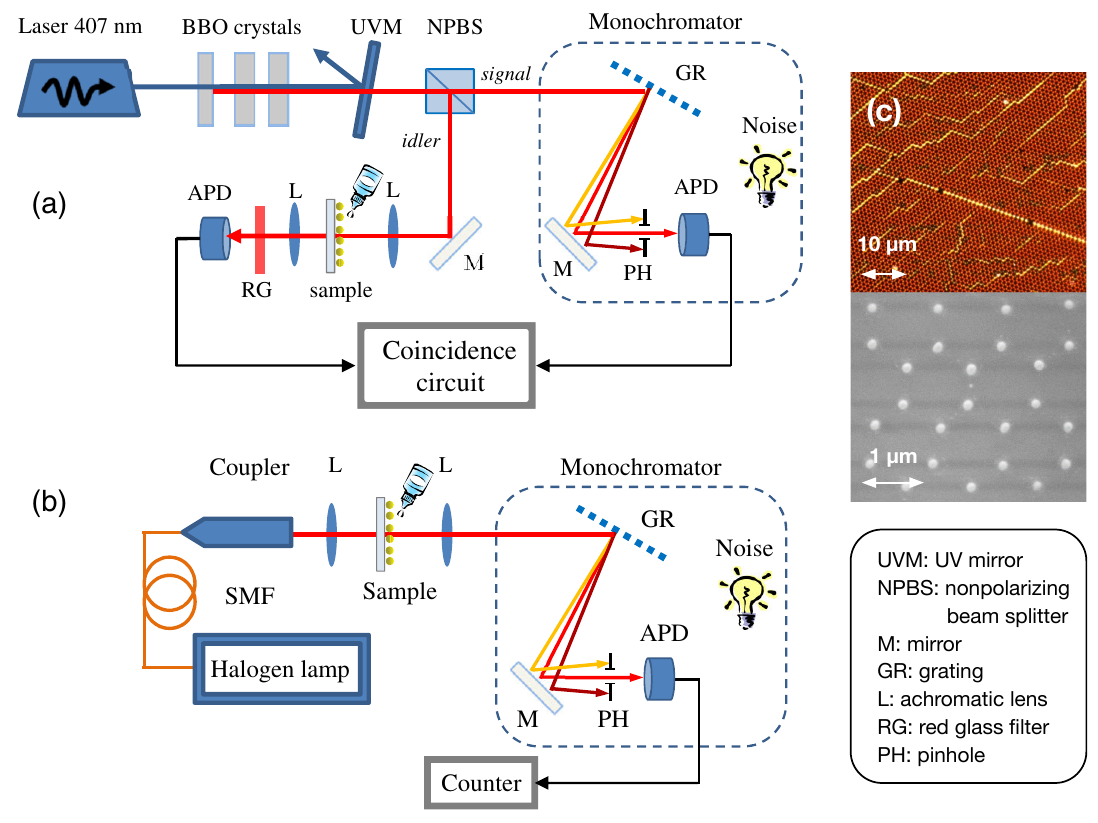}
    \caption{
    (a)~Schematic of quantum spectroscopy with a twin beam (i.e., the TMSV state of eq~\ref{eq:TMSV}) generated from a SPDC source (BBO crystals), where coincidence photon counting is carried out by two APDs (avalanche photodiodes). A monochromator placed in the signal channel consists of a mirror (M), a diffraction grating (GR), a pinhole (PH) and an APD, and selects a particular single mode of interest under investigation, realizing spectroscopy. (b)~Schematic of conventional spectroscopy with classical light generated from a Halogen lamp, where a single APD is used together with the same monochromator from the quantum spectroscopy scheme. (c)~The MNP array under investigation: Dark-field microscope image~(top) and scanning electron microscope image~(bottom). Adapted from ref~\citenum{Kalashnikov2014} (CC BY 3.0 License).}
    \label{fig:PRXKalashnikov1}
\end{figure*}

As described in section~\ref{sec:SPR_imaging}, SPR imaging enables highly parallelized, label-free, real-time, high precision biochemical sensing\cite{rothenhausler1988surface,piliarik2007towards,smith2003surface,steiner2004surface,wark2005long,wong2014surface,jordan1997surface}. However, all of the quantum plasmonic sensors described in this review thus far are single pixel sensors.  Quantum plasmonic sensors that utilize multi-spatial-mode quantum correlations in entangled optical readout fields are increasingly a plausible idea. Early research that focused on the transduction of quantum images by EOT showed that SPP-mediated EOT processes do not maintain spatial information, thus limiting the potential for EOT-based spatially resolved quantum sensors\cite{Altewischer2002}. However, LSP-mediated EOT processes do maintain spatial information, as already discussed in this section and shown in Figure~\ref{EOTfig}~\cite{Lawrie2013a}. Kretschmann-style SPR sensors also maintain spatial information, as discussed in section~\ref{sec:SPR_imaging}.

\begin{figure*}[!t]
    \centering
    \includegraphics[width=0.7\textwidth]{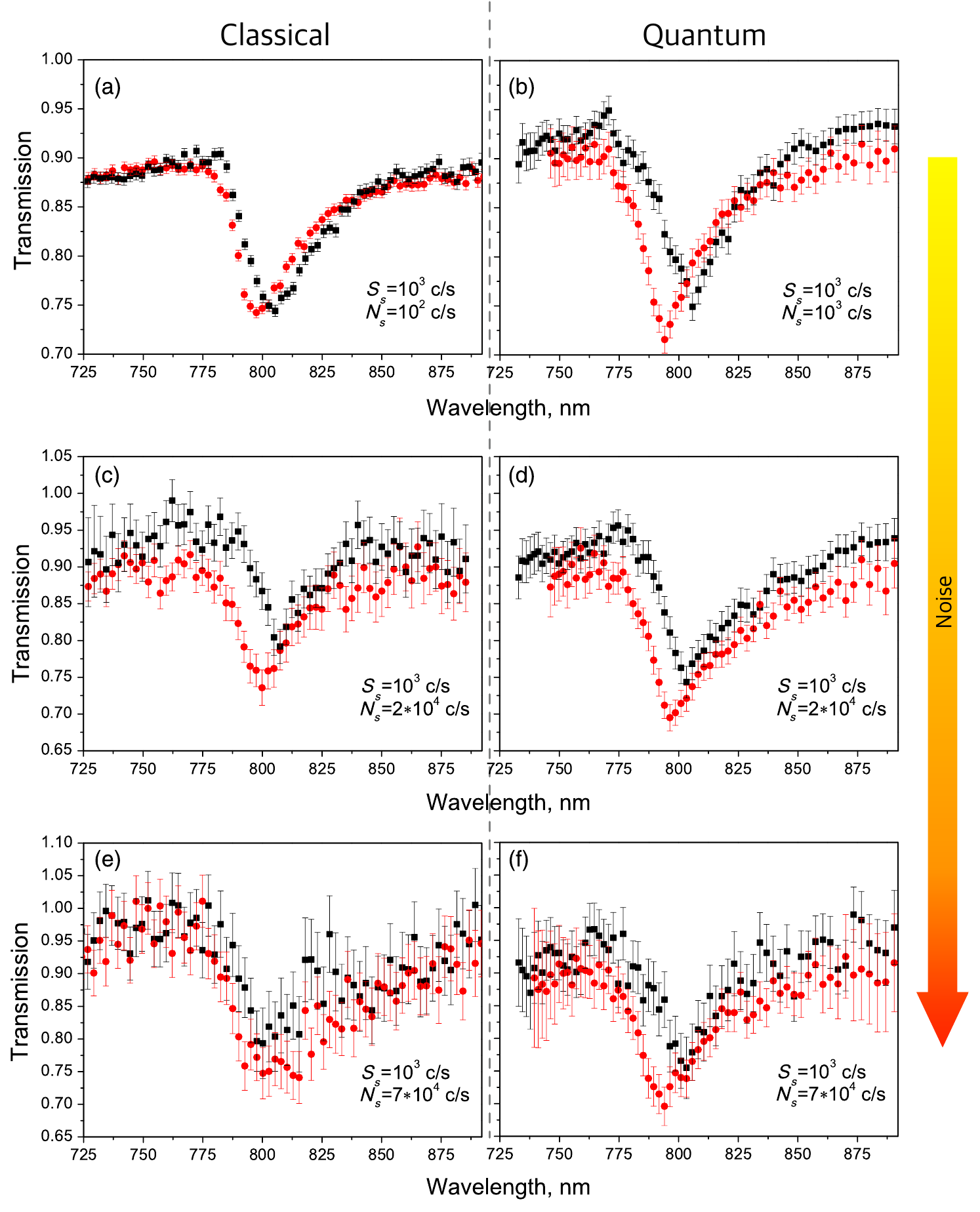}
    \caption{
    Transmission spectra measured by conventional classical (left) and quantum (right) spectroscopy under different noise conditions [$N_{\text{s}}=10^{2}$ (top), $2\times 10^{4}$ (middle), and $7\times 10^{4}~\text{c/s}$ (bottom)]. The measurement is performed for water-glycerin solutions deposited on an array of MNPs with different concentrations [$40\%$ (red dots) and $50\%$ (black dots)]. The photon flux of $10^3$~c/s is kept in the monochromator for all the experiments and the error bars denote the standard deviations of the measured signals. Adapted from ref~\citenum{Kalashnikov2014} (CC BY 3.0 License). 
    }
    \label{fig:PRXKalashnikov2}
\end{figure*}

Furthermore, as discussed in section~\ref{sec:Multi_Quantum_enhanced_intensity_sensing} and highlighted in recent reviews~\cite{magana2019quantum,moreau2019imaging,Genovese2016}, substantial work has gone into the development of quantum imaging platforms for other applications. For SPR imaging sensors that are robust to higher power optical excitation, discrete variable entanglement is less beneficial, but higher power multi-spatial mode squeezing could be leveraged to enable quantum-enhanced SPR imaging. As mentioned above, and as shown in Figure~\ref{EOTfig}, plasmonic images exhibiting intensity-difference squeezing can be efficiently transduced by LSP-mediated EOT processes.  The same effect has been shown more recently for a combination of phase-sum and intensity-difference entanglement~\cite{Holtfrerich2016}. Despite the demonstration of the plasmonic transduction of multi-spatial-mode quantum states of light, there have been no reports of quantum-enhanced SPR imaging.  A key ingredient for such an advance is the multi-pixel readout of an array of plasmonic sensors that can benefit from quantum correlations in the readout field. A growing number of research efforts address exactly this need.  For instance, researchers are now utilizing electron-multiplying CCD cameras for spatio-temporally resolved readout of multi-spatial-mode squeezed states of light~\cite{kumar2017observation,kumar2019spatial,li2020temporal}, and preliminary research has shown that compressive imaging techniques can be used for squeezed light imaging with single pixel detectors~\cite{Lawrie2013b}. Together, multi-spatial-mode quantum light sources, multi-pixel plasmonic sensors, and readout schemes for characterizing multi-spatial-mode quantum signals provide the necessary building blocks for quantum-enhanced SPR imaging.

\subsubsection{Intensity sensing robust to thermal noise}
The strong photon number correlation possessed by the TMSV state of eq~\ref{eq:TMSV}, with a NRF of $\sigma=0$, has also been used to probe an array of gold nanoparticles and measure the refractive index change of a glycerin-water solution~\cite{Kalashnikov2014}. The two-mode scheme shown in Figure~\ref{NRFcompare} was used to carry out quantum spectroscopy with the input state $\vert \Psi\rangle_\text{in}=\vert \text{TMSV}\rangle$, where the measurement is a coincidence detection [see Figure~\ref{fig:PRXKalashnikov1}(a)]. 
The glycerin-water solution surrounds a plasmonic transducer that consists of a hexagonal array of spherical MNPs with a diameter of 130~nm and lattice period of 1.1~$\mu$m, as shown in Figure~\ref{fig:PRXKalashnikov1}(c). The idler mode of the TMSV state is sent through the sample, while the signal mode is sent to a diffraction grating to investigate the spectral response using the coincidence counting that follows. The sensing performance of such a quantum scheme is compared with conventional spectroscopy that uses a classical probe state generated from a lamp, as shown in Figure~\ref{fig:PRXKalashnikov1}(b).

The authors directly compared quantum and classical spectroscopy methods in terms of their robustness to thermal noise influencing the detection. The thermal noise is artificially realized by a lamp inside the monochromater part [see Figs.~\ref{fig:PRXKalashnikov1}(a) and (b)], which induces additional photon counts $N_\text{s}$ in the measurement of the signal photons $S_\text{s}$. The noise level is controlled by changing the brightness of the lamp, i.e., $N_{\text{s}}=10^{2}$, $2\times 10^{4}$, and $7\times 10^{4}~\text{c/s}$ are considered in the experiment. The effect of thermal noise is examined with respect to the distinguishability of the two resonance spectra measured for two concentrations $C$ of glycerin-water solution on top of the array of MNPs.

The results obtained by quantum and classical spectroscopy for the two concentrations under the three noise conditions are presented in Figure~\ref{fig:PRXKalashnikov2}. The red dots represent the spectra for $C=40\%$ and black dots for $C=50\%$. Generally, the resonance curves in both cases become  blurred due to thermal fluctuations that increase with the thermal noise. However, the extent to which the curves become blurred is different for the quantum and classical spectroscopy schemes; the curves for the quantum case are more distinguishable than those for the classical case. This is clearly seen for the case when the noise photocount $N_\text{s}$ is $70$ times larger than the signal photocount $S_\text{s}$ [Figs.~\ref{fig:PRXKalashnikov2}(e) for classical and (f) for quantum]. Such robustness against the thermal noise can be attributed to the strong photon number correlation of the TMSV state, which is absent in the classical scheme.

\begin{figure*}[t]
\centering
\includegraphics[width=0.75\textwidth]{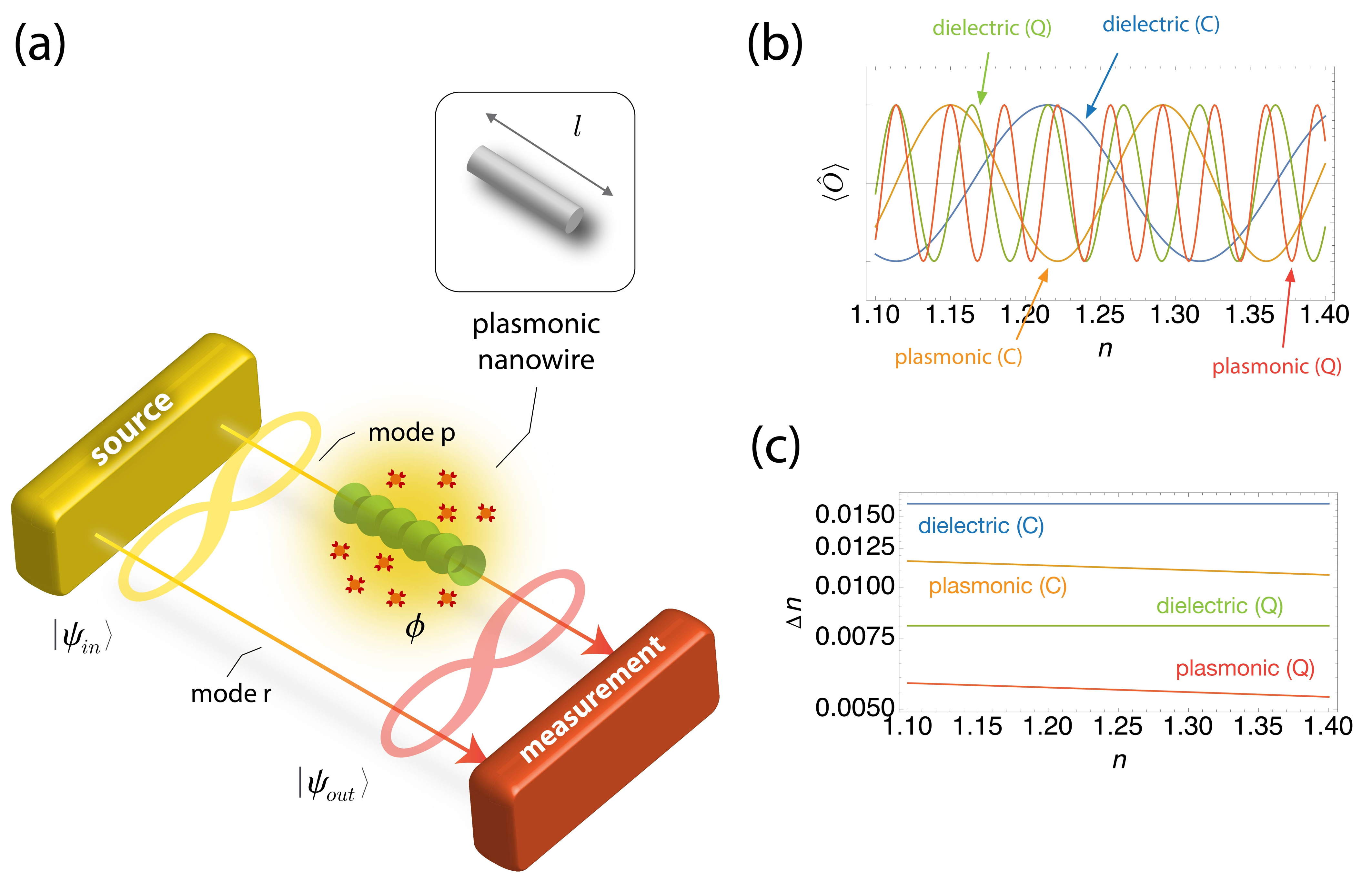}
\caption{\label{figps1} Quantum plasmonic phase sensing using a nanowire. (a) Two-mode scheme with a quantum source, plasmonic nanowire and measurement. (b) Measured signal for the optimal observable used ($\hat{O}=\hat{M}$ for classical (C) or~$\hat{A}$ for quantum (Q)) for a dielectric and plasmonic nanowire. (c) The precision of the refractive index~$\Delta n$ for classical (C) and quantum (Q) states with dielectric and plasmonic nanowires. The photon number on average is~$N=4$, the nanowire radius is~$50$~nm, the propagation length of the nanowire is~$4~\mu\text{m}$ and the free-space wavelength is~$\lambda_0=810$~nm. Reproduced with permission from ref~\citenum{Lee2016}. Copyright 2016 American Chemical Society.}
\end{figure*}

\subsection{Quantum plasmonic phase sensing}
\label{sec:qplasphasesense}

In addition to measuring changes in the refractive index of an optical sample using intensity sensing, it is also possible to use phase sensing, as outlined in section~\ref{sec:Intensity_vs_phase}. In the quantum regime this provides a complementary setting for quantum plasmonic sensing. In order to carry out quantum phase sensing using plasmonics, the physical geometry of the system needs to be modified so that changes in the refractive index are picked up in the phase of the optical signal, rather than the transmission amplitude. In this case, a metal nanowire is one example that can be considered and it provides a compact setting in which to perform sensing below the diffraction limit due to the highly confined field~\citep{Takahara97,Takahara09}. A nanowire also has the potential to be modified into more advanced nanophotonic circuitry for complex functionality~\cite{Bozhevolnyi2006,Gramotnev2010} and has already been considered for classical phase sensing~\cite{Homola1999a}. 

\subsubsection{Phase sensing with discrete variable states}
\label{sec:phasedesc}

The theoretical work by Lee \textit{et al.}~\cite{Lee2016} first studied the use of a nanowire in quantum phase sensing and the model considered is shown in Figure~\ref{figps1}(a). Here, a source produces a two-mode quantum state of light,~$\ket{\Psi_{\rm in}}$, and a silver nanowire is placed in one of the modes as a probe ($p$), while the other mode is used as a reference ($r$). A biological medium is considered to surround the nanowire and for a given change in a dynamical quantity, such as concentration, a change in the refractive index,~$n$, occurs which changes the relative phase between the probe and reference modes by~$\phi(n)=\ell \beta(n)$, where~$\ell$ is the length of the nanowire and~$\beta$ is the propagation constant in the nanowire, which is a function of~$n$. A measurement of the quantum state is then performed in order to obtain an estimate of the phase and therefore an estimate of the refractive index. The dynamical quantity in the biological medium, such as concentration, can then be estimated from the refractive index.

\begin{figure*}[t]
\centering
\includegraphics[width=0.75\textwidth]{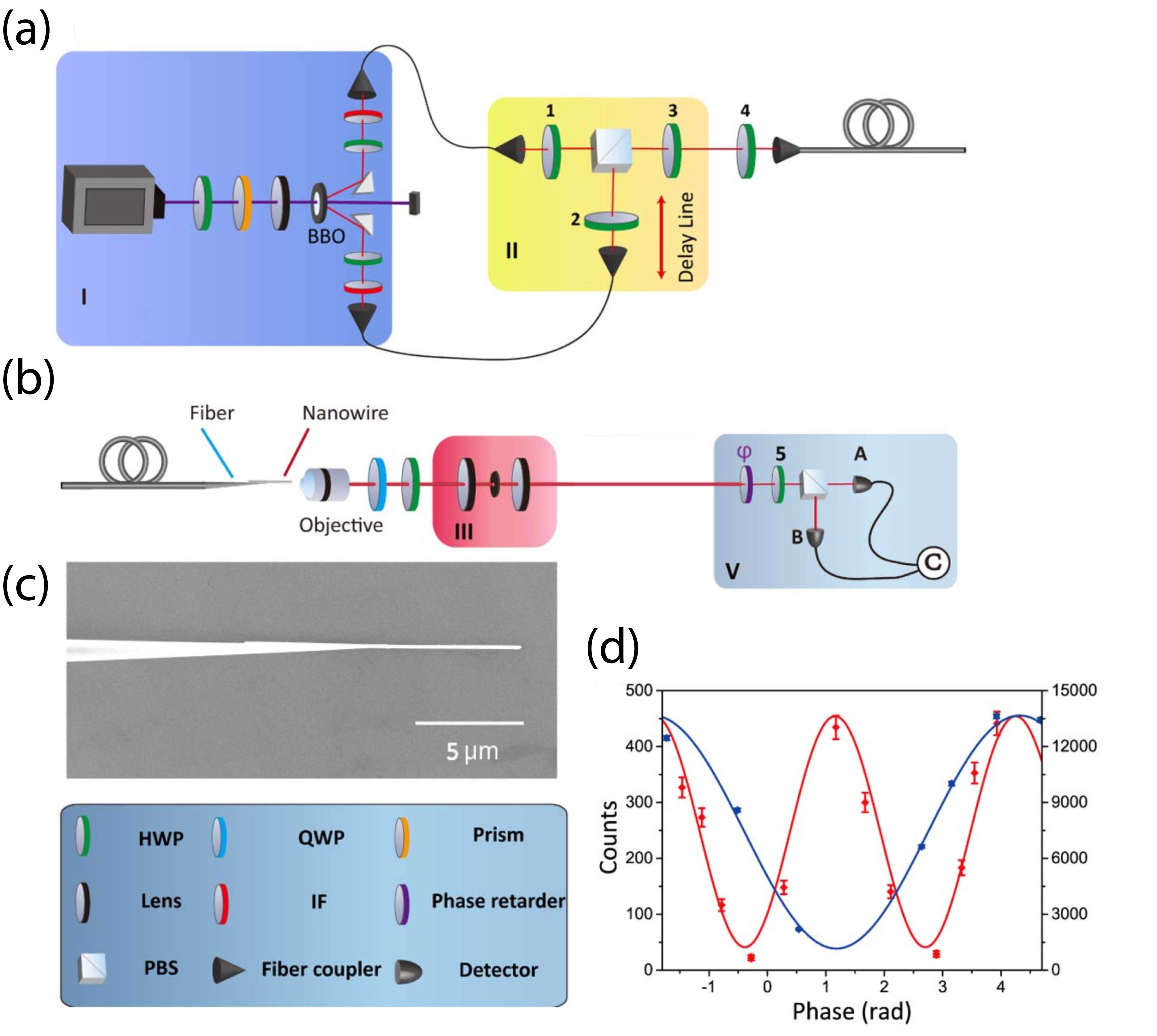}
\caption{\label{figps2} Experimental investigation of quantum plasmonic phase sensing using a silver nanowire. (a) Generation of polarization encoded NOON state ($N=2$) via parametric down-conversion. The two spatial modes of the state (1 and 2) are combined into a single mode (4), while maintaining the two-mode scenario in the polarization degree of freedom. The entangled photons in the single spatial mode are injected into an optical fiber. (b) The fiber is tapered with a silver nanowire fixed to the end. The out-coupled light is collected and sent to a measurement stage. (c) Scanning electron microscope image of the tapered fiber and nanowire. (d) Measurement signal for the two-photon NOON state (red) and classical case (blue). The y-axes on the left (right) is for the NOON state (classical case). Adapted with permission from ref~\citenum{Chen2018}. Copyright 2018 The Optical Society.}
\end{figure*}

In the ideal case when there is no loss, a NOON state is well known to be an optimal state for quantum phase sensing~\cite{Walther04,Crespi12,Afek2010}, as discussed in section~\ref{sec:Phase_sensing_with_NOON_states}, allowing one to reach the HL for the precision. It can be used as the two-mode quantum state produced by the source in Figure~\ref{figps1}(a), i.e.,~$\ket{\Psi_{\rm in}}=\frac{1}{\sqrt{2}}(\ket{N,0}_\text{p,r}+\ket{0,N}_\text{p,r})$, where~$N$ denotes the number of photons in a given mode. The relative phase in the nanowire is picked up by the first term in the state,~$\ket{N,0}\to e^{i N \phi(n)}\ket{N,0}$. The resulting state is measured using the optimal quantum observable~$\hat{A}=\ketbra{0,N}{N,0} +\ketbra{N,0}{0,N}$ (see section~\ref{sec:Phase_sensing_with_NOON_states}). This leads to a measurement signal~$\langle \hat{A} \rangle=A_0 \cos [N \phi(n)]$ and a corresponding precision in the estimation of the refractive index of~$\Delta n = \frac{1}{N}\left|\frac{\text{d} \phi}{\text{d}n}\right|^{-1}$, which is the HL for a single-shot measurement, i.e.,~$\nu=1$, as discussed in section~\ref{sec:Phase_sensing_with_NOON_states}. In contrast, the best classical strategy uses the two-mode classical coherent state~$\ket{\Psi_{\rm in}}=\ket{\alpha/\sqrt{2}}_p\ket{\alpha/\sqrt{2}}_r$ produced by the source, with~$|\alpha|^2=N$, which picks up the relative phase in the first term:~$\ket{\alpha/\sqrt{2}}_p \to \ket{e^{i \phi(n)} \alpha/\sqrt{2}}_p$. The optimal measurement consists of a BS and measurement of the observable~$\hat{M}=\hat{I}_p-\hat{I}_r$, which is the intensity-difference, as exploited in section~\ref{sec:Quantum_noise_reduction_intensity_measurements}. This leads to a measurement signal~$\langle \hat{M} \rangle=M_0 \cos [\phi(n)]$ and a corresponding precision~$\Delta n = \frac{1}{\sqrt{N}}\left|\frac{\text{d} \phi}{\text{d}n}\right|^{-1}$, which is the SNL or SQL. The NOON state therefore provides a~$\sqrt{N}$ improvement in the precision compared to the classical case for a given~$\nu$. The signals~$\langle \hat{A} \rangle$ and~$\langle \hat{M} \rangle$, and their associated precisions are shown in Figures~\ref{figps1}(b) and (c) for~$N=4$, where they are compared with the case in which a conventional dielectic nanowire is used. The results clearly show the benefit of using a plasmonic nanowire with a quantum source and measurement for phase sensing.

The improvement in the phase sensing precision comes from two distinct factors that are revealed by writing the precision (in this case the LOD) as~$\Delta n = \Delta \phi \left|\frac{\text{d} \phi}{\text{d} n}\right|^{-1}$, where~$\Delta \phi=\Delta O \left|\frac{\text{d}\langle \hat{O}\rangle}{\text{d} \phi}\right|^{-1}$ is the precision of the phase estimation and~$\hat{O}$ is the observable in the measurement ($\hat{O}=\hat{A}$ or~$\hat{M}$).~$\Delta \phi$ obviously depends on the observable~$\hat{O}$ and state~$\ket{\Psi_{\rm in}}$, but interestingly not on the nanowire properties, and is therefore quantum in origin. On the other hand,~$\left|\frac{\text{d} \phi}{\text{d} n}\right|$ represents the sensitivity,~$S_{\phi}$, which depends on the length and propagation constant [$\phi(n)=\ell \beta(n)$], i.e., the nanowire properties, and is therefore classical in origin. Both classical and quantum factors are needed to obtain a good precision, however, it is the quantum factor that improves the precision beyond the SQL, as discussed in section~\ref{sec:Limit_of_detection}.

In a more realistic scenario, losses in the nanowire must be considered. In this case, even for moderate losses, the NOON state is no longer the optimal quantum state. A more general analysis using the QFI,~$H$, must then be undertaken in order to find the optimal state to use. In this case, a more general state~$\ket{\Psi_{\rm in}}= \sum_{n=0}^{N} c_n \ket{n,N-n}_{p,r}$ can be considered~\cite{Dorner2009}, with the coefficients~$c_n$ optimized in order to maximize~$H$ and thus minimize the QCR bound~$\Delta \phi_{\rm min}=(\max_{\{ c_n\}}H)^{-1/2}$, which in turn minimizes the estimation uncertainty~$\Delta n$. In Lee \textit{et al.}~\cite{Lee2016} it was found that plasmonic sensors using these optimized states provide a precision beyond the SIL, which is the SNL of standard interferometers when loss is present~\cite{Dorner2009} and discussed in section~\ref{sec:quantum_phase_sensing}, even for a moderate amount of loss in a nanowire. Results were obtained for arbitrary~$N$, where the precision of the optimized states was shown to always be an improvement over the SIL precision.

While the use of the QFI enables an optimization of the quantum state used by the source, it is not clear what the optimal measurement is that allows one to reach the lower bound in the estimation uncertainty of~$\Delta \phi_{\rm min}$, i.e., the QCR bound introduced in section~\ref{sec:parameter_estimation_theory}. Further work in this direction will be needed to uncover practical measurement schemes for discrete variable quantum phase sensing in plasmonics. Furthermore, the use of different types of plasmonic material~\cite{Tassin2012} for the nanowire could be considered in order to reduce loss while maintaining the high field confinement, including the use of metamaterials~\cite{Cai2010,Soukoulis2011,Chen2012,Meinzer2014}, graphene~\cite{Grigorenko2012} and hybrid material systems~\cite{Ozbay2006,Sorger2012,Fang2015b}.

The use of a plasmonic nanowire for phase sensing has recently been experimentally investigated for the simple case of a two-photon NOON state ($N=2$) generated by parametric down-conversion~\cite{Chen2018}, as shown in Figure~\ref{figps2}(a). The two-mode state used takes the form~$\ket{\Psi_{\rm in }}=\frac{1}{\sqrt{2}}(\ket{2_H,0}_{1,2}+\ket{0,2_V}_{1,2})$, where~$\ket{2_H}$ ($\ket{2_V}$) corresponds to two horizontal (vertical) polarized photons in a given spatial mode. The polarization dependence of the photons enables the two spatial modes to be combined into a single spatial mode in an optical fiber, while maintaining an effective `two-mode' setting -- both polarization `modes' co-propagate in the same spatial mode. The vertical (horizontal) polarization represents the probe (reference) mode. The optical fiber is then tapered and a silver nanowire attached to the end, which enables the efficient excitation of SPPs and is shown in Figure~\ref{figps2}(b) and (c). The output of the signal is collected by a microscope objective and sent to a measurement stage with optical elements and single-photon detectors for the measurement of the phase.

In the experiment, the phase was modified using a liquid crystal phase retarder outside the nanowire in order to gain an understanding of the performance of the nanowire for phase sensing in a controlled manner. While not a direct demonstration, the approach is equivalent to the case that the phase is picked up directly in the nanowire, as phase shift operations commute with the loss accumulated in the nanowire~\cite{Demkowicz2015}. A more direct demonstration is yet to be performed, however, this is more challenging due to the requirement of a ligand coating along one of the axes of the nanowire in order for only one polarization mode to interact with the biochemical substance being sensed. Fortunately, this can be achieved in a number of ways, for example using nanografting~\cite{Salaita2007,Leggett2012}. By modifying the phase of the probe mode, then converting the probe and reference modes back into spatial modes and measuring the coincidences, a measurement equivalent to that of the observable~$\hat{A}$ can be performed. The signal of this measurement is shown in Figure~\ref{figps2}(d) as the red curve, along with the corresponding classical signal as the blue curve, equivalent to the observable~$\hat{M}$. The NOON state signal clearly shows an oscillation that occurs over a phase twice as small as the classical case, as expected -- a phenomenon known as `super resolution'~\cite{Resch2007}. 

The coincidence counts in Figure~\ref{figps2}(d) can be normalized to give a probability of a coincidence~\cite{Crespi12}, according to eq~\ref{eq:prob_coincidence} in section~\ref{sec:Phase_sensing_with_NOON_states}, i.e.,~$p_\text{coin}(\phi)=f_2(1+{\cal V}_2 \cos{2 \phi})/2$, where~$f_2$ represents the total proportion of photons that lead to a two-photon coincidence and~${\cal V}_2$ is the two-photon visibility. This coincidence probability can then be used to infer the uncertainty in the measurement of the phase,~$\Delta \phi_\text{coin}$. In this realistic and lossy setting, the SIL for the precision in estimation of the phase is given by~\cite{Demkowicz2009,Thomas-Peter2011}~$\Delta \phi_{\rm SIL}=1/\sqrt{N \eta}$, which is simply the SNL of standard interferometers in the case of no loss multiplied by the factor~$1/\sqrt{\eta}$, where~$\eta$ is the total loss factor. In order to show an improvement, the uncertainty in the experiment must satisfy~$\Delta \phi_\text{coin} < \Delta \phi_{\rm SIL}$, after which it can be called `super sensitive'~\cite{Thomas-Peter2011}. Using the normalized probability function~$p_\text{coin}(\phi)$, the bound can be reformulated as~\cite{Chen2018}~$1<2f_2^2{\cal V}_2^2/\eta$, which can be inverted to give a threshold visibility of~${\cal V}_2^\text{(th)}=\sqrt{\eta/2f_2^2}$. A visibility above this value shows an improvement in the precision beyond the SIL. In the absence of loss,~$f_2=1$ and~$\eta=1$, leading to~${\cal V}_2^\text{(th)}=2^{-1/2}\approx 0.707$. In the experiment, the visibility from the data shown in Figure~\ref{figps2}(d) was calculated to be~${\cal V}_2\approx 0.880$, clearly above the threshold. However, once loss is included,~${\cal V}_2^\text{(th)}$ rises. It was concluded that even if~${\cal V}_2=1$ in the experiment, the threshold could not be surpassed without a reduction in loss in the setup. Fortunately, it was shown that in principle the loss could be reduced by improving collection efficiencies at the various components in the setup, while still accommodating for loss in the plasmonic nanowire. Thus, the precision could be improved beyond the SIL with further experimental improvements.

It is clear from the above discussion that various technical challenges must be overcome for a complete experimental demonstration of quantum phase sensing using a plasmonic system and discrete variable quantum states. Reducing losses in the components of the setup, implementing the phase change at the nanowire itself, as well as the potential use of more robust-to-loss states other than the NOON state are all avenues for further study to address the challenges. These would enable the realization of practical nanoscale quantum plasmonic interferometric sensors.

In light of these challenges, it is interesting to note that the generation of plasmonic NOON states has also been achieved in several other recent experiments. By building on previous demonstrations of two-plasmon quantum interference~\cite{Heeres2013,DiMartino2014,Fakonas2014,Cai2014,Fujii2014}, Fakonas \textit{et al.}~\cite{Fakonas2015} generated NOON states with $N=1$ and $N=2$, and studied the impact of decoherence on these states in low-loss dielectrically loaded SPP waveguides. They found that for the type of waveguide studied, decoherence in the form of phase damping had only a minimal impact on the visibility of the interference fringes. Taken together with the low-loss nature of the waveguides, this demonstrates the ability for plasmonic waveguides to generate and maintain coherence well, highlighting their potential use for practical quantum phase sensing. Other types of plasmonic waveguides have also been used to generate NOON states with $N=2$, including planar~\cite{Vest2018} and stripe~\cite{Dieleman2017} waveguides.

By considering larger photon number NOON states ($N>2$) it may then be possible to demonstrate a Heisenberg scaling in the precision and obtain close to a $\sqrt{N}$ improvement compared to the classical case. Several photonics experiments have already shown how to generate NOON states with $N=3$ photons~\cite{Mitchell04}, $N=4$ photons~\cite{Walther04,Nagata2007} and $N=5$ photons~\cite{Afek2010,Israel2012}. Most recently, the generation of NOON-like states up to $N=8$ photons has been demonstrated~\cite{Thekkadath2020}. Many theoretical schemes have also been proposed for going beyond these experiments to generate NOON states with a higher photon number~\cite{Kok2002,Lee2002,Dowling2008,Afek2010} and other related states~\cite{Huver2008,Tae-Woo_Lee2009,Dorner2009}. In addition to the source of quantum states, e.g., NOON states, and the plasmonic system being used, e.g., plasmonic nanowire, the experimental resources required include single-photon and/or photon number resolving detectors, a coincidence counting interface and quantum statistical measurement acquisition software~\cite{Chen2018}.

\subsubsection{Phase sensing with continuous variable states} 

Despite the substantial developments in classical interferometric plasmonic sensing and the progress in interferometric quantum plasmonic sensing with discrete variable quantum states described above, very little progress has been made in the development of squeezed interferometric plasmonic sensors.  One reason for this is the increased loss in interferometric sensors.  Intensity-based plasmonic sensors are most sensitive at the inflection point of the plasmon absorption spectrum, so the detrimental effect of the plasmon absorption can be mitigated somewhat.  Phase-based plasmonic sensors are most sensitive at the maximum of the plasmon absorption spectrum~\cite{kabashin2009phase}, so as described in section~\ref{sec:Intensity_vs_phase}, little squeezing would remain in a squeezed interferometric plasmonic sensor after accounting for plasmonic absorption.  Nevertheless, some progress has been made in the characterization of plasmons exhibiting squeezing in the phase-sum quadrature, notably in a report that characterized the two-mode squeezing in intensity-difference and phase-sum quadratures for LSPs supported in triangular nanoapertures~\cite{Holtfrerich2016}. Further, substantial progress has been made in the development of squeezed interferometric sensors for other applications~\cite{lawrie2019quantum}. Moreover, as described in section~\ref{sec:nonlinear_interferometer}, nonlinear interferometers offer favorable scaling with loss compared with conventional squeezed light readouts, and could plausibly be integrated with plasmonic sensors in order to obtain a quantum advantage.  Truncated nonlinear interferometers in particular offer favorable loss scaling and the ability to use a high power local oscillator in order to improve shot-noise limited precision without concern for photochemical or photothermal effects~\cite{pooser2020truncated,anderson2017phase,gupta2018optimized}.  Many opportunities still exist to develop loss-resilient squeezed interferometric plasmonic sensors based on lower loss plasmonic sensor designs and new approaches to SU(2) and SU(1,1) interferometry.

\subsection{Quantum plasmonic sensing based on emitter-plasmon coupling} \label{sec:Quantum_plasmonic_sensors_with_emitter}

Outside the field of plasmonic sensing, much attention has been paid to the plasmonic control of emitters~\cite{Hummer2013,Kristensen2014,Lodahl2015,Pelton2015,lawrie2012plasmon, Goncalves2020b}, including semiconductor quantum dots~\cite{Buckley2012a} and color centers~\cite{Aharonovich2011} in the weak and strong coupling regimes~\cite{Tame2013,Hummer2013}. It is increasingly evident that these effects can be leveraged for alternative approaches to quantum plasmonic sensing, building on the PEF and SERS sensors described in section~\ref{sec:PEFandSERS}. The optical field at metallic nanostructures can be significantly enhanced via the excitation of LSPs, enabling strong coupling with matter in the vicinity of metallic structures~\cite{Tame2013}. The effects that originate from strong coupling are inherently nonclassical~\cite{Hummer2013}, and therefore the quantitative analysis and understanding of the various phenomena measured in experiments requires a full-quantum mechanical description, whereby the plasmonic nanostructures supporting LSP excitations can be treated as a lossy and small-mode-volume resonator in cavity quantum electrodynamics~\cite{Hummer2013,torma2014strong}. A result of this is that strong coupling is very sensitive to various structural and material parameters of an analyte being placed in the proximity of a metallic structure, which has led to several types of quantum plasmonic sensors.

\begin{figure*}[t]
\centering
\includegraphics[width=15cm]{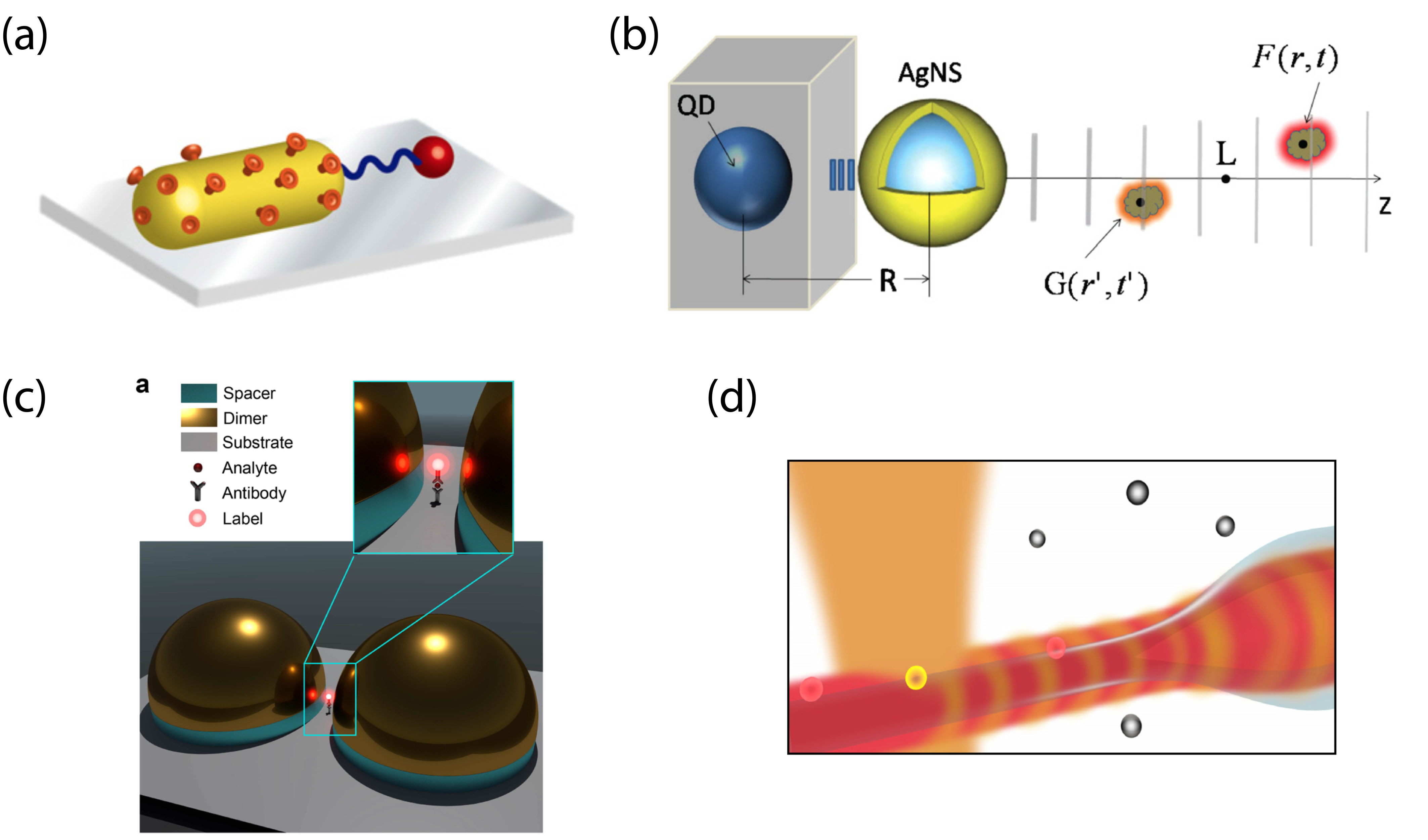}
\caption{\label{figaltsens} Quantum plasmonic sensors based on emitter-plasmon coupling. (a) A single quantum nanosensor in a to-the-end configuration, where a gold nanorod and quantum dot are functionalized. Reproduced with permission from ref~\cite{Hatef2013}. Copyright 2013 IOP Publishing. (b) Illustration of quantum detection and ranging. The antenna includes a quantum dot and a metallic nanoshell.~$F(r,t)$ and~$G(r',t')$ refer to two different time-dependent emission intensity patterns of an optically active nanoscale system at two different locations. Reproduced with permission from ref~\cite{Sadeghi2013a}. Copyright 2013 AIP Publishing. (c) An illustration of a strong-coupling immunoassay setup. A gold hemisphere nanodimer cavity captures an immunoassay complex in the proximity of the plasmonic hotspot. Reproduced with permission from ref~\cite{kongsuwan2019quantum}. Copyright 2019 American Chemical Society. (d) A nanofiber with dark-field heterodyne illumination. Nanoparticles in a droplet of ultrapure water are detected when entering the probe beam waist next to the nanofiber. Reproduced from ref~\cite{Mauranyapin2017}. Copyright 2017 Springer Nature.}
\end{figure*}

In one of the first examples of a quantum sensor based on emitter-plasmon coupling, a coupled system of a quantum dot and metallic nanorod was studied as a means to sense a small change in the local refractive index around the system in the near-infrared regime~\cite{Hatef2013}, as shown in Figure~\ref{figaltsens}(a). The quantum phase-dependent changes in the coherent exciton-plasmon coupling provide the capability of detecting and distinguishing adsorption or detachment of target molecules. A similar structure has also been considered for optical detection and recognition of single biological molecules~\cite{Sadeghi2013b}. Adsorption of a specific molecule to the nanorod results in the ultrafast upheaval of coherent dynamics of the system, that turns off the blockage of energy transfer between the quantum dot and the nanorod. The emission of the system is thus strongly modified depending on the adsorption event. Measuring the emission pattern from a quantum dot and metallic nanoshell system would also enable the remote detection of the spatial coordinates and movement of biological molecules or nanostructures~\cite{Sadeghi2013a}, as shown in Figure~\ref{figaltsens}(b). Refractive index sensing is also possible by analysing the Rabi-splitting spectrum of the exciton-plasmon strong coupling in the gap between a plasmonic nanorod and plasmonic nanowire~\cite{Qian2020}. In another study, a full quantum mechanical theory was developed to model how a small amount of absorbing Trinitrotoluene (TNT) molecules influences the spectrum of a graphene spaser based on a graphene flake with quantum dot emitters~\cite{Nechepurenko2018}. All these studies of the emission spectrum from an emitter-plasmon quantum dynamical system for sensing purposes open up a promising direction for future work, although a careful analysis is required in order to determine whether the quantum effects being exploited provide a quantum advantage in terms of sub-shot-noise sensing, an improvement in the sensitivity, or both.

A related recent theory proposal has suggested the use of quantum dots as quantum labels bonded to individual antibody-antigen-antibody complexes being placed inside or close to a nanoplasmonic dimer~\cite{kongsuwan2019quantum}, as shown in Figure~\ref{figaltsens}(c). As the surface density of the analyte-emitter complexes changes, the extinction cross section spectrum is modified, which not only causes spectrum shifting, but also brings multiple resonance peaks due to the strong coupling between the emitters and LSP at the nanoplasmonic dimer. Through a statistical study of multiple analyte-emitter complexes for the theoretical simulation of realistic conditions, the proposed splitting-type sensing approach using quantum emitter labels has been shown to achieve a nearly 15-fold sensitivity enhancement in comparison with conventional shifting-type label-free plasmonic sensors. Such a study demonstrates the potential of quantum plasmonic sensors to detect a single analyte, which is usually challenging since the size of an analyte (typically~$<100$~nm) is far smaller than the optical wavelength of the light exciting the system. 

Another theoretical work focusing on the strong coupling regime has considered a three-level quantum system, e.g., atom or quantum dot, that by interacting with a plasmonic cavity mode supports a so-called `embedded superstate'~\cite{Nefedkin2020}. The resulting system relies on quantum interference between multiple atomic transition pathways and strong coupling, and exhibits unboundedly narrow emission lines that enable a significant reduction in noise in the scattering spectra. The result thereby provides a means to improve the sensitivity significantly.

On the experimental front, metallic nanorods have been used in a recent study, where an optical nanofiber is immersed in a droplet of water containing nanoparticles, such as silica nanospheres or gold nanorods, and biomolecules, as shown in Figure~\ref{figaltsens}(d). Heterodyne interferometry was used together with a dark-field illumination approach~\cite{Mauranyapin2017}. The experiment demonstrated a quantum-noise limited precision of evanescent single-molecule biosensing, allowing a reduction of four orders of magnitude in the optical intensity that is required to maintain state-of-the-art sensitivity. In a related theoretical study, the fraction of the total spontaneous emission energy from an emitter coupled to SPPs, called the~$\beta$-factor~\cite{Chen2010}, has been investigated for two nitrogen vacancy centers (NVCs) in diamond placed in a plasmonic waveguide~\cite{Rangani_Jahromi2018}. It was shown that a maximally entangled state of NVCs and a product NVC state provide the optimal estimation of the~$\beta$-factor at initial times and at long times, respectively. 

Finally, the use of color center emitters for quantum sensing of magnetic~\cite{Balasubramanian2008,Schirhagl2014} and electric~\cite{Dolde2011} fields has experienced significant attention in the past few years. The use of a plasmonic system as a mediator of the sensing signal is an interesting development. A recent experiment has demonstrated the use of a plasmonic groove waveguide interacting with multiple NVCs, which are positioned at the end of the groove waveguide milled in a thick gold film~\cite{Shalaginov2020}. The gold film carries the microwave control signal for the NVCs, while the groove waveguide acts as a fluorescence collector. The fluorescence signal enables the readout of the NV spin population (or specifically a spin sub-level population) as a function of an applied external magnetic field. This is known as optically detected magnetic resonance (ODMR). It is particularly appealing due to its simplicity and room-temperature operation, as well as scope for also measuring strain or temperature fields.

A related theoretical work has proposed the use of a plasmonic metasurface as a means to achieve an infrared absorption-based readout of the magnetic field resonance for an ensemble of NVCs~\cite{Kim2020a}. This is important as standard fluorescence-based ODMR techniques are limited by low-photon collection efficiency and modulation contrast. The work exploits the enhancement of an infrared probe field via plasmonic excitation, which interacts with an NVC infrared singlet state transition, and whose absorption then carries information about spin state populations that change as a magnetic field is applied. The result is a new kind of microscopic ODMR sensor with infrared readout providing enhanced sensitivity.

\section{Perspective and outlook}

In this review we covered the background and latest developments in the emerging field of quantum plasmonic sensing. This is a research field that sits at the interface between plasmonic sensing and quantum metrology -- the former provides researchers working in the field with decades of extensive knowledge in classical optical sensing across a wide array of plasmonic systems, their commercial development and applications, while the latter provides new concepts and methods intensively developed in recent years that enable the performance of the sensors to be improved by exploiting quantum mechanical features in various quantum systems. 

The review started with a discussion of conventional plasmonic sensors and their basic working principles. The excitation of SPPs was shown to enable enhanced sensing compared to standard optical sensors due to their high electromagnetic field confinement. Two main types of plasmonic sensing were discussed -- intensity and phase sensing -- with the benefits of each described as depending on the specific application. This was followed by a discussion of estimation theory and the limits of sensing in plasmonic systems using classical light, i.e., the SNL. We then showed how the use of quantum states of light and quantum measurements enables sensing beyond this SNL, where we elaborated on single- and two-mode sensing schemes while reviewing recent works on multimode or multiple parameter sensing. Recent works combining these quantum sensing techniques with plasmonic sensing were then reviewed. We covered the basic theory behind the work and highlighted its motivation in relation to applications, including biosensing, monitoring chemical reactions and ultrasound sensing. It was shown that despite the presence of loss in plasmonic systems, one can use techniques from quantum sensing to obtain improvements in sensing performance.

The field of quantum plasmonic sensing has grown steadily over the last five years, with researchers studying many ways in which to combine plasmonic and quantum sensing. However, much work still needs to be done to address the current challenges and bring quantum plasmonic sensors to the same level of maturity as their classical counterparts, including successful commercialization. In the short term (next five years), it is likely that plasmonic sensors based on SPP sensing will provide the most direct and accessible route for incorporating the quantum intensity and phase sensing techniques discussed in section~\ref{sec:Quantum_plasmonic_sensors}. SPP sensing using the Kretschmann configuration is still the main approach used by researchers and in industry. Advancements made in the experimental generation of quantum states of light, such as NOON states and TMSD states, will help bring quantum SPP sensing to maturity. New results from the field of quantum sensing for dealing with loss~\cite{Oh2017,Nair2018,Degen2017,Braun2018,Pirandola2018}, including error correction~\cite{Arrad2014,Kessler2014,Cohen2016,Layden2019,Zhuang2020,Shettell2021} and novel resource states~\cite{Demkowicz2015,Wang2019,Shettell2020,Ouyang2019}, may lead to a widening in the range of plasmonic systems that can be used for quantum sensing and their related applications. 

In the mid to long term (five to ten years) the next steps would be the miniaturization and integration of the sensors. Research currently being done in the field of classical plasmonic sensors~\cite{Hoa2007,Caucheteur2015,Niu2015,Li2020e}, including the use of fibers, waveguides and nanoparticles, would then enable compact quantum plasmonic sensors to be realized for deployment in commercial applications. Furthermore, non-standard material systems, including metamaterials~\cite{Cai2010,Soukoulis2011,Chen2012,Meinzer2014,Wang2019b,Consales2020a,Hassan2020}, graphene~\cite{Grigorenko2012, Rodrigo2015,Heydari2020} and more exotic two-dimensional materials~\cite{Das2015,Zeng2018} may help reduce loss, further improve the level of precision in applications and enable a range of specificity to be achieved, such as in monitoring chemical and biological reactions~\cite{Homola2008}, food safety~\cite{Mello2002}, pathogen detection~\cite{Leonard2003,Pejcic2006}, and environmental monitoring~\cite{Andreescu2004}. Hybrid systems that exploit electro-optic~\cite{Ozbay2006,Sorger2012,Fang2015b}, nonlinear~\cite{Kauranen2012}, nonlocal~\cite{Raza2015}, quantum size effects~\cite{Halperin1986,Fitzgerald2016} and electron tunelling~\cite{Zhu2016} could also offer additional quantum functionality. Another interesting direction is the incorporation of quantum emitters into plasmonic systems, such as quantum dots~\cite{Lodahl2015,Pelton2015} and color centers~\cite{Aharonovich2011}, which may bring further applications, such as magnetic~\cite{Balasubramanian2008,Schirhagl2014} and electric field sensing~\cite{Dolde2011}, and molecular spectroscopy~\cite{Zhan2018}. There is also the natural connection between sensing and imaging~\cite{Rotenberg2014}, and it remains to be seen how the techniques developed for quantum plasmonic sensing can be translated to plasmonic imaging~\cite{Zhang2008,Kawata2009,Willets2017} in order to improve aspects such as image resolution, feature extraction and pattern recognition for applications in the life sciences and medicine. 

The field of quantum plasmonic sensing is likely to develop into a rich subfield of optical science and it is an exciting time to enter this emerging research field. Multidisciplinary collaboration from researchers working in plasmonics, quantum information science, material physics, chemistry, biology and medicine will advance this field significantly, bringing with them new opportunities for sensing in science and industry. For these prospective studies, this review will provide a helpful guidance and inspire novel avenues of research.

\appendix

\section*{Appendix}

\section{Multiparameter QFIM}\label{appendix:A}
\setcounter{equation}{0}
\renewcommand{\theequation}{A\arabic{equation}}
The QFI matrix,~$\boldsymbol{H}$, in eq~\ref{eq:QCRinequality_multi} is defined by
\begin{align}
[\boldsymbol{H}]_{jk}=\text{Tr}[\hat{\rho}_{\boldsymbol{x}} (\hat{\cal L}_j \hat{\cal L}_k+\hat{\cal L}_k \hat{\cal L}_j)/2],
\end{align} 
with~$\hat{\cal L}_j$ being a symmetric logarithmic derivative operator associated with~$j$th parameter~$x_j$~\cite{Braunstein1994a}. Here,~$\boldsymbol{F}^{-1}$ and~$\boldsymbol{H}^{-1}$ are understood as the inverse on their support if the matrices are singular, i.e., not invertible~\cite{Ge2018}.

For single parameter estimation, the optimal measurement setting reaching the QCR bound given by eq~\ref{eq:QCRinequality} always exists and can be constructed with projectors onto the eigenvectors of the SLD operator~\cite{Paris2009, Braunstein1994a, Demkowicz2015}. For multiparameter estimation, on the other hand, the QCR bound has been known to be attained only if the SLD operators commute, i.e.,~$[\hat{\cal L}_j, \hat{\cal L}_{k}]=0$ for all~$j, k$, which is valid even when the generators do not commute, i.e.,~$[\hat{G}_j(x_{j}),\hat{G}_k(x_{k})]\neq0$.
This is the sufficient condition for the saturability of the multiparameter QCR bound, in which case the optimal set of POVMs can be constructed over the common eigenbasis of the commuting SLD operators~\cite{Baumgratz2016}. 
A weaker, but necessary and sufficient condition for the saturability can be found for pure states~$\hat{\rho}_{\boldsymbol{x}} = \ket{\Psi(\boldsymbol{x})}\bra{\Psi(\boldsymbol{x})}$, written as~\cite{Matsumoto2002,Szczykulska2016}
\begin{align}
\bra{\Psi(\boldsymbol{x})}(\hat{\cal L}_j \hat{\cal L}_k-\hat{\cal L}_k \hat{\cal L}_j) \ket{\Psi(\boldsymbol{x})}=0,
\label{eq:saturability_weak_cond}
\end{align}
for all~$j, k$. The condition of eq~\ref{eq:saturability_weak_cond} implies the commutation relation among the SLD operators only on average with respect to the state~$\ket{\Psi(\boldsymbol{x})}$. For commuting generators~$\hat{G}_j(x_j)$ associated with the evolution of a pure probe state, i.e., ~$[\hat{G}_j(x_j),\hat{G}_k(x_k)]=0$ for all~$j,k$,
eq~\ref{eq:saturability_weak_cond} is satisfied, so that the multiparameter QCR bound can be saturated~\cite{Matsumoto2002}.

\section{Calculation of multiparameter QFIM}\label{appendix:B}
\setcounter{equation}{0}
\renewcommand{\theequation}{B\arabic{equation}}
Generally for~$M$ parameters~$\boldsymbol{\phi}=\{\phi_1, \cdots,\phi_M\}$, the relation is given by~\cite{Liu2020}
\begin{align}
\sum_{j,k} H_{jk} \text{d}\phi_j\text{d}\phi_k = 4 {\cal D}_\text{B}^{2}(\hat{\rho}_{\boldsymbol{\phi}}, \hat{\rho}_{\boldsymbol{\phi}+\text{d}\boldsymbol{\phi}}),
\label{eq:Relation_Bures_QFIM}
\end{align} 
where the Bures distance can be written in terms of the quantum fidelity as 
\begin{align}
{\cal D}_\text{B}^{2}(\hat{\rho}_{\boldsymbol{\phi}}, \hat{\rho}_{\boldsymbol{\phi}+\text{d}\boldsymbol{\phi}})
= 2\left[ 1-\sqrt{{\cal F}(\hat{\rho}_{\boldsymbol{\phi}}, \hat{\rho}_{\boldsymbol{\phi}+\text{d}\boldsymbol{\phi}})}\right]
\label{eq:Bures}
\end{align}
and the quantum fidelity~${\cal F}$ is defined as~\cite{Uhlmann1976,Jozsa1994}
\begin{align}
{\cal F}(\hat{\rho}_{\boldsymbol{\phi}}, \hat{\rho}_{\boldsymbol{\phi}+\text{d}\boldsymbol{\phi}}) = \left(\text{Tr}\sqrt{\sqrt{\hat{\rho}_{\boldsymbol{\phi}}} \hat{\rho}_{\boldsymbol{\phi}+\text{d}\boldsymbol{\phi}}\sqrt{\hat{\rho}_{\boldsymbol{\phi}}}}\right)^2.
\end{align}
Thus, the calculation of the quantum fidelity leads to the calculation of QFIM for any pair~$(\phi_j, \phi_k)$.

In particular, the input state~$\vert \Psi\rangle_\text{in}=\vert \alpha\rangle_\text{a}\vert \xi\rangle_\text{a}$ considered in the MZI of Figure~\ref{Fig:Sec3:Basic_Scheme}(b) is called a Gaussian state as its characteristic Wigner function can be described by a Gaussian distribution. Because the BS operation~$\hat{B}$, the phase operation~$\hat{U}(\phi)$, and linear photonic loss channel, including inefficient detectors, are a Gaussian map, the output state~$\hat{\rho}_\text{out}$ is kept in the form of a Gaussian state. This enables us to describe the output state~$\hat{\rho}_\text{out}$ in terms of the first-order moment vector~$\boldsymbol{d}$, and the second-order moment matrix~$\boldsymbol{V}$~\cite{Braunstein2005,Weedbrook2012}. 
The displacement vector~$\boldsymbol{d}$ is defined as~$d_{j}=\text{Tr}[\hat{\rho}_{\boldsymbol{\phi}}\hat{Q}_{j}]$, 
whereas the covariance matrix~$\boldsymbol{V}$ is defined by~$V_{jk}=\text{Tr}[\hat{\rho}_{\boldsymbol{\phi}}\{ \hat{Q}_{j}-d_{j},\hat{Q}_{k}-d_{k}\}/2]$, where~$\{\hat{A},\hat{B}\}\equiv \hat{A}\hat{B}+\hat{B}\hat{A}$. Here,~$\hat{\boldsymbol{Q}}$ denotes a quadrature operator vector for a two-mode continuous variable quantum system and is written as~$\hat{\boldsymbol{Q}}=(\hat{x}_1,\hat{p}_{1},\hat{x}_{2},\hat{p}_{2})^\text{T}$, satisfying the canonical commutation relation,~$[\hat{Q}_j,\hat{Q}_k]=i\Omega_{jk}$, where
\begin{align}
\boldsymbol{\Omega}=\begin{pmatrix} 0 & 1 \\ -1 & 0 \end{pmatrix} \times \mathbb{I}_2
\end{align} 
and ~$\mathbb{I}_n$ is the~$n\times n$ identity matrix.

Due to the input state remaining a Gaussian state throughout the evolution, the analytical form of the quantum fidelity can be found for the displacement vectors and covariance matrices~\cite{Marian2012, Banchi2015}, so one can readily calculate the quantum fidelity. For two Gaussian states~$\hat{\rho}_{j=1,2}$ with~$\boldsymbol{d}_{j=1,2}$ and~$\boldsymbol{V}_{j=1,2}$, the quantum fidelity can be calculated via~\cite{Marian2012, Banchi2015}
\begin{align}
{\cal F}(\hat{\rho}_{1}, \hat{\rho}_{2})=\frac{\exp\left[-\frac{1}{2}\delta\boldsymbol{d}^{\text{T}}(\boldsymbol{V}_1+\boldsymbol{V}_2)^{-1}\delta\boldsymbol{d}\right]}{(\sqrt{\Gamma}+\sqrt{\Lambda}) - \sqrt{(\sqrt{\Gamma}+\sqrt{\Lambda})^2-\Delta}},
\label{eq:Quantum_Fidelity_Guassian}
\end{align}
where~$\Delta = \text{det}(\boldsymbol{V}_1+\boldsymbol{V}_2)$,~$\Gamma =16~\text{det}(\boldsymbol{\Omega} \boldsymbol{V}_1 \boldsymbol{\Omega} \boldsymbol{V}_2 -\mathbb{I}_4/4)$,~$\Lambda =16~\text{det}(\boldsymbol{V}_1+i\boldsymbol{\Omega}/2)\text{det}(\boldsymbol{V}_2+i\boldsymbol{\Omega}/2)$, and~$\delta\boldsymbol{d}=\boldsymbol{d}_2-\boldsymbol{d}_1$.

Using eqs~\ref{eq:Relation_Bures_QFIM},~\ref{eq:Bures} and~\ref{eq:Quantum_Fidelity_Guassian}, one can calculate the QFIM~$\boldsymbol{H}$ generally for multiple parameters~$\boldsymbol{\phi}=\{\phi_1,\cdots,\phi_M\}$. 
In the case when only a single parameter~$\phi$ encoded by~$\hat{U}(\phi)$ is of interest, the QFI is calculated by
$H=8\left[ 1-\sqrt{{\cal F}(\hat{\rho}_{\phi}, \hat{\rho}_{\phi+\text{d}\phi})}\right]/\text{d}\phi^2$.

\section{CR bound in lossy MZIs}\label{appendix:C}
\setcounter{equation}{0}
\renewcommand{\theequation}{C\arabic{equation}}
Homodyne detection is particularly useful, as the measurement bases are represented by Gaussian states, i.e., the POVM element is written as~$\hat{\Pi}_x=\vert x_{\phi_\text{HD}}\rangle \langle x_{\phi_\text{HD}} \vert$, where~$\vert x_{\phi_\text{HD}}\rangle$ denotes a quadrature variable state. Projection of the reduced output state~$\hat{\rho}_\text{out,a}=\text{Tr}_\text{b}[\hat{\rho}_\text{out}]$ into the measurement basis~$\vert x_{\phi_\text{HD}}\rangle$ leads to the probability density function of the quadrature variable outcomes~$\{x\}$, which is written as
\begin{align}
p(x\vert \phi) = \langle x_{\phi_\text{HD}} \vert \hat{\rho}_\text{out,a} \vert x_{\phi_\text{HD}}\rangle. 
\label{eq:p_xy}
\end{align}
It can be shown that the probability density function of eq~\ref{eq:p_xy} follows a Gaussian distribution for the measurement outcomes, leading to homodyne detection often being called a Gaussian measurement~\cite{Weedbrook2012, Giedke2002}. Hence, the probability density function of eq~\ref{eq:p_xy} can be described by the first-order moment vector~$\tilde{\boldsymbol{d}}$ and the second-order moment covariance matrix~$\tilde{\boldsymbol{V}}$, for which the FI can be calculated via~\cite{Kay1993,Porat1986}
\begin{align}
F(\phi) &= \frac{\partial \tilde{\boldsymbol{d}}^\text{T}}{\partial \phi } \tilde{\boldsymbol{V}}^{-1} \frac{\partial \tilde{\boldsymbol{d}}}{\partial \phi } +\frac{1}{2}\text{Tr}\left[ \tilde{\boldsymbol{V}}^{-1}\frac{\partial \tilde{\boldsymbol{V}}}{\partial \phi }\tilde{\boldsymbol{V}}^{-1}\frac{\partial \tilde{\boldsymbol{V}}}{\partial \phi }\right].
\end{align}
For an optimally chosen homodyne angle~$\phi_\text{HD}$ with respect to the phase~$\phi$ being estimated, one can show that the CR bound reads~\cite{Gard2017}
\begin{align}
\Delta \phi_\text{CR}=\frac{1}{\sqrt{\nu}}\sqrt{\frac{1}{\vert \alpha\vert^2 e^{2r}}+\frac{1-\eta}{\eta \vert \alpha \vert^2}}.
\label{eq:QFI_phase_lossy_quantum2}
\end{align}



\section*{Acknowledgments}
At KIT, this work was partially supported by the Volkswagen Foundation and by the VIRTMAT project. At KIAS, CL was supported by a KIAS Individual Grant (QP081101) via the Quantum Universe Center. At ORNL, BL was supported by the U. S. Department of Energy, Office of Science, Basic Energy Sciences, Materials Sciences and Engineering Division. At HU, KGL was supported by the Basic Science Research Program through the National Research Foundation (NRF) of Korea and funded by the Ministry of Science and ICT (Grants No.  2020R1A2C1010014) and Institute of Information \& Communications Technology Planning \& Evaluation (IITP) grant funded by the Korea government (MSIT) (No. 2019-0-00296). At SU, MST was supported by the South African National Research Foundation, the Council for Scientific and Industrial Research National Laser Centre and the South African Research Chair Initiative of the Department of Science and Innovation and National Research Foundation.

\section*{Glossary}

\begin{itemize}

\item Bias - The difference between an estimate $x_\text{est}$ and the true value $x$ of the parameter being estimated on average, i.e., $\bar{x}_\text{est}-x$. An unbiased estimator has a bias of zero.

\item Cram\'er-Rao bound - The lower bound of the Cram\'er-Rao inequality.

\item Cram\'er-Rao inequality - The inequality that the standard deviation of an unbiased estimator of a parameter should obey for a given measurement setting. The inequality reads $\Delta x \geq 1/\sqrt{\nu F(x)}$, where $\Delta x$ is the standard deviation, $\nu$ is the number of measurements in a sample and $F(x)$ is the Fisher information. The equality holds when the optimal unbiased estimator is chosen. It can also be asymptotically saturated in the limit of large $\nu$ when maximum-likelihood-method is employed as an estimator although it is not optimal. 

\item Estimate - The value $x_\text{est}$ obtained from the estimator $\hat{x}(y_1,...y_\nu)$ for a given sample (single set of data observed). It is also known as a point estimate.

\item Estimation accuracy - The interpretation of the bias $\bar{x}_\text{est}-x$. The accuracy becomes better as the bias decreases. 

\item Estimation precision - The interpretation of the standard deviation, $\Delta x$, of the estimate $x_\text{est}$, each taken from a sample made up of a finite set of $\nu$ measurements of the parameter $x$. The smaller $\Delta x$ is, the better the precision is. It is commonly known as the fluctuation or uncertainty in estimation, or can sometimes be interpreted as the resolution. It can also be understood as the reproducibility.

\item Estimation uncertainty - The quantity directly given by the standard deviation $\Delta x$. The smaller $\Delta x$ is, the smaller the uncertainty is. 

\item Estimator - A rule, $\hat{x}$, that yields an estimate $x_\text{est}$ of an unknown parameter $x$ from an underlying probability distribution based on the observed data $\boldsymbol{y}$ from a sample of $\nu$ measurements, $\boldsymbol{y}\equiv (y_1,...y_\nu)$. For example, $\hat{x}$ can be expressed by some function $f$, i.e.,  $\hat{x}=f(\boldsymbol{y})$. The most commonly used estimator is the sample mean. 

\item Extraordinary optical transmission - the transmission of light through a structured medium, where the transmission is due to the excitation of surface plasmons. A similar structured medium that does not support surface plasmons has a reduced transmission.

\item Fisher information - A quantification of the amount of information about a parameter $x$ contained in the measurement results for a given measurement setting. We express it by $F(x)$.

\item Heisenberg limit - The ultimate quantum limit achievable using the optimal quantum state and measurement, or the precision scaled with $N^{-1}$ that is often the case of interferometeric sensing. A scaling of $N^{-1}$ is called Heisenberg scaling, where $N$ is the average number of particles in the resource.

\item Limit of detection - An overall figure of merit for sensing quality that takes into account both the sensitivity~${\cal S}_y$ and the minimum detectable range~$\Delta y_\text{min}$, or equivalently the value of the noise level. This is often interpreted as a resolution.

\item Mean-squared-error - The average square distance between the estimator $\hat{x}(y_1,...y_\nu)$ and true value of $x$ as the data $y_1,...y_\nu$ varies according to the underlying probability distribution. It is a measure of average closeness of an estimator $\hat{x}$ to the true value $x$, and depends on both the standard deviation $\Delta x$ and the bias $\bar{x}_\text{est}-x$. If the estimator is unbiased, then the mean-square error is equivalent to $\Delta x^2$, allowing to refer to the standard deviation $\Delta x$ as the `estimation error', `estimation precision', or simply `precision'.

\item Multi-parameter estimation - This is categorized into three kinds:
\begin{enumerate}
\item Individual estimation - Multiple parameters $(x_1, x_2, \cdots, x_n)$ are estimated individually. This scheme is also called a `local strategy'. 

\item Simultaneous estimation - Multiple parameters $(x_1, x_2, \cdots, x_n)$ are estimated simultaneously. This scheme is also called a `global strategy'. 

\item Distributed estimation - A global parameter $X$, which is a function of multiple parameters, i.e.,  $X=f(x_1, x_2, \cdots, x_n)$, is estimated.
\end{enumerate}

\item Noise reduction factor - In an intensity difference measurement, it is the ratio of the variance of the photon number difference between the signal and reference mode for a given quantum state to that of coherent states with matching average photon number.

\item $N$-mode - A sensing scheme where $N$ spatial/temporal/angular/spectral modes -- the eigenvectors of the wave equation -- are used. Here, $N$ can be ``single'', ``two'', or ``multi''. 

\item Quantum Cram\'er-Rao bound - The lower bound of the quantum Cram\'er-Rao inequality.

\item Quantum Cram\'er-Rao inequality - The same as Cram\'er-Rao inequality but with the Fisher information replaced by quantum Fisher information, i.e.,  $\Delta x \ge1/\sqrt{\nu H}$. 

\item Quantum Fisher information - The maximized Fisher information over all possible physical measurement settings $\{ \hat{\Pi}\}$, i.e.,  $H=\max_{\{\hat{\Pi}\}} F(x)$.

\item Sensitivity - The derivative of a quantity $y$ being monitored, which is used to obtain an estimation of some parameter $x$ that changes with $y$. Mathematically, the sensitivity is $S_y = |{\rm d}y /{\rm d} x|$. By increasing the sensitivity a sensor becomes more sensitive to changes in $x$.

\item Shot noise - Pure noise originating from the underlying statistical properties of a resource to be analyzed or a measurement involving the random fluctuation of electric signals. It has nothing to do with fundamental noise, which in an ideal theory model can be assumed to be completely absent. The term originally comes from electronics, where a current consists of a stream of discrete charges, i.e.,  electrons. The charges are randomly distributed in space and time, thereby following a Poisson distribution. The same random feature is present in a coherent state of light that consists of discretized particles, i.e.,  photons, that follow a Poisson distribution in photon number. This leads to the shot noise when the coherent state's intensity, i.e.,  photon number, is measured.

\item Shot-noise limit - The best possible precision obtained using the classical resource of a coherent state. It originates from statistical noise in the resource that is mapped into noise in the quantity being measured, {\it e.g.} for an intensity measurement the shot-noise limit comes from Poissonian noise that is intrinsic to the coherent state's photon number, or energy. 

\item Single-to-noise ratio - The ratio of the measurement signal to the noise in the measurement signal.

\item Standard quantum limit - The fundamental limit to precision achievable in standard interferometers made of standard devices employing only classical resources. The term is now generally used to denote the ultimate precision limit achievable in arbitrary interferometric sensing when only classical resources are used. 

\item Unbiased estimator - The estimator has a bias of zero. When applied to many samples from the underlying probability distribution it equals the true value of the parameter being estimated on average. Equivalently, the sampling distribution of the estimator fluctuates around the true value of the parameter.

\end{itemize}

\bibliography{referencev2}

\end{document}